\documentclass[11pt]{article}
\usepackage[T1]{fontenc}
\usepackage{amssymb,amsmath,amsfonts,dsfont,amsthm,bm,bbm}
\usepackage{float,rotfloat} 
\usepackage{mathrsfs}
\usepackage{multirow}
\usepackage{enumerate}
\usepackage{enumitem}
\usepackage{mathtools}
\usepackage{hyperref,url}
\usepackage{graphicx}
\usepackage[hang,small,bf]{caption}
\usepackage[top=1in, bottom=1in, left=1.0in, right=1.0 in]{geometry}
\usepackage{longtable}
\usepackage{color,xcolor}
\usepackage[round,sort,comma]{natbib}
\usepackage{url}

\usepackage[title]{appendix}
\usepackage{etoolbox}
\usepackage[english]{babel}
\usepackage[autostyle, english = american]{csquotes}
\MakeOuterQuote{"}
\usepackage{orcidlink}

\usepackage[linesnumbered,ruled,vlined]{algorithm2e}

\patchcmd{\appendices}{\quad}{: }{}{} 

\newtheorem{thm}{Theorem}
\newtheorem{lemma}{Lemma}

\newtheorem{prop}{Proposition}
\newtheorem{cor}{Corollary}
\newtheorem{note}{Note}

\newcommand{\E}{\mathbb{E}}

\newcommand{\Var}{\mathbb{V}ar}

\newcommand{\ol}{\overline}
\newcommand{\lb}{\left[}
\newcommand{\rb}{\right]}
\newcommand{\lp}{\left(}
\newcommand{\rp}{\right)}

\definecolor{darkblue}{rgb}{0,0,0.55}

\hypersetup{,colorlinks=true,allcolors=darkblue}

\allowdisplaybreaks

\begin{document}   
	\let\nofiles\relax
	
	\baselineskip 5mm
	
	\thispagestyle{empty}
	
	\begin{center}
		{\Large 
			Robust and Computationally Efficient \\[8pt] 
			Trimmed $L$-Moments Estimation for Parametric Distributions
		}
		
		\vspace{5mm}
		
		{\large\sc
			Chudamani Poudyal\footnote[1]{{\sc Corresponding Author}: 
				Chudamani Poudyal, PhD, ASA, 
				is an Assistant Professor 
				in the Department of Statistics and Data Science,
				University of Central Florida, 
				Orlando, FL 32816, USA. 
				~~ {\em e-mail\/}: ~{\tt Chudamani.Poudyal@ucf.edu}}}%
		\orcidlink{0000-0003-4528-867X}
		
		\vspace{0mm}
		
		{\large\em University of Central Florida}
		
		\vspace{3mm}
		
		{\large\sc
			Qian Zhao\footnote[2]
			{Qian Zhao, PhD, ASA,
				is an Associate Professor 
				in the Department of Mathematics, 
				Robert Morris University,
				Moon Township, 
				PA 15108, USA. 
				~~{\em e-mail\/}: ~{\tt zhao@rmu.edu}}}% 
		\orcidlink{0009-0009-9476-9112}  
		
		\vspace{0mm}
		
		{\large\em Robert Morris University}
		
		\vspace{3mm}
		
		{\large\sc
			Hari Sitaula\footnote[3]
			{Hari Sitaula, PhD,
				is an Assistant Professor 
				in the Department of Mathematical Sciences, 
				Montana Technological University,
				Butte, 
				MT 59701, USA. 
				~~{\em e-mail\/}: ~{\tt Hsitaula@mtech.edu}}}  
		
		\vspace{0mm}
		
		{\large\em Montana Technological University}
		
		\vspace{5mm}
		
		\copyright \
		Copyright of this Manuscript is held by the Authors! 
		
		\vspace{0.05in}
		
	\end{center}
	
	\vspace{0.02in}
	
	\begin{quote}
		{\bf\em Abstract\/}.
		This paper proposes a robust and computationally efficient
		estimation framework for fitting parametric distributions
		based on trimmed \( L \)-moments. 
		Trimmed \( L \)-moments extend classical \( L \)-moment 
		theory by downweighting or excluding extreme order statistics, 
		resulting in estimators that are less sensitive to outliers 
		and heavy tails. 
		We construct estimators for both location-scale and shape 
		parameters using asymmetric trimming schemes tailored
		to different moments, 
		and establish their asymptotic properties
		for inferential justification 
		using the general structural theory of \( L \)-statistics, 
		deriving simplified single-integration expressions 
		to ensure numerical stability. 
		State-of-the-art algorithms are developed to resolve the 
		sign ambiguity in estimating the scale parameter for 
		location-scale models and the tail index for the Fr{\'e}chet model. 
		The proposed estimators offer improved efficiency 
		over traditional robust alternatives for selected
		asymmetric trimming configurations, 
		while retaining closed-form expressions for 
		a wide range of common distributions,
		facilitating 
		fast and stable computation. 
		Simulation studies demonstrate strong finite-sample
		performance. 
		An application to financial claim severity
		modeling highlights the practical relevance
		and flexibility of the approach.
		
		\vspace{4mm}
		
		{\bf\em Keywords\/}. 
		Computational efficiency; 
		Location-scale models;
		$L$-statistics; 
		Robust estimation; 
		Tail index estimation;
		Trimmed $L$-moments.
	\end{quote}
	
	\newpage
	
	\setcounter{page}{1} 
	
	\baselineskip 7mm
	
	\section{Introduction}
	\label{sec:Introduction}
	
	Robust and computationally efficient parameter estimation is essential in statistical learning,
	especially under data contamination,
	measurement error, or heavy tails.
	While the maximum likelihood estimator (MLE) is asymptotically efficient under correct model assumptions, it is well known to be highly sensitive to deviations from ideal conditions and to contamination in the data \citep{MR829458, MR2488795}. These issues frequently arise in real-world settings such as biomedical studies, engineering reliability analysis, and environmental monitoring, where irregularities and extreme values are often present. Robust statistics provides a formal framework for addressing such challenges by developing estimators that maintain good performance even when standard assumptions are violated. These procedures aim to reduce the influence of atypical data while preserving efficiency, ensuring stability and interpretability in practical applications.
	
	The theory of robust statistics has been systematically developed since the mid-1960s \citep{MR0120720}, with foundational work introducing methods to reduce the sensitivity of estimators to model misspecification and data contamination \citep{MR2488795}. Within this framework, the method of trimmed moments (MTM)—a subclass of $L$-moments \citep{MR0203874}—has gained attention as a practical and computationally efficient alternative to likelihood-based estimation. MTM replaces classical moments with trimmed moments computed by excluding extreme order statistics, yielding estimators that are finite, stable, and robust to outliers. These estimators have been developed for various distributional families, including location-scale models, exponential-type distributions, and heavy-tailed models \citep{MR2497558, MR2591318, MR2994446}. 
	MTMs have also been applied in several domains:
	\citet{oc12} used them in operational risk modeling, 
	\citet{MR3081460} applied them in credibility theory, 
	and \citet{hyl14} used them to construct bootstrap confidence 
	intervals for reliability data.
	Their application to incomplete loss data has been further examined by \citet{MR4263275} and \citet{MR4602526}, demonstrating that MTM provides an effective framework for enhancing robustness by mitigating the influence of heavy point masses at left truncation and right censoring boundaries \citep{MR4887633}.
	
	Despite the existence of general asymptotic distributional results for MTM, independently established by \citet{MR0203874} and \citet{MR2497558}, and their equivalence formally shown by \citet{cp24}, most existing implementations adopt a single trimming configuration uniformly across all moments, regardless of the number of parameters to be estimated or the distributional features of the data. This constraint can lead to different or suboptimal subsets of data being used for estimating each moment, reducing efficiency or discarding relevant information. To address this limitation, 
	\citet{cp24} developed computationally tractable expressions for the asymptotic distribution of MTM estimators that allow for distinct trimming proportions for each moment. 
	This extension maintains the foundational asymptotic structure of $L$-statistics \citep{MR0203874} while allowing practitioners to achieve a more effective robustness-efficiency trade-off tailored to the nature of the parameter being estimated.
	
	Therefore, this work introduces a general and flexible framework for robust estimation using trimmed $L$-moments that allows for \emph{distinct trimming proportions across different moments}. 
	
	This framework tailors trimming to moment-specific sensitivity, accommodating asymmetry where needed and enabling a more effective balance between robustness and efficiency.
	The proposed framework includes previously studied symmetric-trimming methods as special cases, while extending the modeling flexibility to better adapt to real-world data structures. We develop closed-form estimators and their asymptotic distributional properties for location-scale and Fr{\'e}chet models under the general $L$-statistic framework, along with computational expressions for analytic variance. The estimators do not require iterative optimization, making them computationally efficient and scalable to large datasets. The explicit nature of the estimators ensures numerical stability and reproducibility—key requirements for scalable statistical methods in scientific computing and industrial analytics. We also address the sign ambiguity in scale and shape estimation through a principled selection strategy based on proximity to the full-sample estimate. The proposed methods are validated through extensive simulation studies and real-data applications, confirming their strong finite-sample performance and practical relevance.
	
	The remainder of the paper is organized as follows. 
	Section~\ref{sec:TW} introduces the general $L$-estimator framework and derives its theoretical properties, specifically focusing on MTM. 
	Section~\ref{sec:ParExamples} is the section where we implement 
	the designed methodology for specific parcomprehensive simulation study, 
	and Section~\ref{sec:RealData} provides real-data applications. 
	Section~\ref{sec:Conclusion} concludes with a discussion and future directions.
	
	\section{General Method of Trimmed
		{\em L}-Moments}
	\label{sec:TW}
	
	Consider a random variable 
	$X$ with the cdf 
	$F(x \, | \, \bm{\theta})$,
	where 
	$
	\bm{\theta}
	=
	\lp 
	\theta_{1}, 
	\ldots, 
	\theta_{k}
	\rp,
	$
	for some positive integer $k$,
	is the parameter vector to be estimated. 
	Consider a random sample 
	$X_{1}, \ldots, X_{n}$ of size $n$
	with the corresponding order 
	statistics 
	$X_{1:n} \le \ldots \le X_{n:n}$. 
	Then the {\em method of trimmed moments} (MTM)
	estimators of 
	$\theta_{1}, \theta_{2},...,\theta_{k}$
	are found as follows:
	
	\begin{itemize}
		\item 
		Compute the sample trimmed moments 
		\begin{equation} 
			\label{eq:sample_mtmG}
			\widehat{T}_{j} 
			=
			\frac{1}{n-\lfloor na_{j} \rfloor - \lfloor nb_{j} \rfloor}
			\sum_{i=\lfloor na_{j} \rfloor+1}^{n-\lfloor nb_{j} \rfloor}{h_{j}(X_{i:n})},
			\quad 1 \leq j \leq k.
		\end{equation}
		The $h_{j}'s$ in (\ref{eq:sample_mtmG}) 
		are specially chosen functions for
		mathematical convenience and are typically 
		specified by the data analyst.
		For a detailed discussion, 
		we refer the reader to 
		\citet{MR2497558} and \citet{MR4263275},
		The proportions 
		\( 0 \le a_{j}, b_{j} \le 1 \) 
		should be selected based on the
		desired balance between efficiency and robustness.
		
		\item
		Compute the corresponding population trimmed moments 
		\begin{equation} 
			\label{eq:pop_mtmG}
			T_{j} 
			=
			\frac{1}{1-a_{j}-b_{j}}\int_{a_{j}}^{\ol{b}_{j}}
			H_{j}
			\lp 
			u
			\rp \, du,\ \ \ 1\leq{j}\leq{k},
			\quad 
			\overline{v} := 1-v, \ 
			v \in [0,1].
		\end{equation} 
		In (\ref{eq:pop_mtmG}), $F^{-1}(u|\bm{\theta})=\inf\:\{x:F(x|\bm{\theta})\geq{u}\}$ is the quantile function, 
		and $H_{j} := h_{j} \circ F^{-1}$.
		
		\item 
		Match the sample and population trimmed moments from (\ref{eq:sample_mtmG}) and (\ref{eq:pop_mtmG}) to get the following system of equations for $\theta_{1},\theta_{2},...,\theta_{k}:$
		\begin{equation} 
			\label{eq:match_mtm2}
			\left\{
			\begin{array}{lcl}
				T_1 (\theta_1, \ldots, \theta_k) 
				& = & 
				\widehat{T}_1, \\
				& \vdots & \\
				T_k (\theta_1, \ldots, \theta_k) 
				& = & 
				\widehat{T}_k.  \\
			\end{array} \right.
		\end{equation}
	\end{itemize}
	
	A solution, say 
	$\widehat{\bm{\theta}}_{\text{\tiny T}}
	=
	\lp 
	\widehat{\theta}_{1},\widehat{\theta}_{2},...,\widehat{\theta}_{k}
	\rp$, 
	if it exists, 
	to the system of equations (\ref{eq:match_mtm2}) 
	is called the {\textit{method of trimmed moments (MTM)}} 
	estimator of $\bm{\theta}$. 
	Thus, $\widehat{\theta}_{j}
	=:
	g_{j}
	\left(
	\widehat{T}_{1},\widehat{T}_{2},..., \widehat{T}_{k}
	\right)$, $1\leq{j}\leq{k}$, 
	are the MTM estimators of 
	$\theta_{1},\theta_{2},...,\theta_{k}$.
	
	Asymptotically, 
	Eq.~\eqref{eq:sample_mtmG} is equivalent to a general 
	structure of $L$-statistics 
	under the condition 
	\( 0 \le a_{j} < \ol{b}_{j} \le 1 \) with 
	\( a_{j} + b_{j} < 1 \); 
	see \citet[][p.~264]{MR595165}.
	Specifically,
	\begin{align}
		\label{eqn:S1}
		\widehat{T}_{j} 
		& :=
		\dfrac{1}{n}
		\sum_{i=1}^{n}
		J_{j} \left( \dfrac{i}{n+1} \right) 
		h_{j}(X_{i:n}),
		\quad 
		1 \le j \le k,
	\end{align}
	with the specified weights-generating function:
	\begin{align}
		\label{eqn:J_Fun1}
		J_{j}(s) 
		& =
		\left\{ 
		\begin{array}{ll}
			(1-a_{j}-b_{j})^{-1}; & a_{j} < s < \ol{b}_{j}, \\
			0; & \text{otherwise}. \\
		\end{array}
		\right. 
		\quad 
		1 \le j \le k,
	\end{align}
	Similarly, Eq.~\eqref{eq:pop_mtmG} is
	equivalent to 
	\begin{align}
		T_{j}
		& \equiv 
		T_{j} 
		\left( 
		\bm{\theta}
		\right) 
		\equiv 
		T_{j} 
		\left( 
		\theta_1, \ldots, \theta_k
		\right) 
		= 
		\int_{0}^{1}
		J_{j}(u)H_{j}(u) \, du.
		\label{eqn:P1}
	\end{align}
	
	We define 
	\begin{align}
		\label{eqn:MeanVectors1}
		\widehat{\bm T}
		& :=
		\left( 
		\widehat{T}_{1}, 
		\widehat{T}_{1}, 
		\ldots, 
		\widehat{T}_{k}
		\right)
		\quad 
		\mbox{and}
		\quad 
		\bm{T}
		:=
		\left(
		T_{1}, 
		T_{2}, 
		\ldots, 
		T_{k}
		\right).
	\end{align}
	
	As sample statistics,
	the vector \(\widehat{\bm{T}}\) is expected to converge 
	in distribution to the corresponding population parameter 
	\(\bm{T}\). 
	In total,
	there are six possible combinations of trimming 
	proportions \((a_{i}, b_{i})\) and \((a_{j}, b_{j})\) 
	for \(1 \leq i, j \leq k\); 
	see \citet{cp24} for a detailed discussion. 
	Among these,
	we focus on the case defined 
	by the following inequality:
	\begin{align}
		\label{eqn:abCondition1}
		0 
		\le 
		a_{j} 
		\le 
		a_{i} 
		< 
		\ol{b}_{j} 
		\le 
		\ol{b}_{i} 
		\le 
		1.
	\end{align}
	
	Under the trimming inequality \eqref{eqn:abCondition1}, 
	and following the general asymptotic distributional 
	results of \citet{MR0203874}, 
	the asymptotic distribution of the vector
	\(\widehat{\bm{T}}\), 
	along with the computational expressions
	developed by \citet{cp24}, 
	is summarized in Theorem~\ref{thm:MTM_Var1}.
	The result is based on the following
	two integral quantities and a kernel function.
	
	Following \citet{cp24},
	for \( 1 \le i \le k \) and 
	\( 0 \le a \le b \le 1 \), 
	define
	\begin{align}
		\label{eqn:I_Integrals1}
		\begin{cases}
			I_{i}(a,b) 
			: = 
			b H_{i}(b) 
			-
			a H_{i}(a)
			- 
			\int_{a}^{b} H_{i}(v) \, dv, \\[5pt] 
			\ol{I}_{i}(a,b)
			: = 
			\ol{b} H_{i}(b) 
			-
			\ol{a} H_{i}(a)
			+ 
			\int_{a}^{b} H_{i}(v) \, dv,
		\end{cases}
	\end{align} 
	and define the kernel function \( K(w,v) \) as
	\begin{align}
		\label{eqn:kFun1}
		K(w,v) 
		& := 
		K(v,w)
		= 
		\min \{w, v \} - wv,
		\quad 
		\mbox{for} 
		\quad 
		0 \le w, v \le 1.
	\end{align}
	
	\begin{thm}
		\label{thm:MTM_Var1}
		With the trimming proportions satisfying 
		inequality \eqref{eqn:abCondition1}, 
		it follows that 
		\begin{align}
			\label{eqn:MTMAsym1}
			\sqrt{n}
			\lp 
			\widehat{\bm{T}} - \bm{T}
			\rp 
			& \sim 
			{\cal{AN}}
			\left( 
			\mathbf{0}, 
			\mathbf{\Sigma}_{\text{\tiny T}}
			\right), \
			\mathbf{\Sigma}_{\text{\tiny T}}
			= 
			\left[
			\sigma_{ij}^{2}
			\right]_{i,j=1}^{k},  \ 
			\sigma_{ij}^{2}
			=
			\Gamma(i,j) \, V(i,j),
		\end{align}
		where 
		\begin{eqnarray}
			\Gamma 
			& \equiv &
			\Gamma(i,j) 
			= 
			\prod_{r = i, j}
			\left(
			1-a_{r}-b_{r}
			\right)^{-1}, 
			\label{eqn:DefnGamma1} \\
			V(i,j) 
			& = &
			\int_{a_{i}}^{\ol{b}_{i}}\int_{a_{j}}^{\ol{b}_{j}}
			{K(v,w)} 
			H_{j}'(v) H_{i}'(w) \, dv \, dw
			\label{eqn:DefnV110} \\
			& = & 
			I_{j}(a_{j},a_{i}) \, 
			\ol{I}_{i}(a_{i},\ol{b}_{i})
			+
			b_{i} H_{i}(\ol{b}_{i}) 
			I_{j}(a_{i},\ol{b}_{j})
			-
			a_{i} H_{i}(a_{i})
			\ol{I}_{j}(a_{i},\ol{b}_{j})
			+ 
			\int_{a_{i}}^{\ol{b}_{j}}
			H_{i}(v) H_{j}(v) \, dv 
			\nonumber \\
			& & 
			+ \,
			\lb 
			\ol{b}_{j}H_{j}(\ol{b}_{j}) 
			-
			a_{i} 
			H_{j}(a_{i})
			\rb 
			\int_{\ol{b}_{j}}^{\ol{b}_{i}}
			H_{i}(v) \, dv
			- 
			\lb 
			a_{i} H_{j}(a_{i})
			+
			b_{j} H_{j}(\ol{b}_{j})
			\rb 
			\int_{a_{i}}^{\ol{b}_{j}}
			H_{i}(v) \, dv 
			\nonumber \\ 
			& & 
			-
			\lp 
			\int_{a_{i}}^{\ol{b}_{j}} 
			H_{j}(v) \, dv 
			\rp 
			\lp 
			\int_{a_{i}}^{\ol{b}_{j}} 
			H_{i}(v) \, dv 
			\rp
			-
			\lp 
			\int_{a_{i}}^{\ol{b}_{j}}
			H_{j}(v) \, dv
			\rp 
			\lp 
			\int_{\ol{b}_{j}}^{\ol{b}_{i}}
			H_{i}(v) \, dv
			\rp.
			\label{eqn:DefnV11}
		\end{eqnarray}
	\end{thm}
	
	Different trimming proportions for different moments 
	were used by \cite{MR2591318} to estimate the parameters 
	of generalized Pareto distributions. 
	While they approximated the entries 
	$\sigma_{ij}^{2}$ of the variance-covariance matrix 
	\( \mathbf{\Sigma}_{\text{\tiny T}} \) 
	directly from Eq.~\eqref{eqn:DefnV110}
	using a numerical bivariate trapezoidal rule 
	\citep[Appendix A.2]{MR2591318}, 
	our approach derives the simplified closed-form expression 
	in Eq.~\eqref{eqn:DefnV11}, 
	as presented in Theorem~\ref{thm:MTM_Var1}.
	
	\begin{note}
		\label{note:ChangeInequality1}
		If the trimming inequality 
		\eqref{eqn:abCondition1} 
		is replaced by  
		\begin{align}
			\label{eqn:abCondition1Flip}
			0 
			\le 
			a_{i} 
			\le 
			a_{j} 
			< 
			\ol{b}_{i} 
			\le 
			\ol{b}_{j} 
			\le 
			1,
		\end{align}
		then the asymptotic result in Theorem~\ref{thm:MTM_Var1} 
		remains still valid by simply interchanging the indices 
		\( i \) and \( j \).
		\qed 
	\end{note}
	
	Using the delta method 
	\citep[see, e.g.,][Theorem A, p. 122]{MR595165},
	along with 
	$\bm{\widehat{T}}
	=
	\left(\widehat{T}_{1},\ldots,\widehat{T}_{k} \right)$ 
	and
	$\theta_j 
	=
	g_j(T_1(\boldsymbol{\theta}),
	\ldots, 
	T_k(\boldsymbol{\theta})),
	$
	we present the following asymptotic result for
	$\widehat{\bm{\theta}}_{\text{\tiny T}}$.
	
	\begin{thm}[Delta method]
		\label{thm:CGJ3}
		The MTM-estimator of $\bm{\theta}$, 
		denoted by
		$\widehat{\bm{\theta}}_{\text{\tiny T}}$, 
		has the following asymptotic distribution: 
		\begin{eqnarray}
			\label{eqn:DM2}
			\widehat{\bm{\theta}}_{\text{\tiny T}}
			& = &
			\left(\widehat{\theta}_{1},
			\ldots,\widehat{\theta}_{k}\right) 
			\sim 
			\mathcal{AN}
			\left(\bm{\theta},
			\frac{1}{n}
			\bm{S}_{\text{\tiny T}}
			\right),
			\quad 
			\bm{S}_{\text{\tiny T}}
			:= 
			{\bm{D_{\text{\tiny T}}
					\Sigma_{\text{\tiny T}}
					D_{\text{\tiny T}}'}},
		\end{eqnarray}
		where the Jacobian $\bm{D}_{\text{\tiny T}}$ 
		is given by
		$
		\bm{D}_{\text{\tiny T}}
		=
		\left[\left. \frac{\partial g_{i}}{\partial \widehat{T}_{j}}\right\vert_{\widehat{\bm{T}}
			=
			\bm{T}}\right]_{k\times k} 
		=:
		\left[d_{ij}\right]_{k\times k}
		$ and the variance-covariance matrix 
		$\bm{\Sigma}_{\text{\tiny T}}$ has 
		the same form as in Theorem \ref{thm:MTM_Var1}.
	\end{thm}
	
	With the asymptotic distributional 
	properties from Theorem \ref{thm:CGJ3},
	the asymptotic performance of MTM-estimators 
	are assessed by computing their 
	asymptotic relative efficiency (ARE)
	in relation to the maximum likelihood estimator (MLE). 
	For a model parameterized by $k$ parameters, 
	the ARE is defined as \citep[see, e.g.,][]{MR595165}:
	\begin{equation} 
		\label{eqn:ARE1}
		ARE(\mathcal{C}, \text{MLE}) 
		=
		\left( 
		\dfrac{\det
			\left(\bm{\Sigma}_{\text{\tiny MLE}}\right)}
		{\det
			\left(\bm{\Sigma}_{\text{\tiny $\mathcal{C}$}}\right)}
		\right)^{1/k},
	\end{equation}
	where 
	$\bm{\Sigma}_{\text{\tiny MLE}}$
	and 
	$\bm{\Sigma}_{\text{\tiny $\mathcal{C}$}}$
	are the asymptotic covariance matrices of the MLE and the 
	MTM-estimator $\mathcal{C}$, respectively, with $\det$ denoting the determinant operation on a square matrix. The MLE serves as a reference due to its superior asymptotic efficiency regarding variance, contingent on specific regularity conditions being met. For additional insights, consult \cite{MR595165}, Section 4.1.
	
	\begin{note}
		\label{note:RobustnessBreakdown}
		With trimming proportions satisfying 
		\( 0 \le a_i \le \ol{b}_i \le 1 \),
		if \( a_i > 0 \), \( b_i > 0 \), and \( a_i + b_i < 1 \) for \( 1 \le i \le k \), 
		then the resulting estimators are globally robust with lower and upper breakdown points
		\[
		\text{LBP} = \min\{a_1, a_2, \ldots, a_k\} \quad \mbox{and} \quad 
		\text{UBP} = \min\{b_1, b_2, \ldots, b_k\}.
		\]
		These breakdown points quantify resistance to outliers: observations with order less than \( n \times \text{LBP} \) or greater than \( n \times (1 - \text{UBP}) \) are effectively excluded from the estimation procedure. 
		Such trimming ensures robustness against extreme values on both ends of the distribution. 
		For a rigorous treatment of breakdown points
		and robust estimation under heavy-tailed models, 
		see \citet{MR829458} and \citet{MR1987777}.
		\qed 
	\end{note}
	
	\section{Parametric Examples}
	\label{sec:ParExamples}
	
	In this section, 
	we derive general MTM estimators for the location and scale 
	parameters of broad location-scale families, 
	which are not necessarily symmetric, 
	while also highlighting the advantages of the general
	MTM framework when the distribution is symmetric about zero.
	We also obtain the entries of the corresponding asymptotic 
	variance-covariance matrix. 
	For numerical illustrations, 
	we consider the lognormal and Fr{\'e}chet distributions
	and present state-of-the-art algorithms designed 
	to resolve the sign ambiguity in estimating
	the scale parameter for location-scale models 
	and the tail index for the Fr{\'e}chet model.
	Additionally, 
	we evaluate the asymptotic relative efficiency (ARE) 
	of the MTM estimators with respect to the maximum 
	likelihood estimator (MLE), as defined in Eq.~\eqref{eqn:ARE1}.
	
	\subsection{Location Scale Model}
	\label{sec:LS1}
	
	Consider
	$X_{1}, X_{2}, \ldots, X_{n} \stackrel{iid}{\sim} X$,
	where $X$ is a location-scale random variable with the CDF
	\begin{align}
		\label{eqn:LC1}
		F(x) 
		& = 
		F_{0} 
		\left( 
		\dfrac{x - \theta}{\sigma}
		\right), 
		\quad 
		-\infty < x < \infty,
	\end{align}
	where $-\infty < \theta < \infty$ 
	and $\sigma > 0$ are, respectively, 
	the location and scale parameters of $X$,
	and $F_{0}$ is the standard parameter-free
	version of $F$, i.e., with $\theta = 0$ and $\sigma = 1$.
	The corresponding percentile/quantile function of $X$ is
	given by
	\begin{align}
		\label{eqn:LC2} 
		F^{-1}(u) 
		& = 
		\theta + \sigma F_{0}^{-1}(u).
	\end{align}
	Since we are estimating two unknown parameters, 
	$\theta$ and $\sigma$, 
	we equate the first two sample trimmed-moments with
	their corresponding population trimmed-moments.
	Further, knowing 
	$-\infty < \theta < \infty$ and $\sigma > 0$,
	we choose 
	\begin{align}
		\label{eqn:LS_hFun_Def1}
		h_{1}(x)
		& =
		x
		\quad 
		\mbox{and}
		\quad 
		h_{2}(x) 
		= 
		x^{2}.
	\end{align}
	
	From Eq.~\eqref{eqn:P1}, 
	we note that $H_{j} := h_{j} \circ F^{-1}$.
	Then, from Eq.~\eqref{eqn:LC2} and 
	Eq.~\eqref{eqn:LS_hFun_Def1},
	we have 
	\begin{eqnarray}
		& & 
		H_{1}(u) 
		=
		h_{1}
		\left(
		F^{-1}(u)
		\right) 
		= 
		F^{-1}(u)
		= 
		\theta + \sigma F_{0}^{-1}(u), 
		\label{eqn:LSH1} \\
		\implies 
		& & 
		dH_{1}(u) 
		= 
		\sigma \, d F_{0}^{-1}(u),
		\label{eqn:LSH1D} \\ 
		& & 
		H_{2}(u) 
		=
		h_{2}
		\left(
		F^{-1}(u)
		\right) 
		= 
		\theta^{2} 
		+ 
		2 \theta \sigma \, 
		F_{0}^{-1}(u) 
		+ 
		\sigma^{2} 
		\left[ F_{0}^{-1}(u) \right]^{2}, 
		\label{eqn:LSH2}\\
		\implies & & 
		dH_{2}(u) 
		= 
		2 \theta \sigma \, d F_{0}^{-1}(u)
		+ 
		2 \sigma^{2} F_{0}^{-1}(u) \, 
		dF_{0}^{-1}(u).
		\label{eqn:LSH2D}
	\end{eqnarray}
	
	With only two parameters,
	$\theta$ and $\sigma$, 
	to estimate,
	we focus on the trimming inequality 
	derived from \eqref{eqn:abCondition1}, 
	stated explicitly as:
	\begin{align}
		\label{eqn:abCondition2}
		0 
		\le 
		a_{2} 
		\le 
		a_{1} 
		< 
		\ol{b}_{2} 
		\le 
		\ol{b}_{1} 
		\le 
		1.
	\end{align}
	
	With the trimming proportions 
	$(a_{1}, b_{1})$ and $(a_{2},b_{2})$
	as given in \eqref{eqn:abCondition2},
	then from Eq.~\eqref{eq:sample_mtmG}, 
	the first two sample trimmed-moments 
	are given by:
	\begin{align}
		\label{eqn:S2}
		\begin{cases}
			\displaystyle 
			\widehat{T}_{1} 
			=
			\frac{1}{n-\lfloor na_{1} \rfloor - \lfloor nb_{1} \rfloor}
			\sum_{i=\lfloor na_{1} \rfloor+1}^{n-\lfloor nb_{1} \rfloor}{h_{1}(X_{i:n})}
			= 
			\frac{1}{n-\lfloor na_{1} \rfloor - \lfloor nb_{1} \rfloor}
			\sum_{i=\lfloor na_{1} \rfloor+1}^{n-\lfloor nb_{1} \rfloor}
			X_{i:n}, \\[15pt]
			\displaystyle 
			\widehat{T}_{2} 
			=
			\frac{1}{n-\lfloor na_{2} \rfloor - \lfloor nb_{2} \rfloor}
			\sum_{i=\lfloor na_{2} \rfloor+1}^{n-\lfloor nb_{2} \rfloor}{h_{2}(X_{i:n})}
			= 
			\frac{1}{n-\lfloor na_{2} \rfloor - \lfloor nb_{2} \rfloor}
			\sum_{i=\lfloor na_{2} \rfloor+1}^{n-\lfloor nb_{2} \rfloor}
			X_{i:n}^{2}.
		\end{cases}
	\end{align}
	
	The corresponding first two population 
	trimmed-moments using Eq.~\eqref{eq:pop_mtmG}
	takes the form:
	\begin{align}
		\label{eqn:P2}
		\begin{cases}
			T_{1}
			\equiv 
			T_{1} 
			\left( 
			\theta, \sigma
			\right)  
			= 
			\displaystyle 
			\frac{1}{1-a_{1}-b_{1}}
			\int_{a_{1}}^{\ol{b}_{1}}
			H_{1}
			\lp 
			u
			\rp \, du
			= 
			\theta + \sigma c_{1}(a_{1},\ol{b}_{1}), \\[10pt]
			T_{2}
			\equiv 
			T_{2} 
			\left( 
			\theta, \sigma
			\right)  
			= 
			\displaystyle 
			\frac{1}{1-a_{2}-b_{2}}
			\int_{a_{2}}^{\ol{b}_{2}}
			H_{2}
			\lp 
			u
			\rp \, du
			=
			\theta^{2} 
			+ 
			2 \theta \sigma c_{1}(a_{2},\ol{b}_{2}) 
			+ 
			\sigma^{2} c_{2}(a_{2}, \ol{b}_{2}),
		\end{cases}
	\end{align}
	where 
	\begin{align}
		\label{eqn:ConstantC1}
		c_{k}(a,b) 
		& \equiv 
		c_{k}\left( F_{0}, a, b \right)
		=
		\dfrac{1}{b-a}
		\int_{a}^{b}  
		\left[ F_{0}^{-1} (u) \right]^{k} du, 
		\quad 
		k \ge 1.
	\end{align}
	
	Equating 
	\( T_{1} = \widehat{T}_{1} \) 
	and 
	\( T_{2} = \widehat{T}_{2} \),
	and solving the resulting system of equations, 
	we obtain the explicit expressions for 
	\( \theta \) and \( \sigma \) as:
	\begin{align}
		\label{eqn:MTMSol1}
		\left\{
		\begin{array}{lll}
			\widehat{\theta}_{\text{\tiny T}} 
			& = &
			\widehat{T}_{1} 
			- 
			c_{1}(a_{1},\ol{b}_{1})
			\widehat{\sigma}_{\text{\tiny T}} 
			=: 
			g_{1}
			\left(
			\widehat{T}_{1},\widehat{T}_{2}
			\right), \\[10pt]
			\widehat{\sigma}_{\text{\tiny T}}
			& = & 
			\dfrac{\pm \, 1}
			{
				\sqrt{
					\eta(a_{1},\ol{b}_2)}}
			\lp 
			\widehat{T}_{2} 
			-  
			\eta_{r} \,
			\widehat{T}_{1}^{2}
			\rp^{1/2}
			+ 
			\dfrac{
				\widehat{T}_{1}
				\lp 
				c_1(a_{1},\ol{b}_{1})
				- 
				c_1(a_{2}, \ol{b}_{2})
				\rp 
			}
			{\eta(a_{1},\ol{b}_{2})}
			=: 
			g_{2}
			\left(
			\widehat{T}_{1},\widehat{T}_{2}
			\right),
		\end{array}
		\right.
	\end{align}
	where for $1\le i, \, j \le 2$,
	\begin{align}
		\label{eqn:EtaDefn1}
		\eta(a_{i},\ol{b}_{j}) 
		& := 
		c_{1}^{2}(a_{i},\ol{b}_{i})
		- 
		2 c_{1}(a_{i},\ol{b}_{i}) c_1(a_{j},\ol{b}_{j}) 
		+ 
		c_{2}(a_{j},\ol{b}_j), 
		\quad 
		\eta_{r} 
		: = 
		\dfrac{\eta(a_j,\ol{b}_{j})}
		{\eta(a_{i},\ol{b}_j)}.
	\end{align}
	
	\begin{note}
		\label{note:DiffRel2}
		The motivation for imposing the trimming inequality
		\eqref{eqn:abCondition2} is to ensure that 
		the same data points are trimmed from 
		both tails for the first and second moments, 
		particularly when the sample data 
		are approximately symmetric about zero,
		an approach that,
		to our knowledge, 
		has not been fully addressed in the existing literature.
		
		For illustration, consider the sample dataset:
		\[
		-15,\ -13,\ -8,\ -4,\ -2,\ 3,\ 5,\ 7,\ 9,\ 12.
		\]
		
		\begin{itemize}
			\item 
			Under equal trimming proportions, say 
			$(a_{1}, b_{1}) = (a_{2}, b_{2}) = (0.2, 0.2)$, the trimmed samples are:
			\begin{align*}
				\text{Trimmed sample for the first moment:} 
				& \quad -8,\ -4,\ -2,\ 3,\ 5,\ 7. \\
				\text{Trimmed sample for the second moment:} 
				& \quad (-4)^2,\ 5^2,\ 7^2,\ (-8)^2,\ 9^2,\ 12^2.
			\end{align*}
			In this case, observations $-15$, $-13$, $9$, and $12$ are trimmed for the first moment, whereas $-15$, $-13$, $-2$, and $3$ are trimmed for the second moment. Hence, different data points are being trimmed for the first and second moments.
			
			\item 
			Now consider unequal trimming proportions, e.g., 
			$(a_{1}, b_{1}) = (0.2, 0.2)$ and 
			$(a_{2}, b_{2}) = (0, 0.4)$, which yield:
			\begin{align*}
				\text{Trimmed sample for the first moment:} 
				& \quad -8,\ -4,\ -2,\ 3,\ 5,\ 7. \\
				\text{Trimmed sample for the second moment:} 
				& \quad (-8)^2,\ (-4)^2,\ (-2)^2,\ 3^2,\ 5^2,\ 7^2.
			\end{align*}
			In this case, the same observations $-15$, $-13$, $9$, and $12$ are trimmed from both the first and second moments.
		\end{itemize}
		
		Therefore, if the data arise from a distribution symmetric about zero, ensuring that the same observations are trimmed from both the first and second moments generally requires using different trimming proportions, i.e., $(a_{1}, b_{1}) \ne (a_{2}, b_{2})$. 
		
		Even when all observations are negative, equal trimming proportions—e.g., $(a_{1}, b_{1}) = (a_{2}, b_{2}) = (0.1, 0.2)$—can still result in trimming different observations for the first and second moments. In contrast, if all data points are positive, then equal trimming proportions typically lead to trimming the same observations for both moments.
		
		Nevertheless, to enhance robustness, one may deliberately choose different trimming proportions that result in trimming different observations for the first and second moments—even when the underlying model assumes strictly positive support. In such cases, it is advisable to select trimming parameters that satisfy the condition
		\begin{align}
			\label{eqn:abCondition3}
			0 
			\le 
			a_{1} 
			\le 
			a_{2} 
			< 
			\ol{b}_{1} 
			\le 
			\ol{b}_{2} 
			\le 
			1,
		\end{align}
		which allows greater flexibility in capturing potential distributional asymmetries and tail behaviors.
		
		Therefore, our motivation for adopting the trimming inequality \eqref{eqn:abCondition2} is to ensure consistent trimming across both moments when the underlying distribution is symmetric about zero, whereas inequality \eqref{eqn:abCondition3} offers an alternative strategy to enhance robustness when the random variable is strictly positive.
		\qed 
	\end{note}
	
	\begin{prop}
		\label{prop:EtaRealtion1}
		With the trimming proportions
		$\lp a_{i}, b_{i} \rp$ and 
		$\lp a_{j}, b_{j} \rp$
		satisfying the inequality \eqref{eqn:abCondition1}, 
		it follows that 
		\begin{itemize}
			\item[(i)]
			\(
			\eta 
			\lp 
			a_{i},\ol{b}_{j}
			\rp 
			= 
			c_{1}^{2} 
			\lp 
			a_{i}, \ol{b}_{i}
			\rp 
			-
			2 
			c_{1} 
			\lp 
			a_{i}, \ol{b}_{i}
			\rp 
			c_{1} 
			\lp 
			a_{j}, \ol{b}_{j}
			\rp 
			+ 
			c_{2} 
			\lp 
			a_{j}, \ol{b}_{j}
			\rp 
			> 
			0.
			\)
			
			\item[(ii)]
			\(
			c_{k}
			\lp 
			a_{i}, \ol{b}_{i}
			\rp 
			\ge 
			c_{k}
			\lp 
			a_{j}, \ol{b}_{j}
			\rp,
			\)
			for any odd positive integer $k$.
			
			\item[(iii)]
			\(
			0
			< 
			\eta_{r} 
			= 
			\dfrac{\eta(a_j,\ol{b}_{j})}
			{\eta(a_{i},\ol{b}_j)}
			\le 
			1.
			\)
			
			\item[(iv)]
			\(
			c_2
			\lp 
			a_j, \overline{b}_{j}
			\rp 
			\ge  
			\eta_r \, 
			c_1^2
			\lp 
			a_i, \overline{b}_i
			\rp.
			\)
			
			\item[(v)]
			\(
			T_{2} - \eta_{r} T_{1}^{2} \ge 0.
			\)
		\end{itemize}
		
		\begin{proof}
			See Appendix \ref{sec:Appendix1}.
		\end{proof}
	\end{prop}
	
	\begin{note}
		\label{note:DiffRel1}
		Under the inequality condition \eqref{eqn:abCondition2}, no fixed ordering can be established between $c_{2}(a_{2}, \ol{b}_{2})$ and $c_{1}^{2}(a_{1}, \ol{b}_{1})$, or equivalently between $T_{2}$ and $T_{1}^{2}$. To illustrate both possible directions, consider the standard normal distribution, i.e., $F_{0} = \Phi$, with $\theta = 0$ and $\sigma = 1$. 
		Then:
		\begin{enumerate}
			\item 
			For 
			\( a_{2} = 0.02, \ \ol{b}_{2} = 0.75 \) 
			and 
			\( a_{1} = 0.05, \ \ol{b}_{1} = 0.99 \),
			\[
			T_{2} = c_{2}(a_{2}, \ol{b}_{2}) = 0.5702 
			> 
			0.0066 = c_{1}^{2}(a_{1}, \ol{b}_{1}) = T_{1}^{2}.
			\]
			
			\item 
			For 
			\( a_{2} = 0.02, \ \ol{b}_{2} = 0.75 \) 
			and 
			\( a_{1} = 0.50, \ \ol{b}_{1} = 0.99 \),
			\[
			T_{2} = c_{2}(a_{2}, \ol{b}_{2}) = 0.5702 
			< 
			0.5773 = c_{1}^{2}(a_{1}, \ol{b}_{1}) = T_{1}^{2}.
			\]
		\end{enumerate}
		
		However, 
		by Proposition~\ref{prop:EtaRealtion1}~(iv), 
		the inequality is restored as below
		\[
		c_{2}(a_{2}, \ol{b}_{2}) 
		\ge 
		\eta_{r} \, c_{1}^{2}(a_{1}, \ol{b}_{1}),
		\quad \text{equivalently} \quad
		T_{2} - \eta_{r} \, T_{1}^{2} \ge 0.
		\]
		
		Similarly,
		for a nonzero location parameter,
		say $\theta = 5$ and $\sigma = 2$,
		we observe that 
		\begin{align}
			T_{2} &= 19.9010 
			\le T_{1}^{2} = 42.5046, 
			\quad \text{but} \quad 
			T_{2} = 19.9010 
			\ge \eta_{r} \, T_{1}^{2} = 10.8001.
		\end{align}
		\qed
	\end{note}
	
	\begin{cor}
		\label{cor:InequalityFlip1}
		Under the trimming inequality 
		\eqref{eqn:abCondition1Flip},
		that is, 
		\eqref{eqn:abCondition3} for location-scale models,
		all results of Proposition~\ref{prop:EtaRealtion1} 
		remain valid except for Part~\((ii)\),
		which instead takes the form
		\(
		c_{k}
		\lp 
		a_{i}, \ol{b}_{i}
		\rp 
		\le  
		c_{k}
		\lp 
		a_{j}, \ol{b}_{j}
		\rp,
		\)
		for any odd positive integer \( k \).
	\end{cor}
	
	\begin{note}
		\label{note:LS_Sigma_Terms1}
		Define 
		\begin{align}
			\label{eqn:SigmaTwoTerms1}
			\sigma_{\mbox{\tiny FT}}
			& = 
			\dfrac{1}
			{
				\sqrt{
					\eta(a_{1},\ol{b}_2)}}
			\lp 
			{T}_{2} 
			-  
			\eta_{r} \,
			{T}_{1}^{2}
			\rp^{1/2}
			\quad 
			\mbox{and} 
			\quad 
			\sigma_{\mbox{\tiny ST}}
			=
			\dfrac{
				{T}_{1}
				\lp 
				c_1(a_{1},\ol{b}_{1})
				- 
				c_1(a_{2}, \ol{b}_{2})
				\rp 
			}
			{\eta(a_{1},\ol{b}_{2})}.
		\end{align}
		Under the trimming inequality 
		\eqref{eqn:abCondition2}, 
		a consistent directional relationship between
		\(\sigma_{\text{\tiny FT}}\) and 
		\(\sigma_{\text{\tiny ST}}\) is not guaranteed.
		
		For illustration, consider 
		\( X \sim N(\theta = 10, \sigma = 3) \). 
		Then, for
		\[
		(a_{1}, b_{1}) = (0.02, 0.02) 
		\quad \text{and} \quad 
		(a_{2}, b_{2}) = (0.00, 0.03),
		\]
		we have
		\[
		\sigma_{\mbox{\tiny FT}}
		= 
		2.192
		> 
		0.808
		=
		\sigma_{\mbox{\tiny ST}}.
		\]
		
		However, for
		\[
		(a_{1}, b_{1}) = (0.02, 0.02) 
		\quad \text{and} \quad 
		(a_{2}, b_{2}) = (0.00, 0.10),
		\]
		we obtain
		\[
		\sigma_{\mbox{\tiny FT}}
		= 
		0.400
		< 
		2.600
		=
		\sigma_{\mbox{\tiny ST}}.
		\]
		\qed 
	\end{note}
	
	Since the trimmed estimators 
	\( 
	\lp 
	\widehat{\theta}_{\mbox{\tiny T}},
	\widehat{\sigma}_{\mbox{\tiny T}}
	\rp 
	\)
	in Eq.~\eqref{eqn:MTMSol1} 
	are derived from Eq.~\eqref{eqn:P2}, 
	it is guaranteed in theory that one 
	of the solutions for
	\(\sigma\) from the \(\pm\) branch must be positive. 
	However, in finite samples, it is possible that 
	\(
	\widehat{\sigma}_{\mbox{\tiny T}}
	< 
	0,
	\)
	in which case the proposed trimmed
	\(L\)-moment estimation fails. 
	Thus, 
	to determine the appropriate sign of 
	$\widehat{\sigma}_{\text{\tiny T}}$ 
	in Eq.~\eqref{eqn:MTMSol1}, 
	the following strategy is proposed. 
	First, 
	as established in Proposition~\ref{prop:EtaRealtion1}, 
	it holds theoretically that 
	\(
	T_{2} - \eta_{r} T_{1}^{2} \ge 0.
	\)
	Therefore, assuming 
	\(
	\lp 
	\widehat{T}_{2} 
	-  
	\eta_{r} \,
	\widehat{T}_{1}^{2}
	\rp^{1/2} 
	\ge 
	0,
	\)
	we consider  
	\begin{align} 
		\label{eqn:SigmaTwoTerms1Hat}
		\widehat{\sigma}_{\mbox{\tiny FT}}
		& = 
		\dfrac{1}
		{
			\sqrt{
				\eta(a_{1},\ol{b}_2)}}
		\lp 
		\widehat{T}_{2} 
		-  
		\eta_{r} \,
		\widehat{T}_{1}^{2}
		\rp^{1/2} 
		\quad 
		\mbox{and} 
		\quad 
		\widehat{\sigma}_{\mbox{\tiny ST}}
		=
		\dfrac{
			\widehat{T}_{1}
			\lp 
			c_1(a_{1},\ol{b}_{1})
			- 
			c_1(a_{2}, \ol{b}_{2})
			\rp 
		}
		{\eta(a_{1},\ol{b}_{2})}. 
	\end{align}
	
	But for $a_{1} \ne a_{2}$ or $b_{1} \ne b_{2}$,
	it is possible to have
	\(
	\widehat{T}_{2} 
	-  
	\eta_{r} \,
	\widehat{T}_{1}^{2}
	< 
	0,
	\)
	i.e., 
	\(
	\widehat{T}_{2} / \widehat{T}_{1}^{2}
	<
	\eta_{r},
	\)
	and in that case we consider 
	\begin{align} 
		\label{eqn:SigmaTwoTerms2Hat}
		\widehat{\sigma}_{\mbox{\tiny FT}}
		& =
		\dfrac{1}
		{
			\sqrt{
				\eta(a_{1},\ol{b}_2)}}
		\lp 
		\left| 
		\widehat{T}_{2} 
		-  
		\eta_{r} \,
		\widehat{T}_{1}^{2}
		\right|
		\rp^{1/2}
		\quad 
		\mbox{and}
		\quad 
		\widehat{\sigma}_{\mbox{\tiny ST}}
		=
		\dfrac{
			\widehat{T}_{1}
			\lp 
			c_1(a_{1},\ol{b}_{1})
			- 
			c_1(a_{2}, \ol{b}_{2})
			\rp 
		}
		{\eta(a_{1},\ol{b}_{2})}.
	\end{align}
	
	Thus, 
	for location-scale models, 
	it is guaranteed that 
	\(
	\widehat{\sigma}_{\mbox{\tiny FT}} 
	\ge 0.
	\) 
	However, 
	\(
	\widehat{\sigma}_{\mbox{\tiny ST}} 
	\in \mathbb{R}.
	\)
	
	If 
	\(
	\widehat{\sigma}_{\mbox{\tiny FT}} 
	< 
	\widehat{\sigma}_{\mbox{\tiny ST}},
	\)
	we define 
	\(
	\widehat{\sigma}_{-}
	:= 
	- 
	\widehat{\sigma}_{\mbox{\tiny FT}} 
	+ 
	\widehat{\sigma}_{\mbox{\tiny ST}} > 0.
	\) 
	Similarly, if 
	\(
	\widehat{\sigma}_{\mbox{\tiny FT}} 
	> 
	\left| 
	\widehat{\sigma}_{\mbox{\tiny ST}} 
	\right|,
	\)
	we define 
	\(
	\widehat{\sigma}_{+}
	:= 
	\widehat{\sigma}_{\mbox{\tiny FT}} 
	+ 
	\widehat{\sigma}_{\mbox{\tiny ST}} > 0.
	\)
	Finally, if 
	\(
	\widehat{\sigma}_{\mbox{\tiny FT}} 
	\le 
	\left| 
	\widehat{\sigma}_{\mbox{\tiny ST}} 
	\right|
	\)
	and 
	\(
	\widehat{\sigma}_{\mbox{\tiny ST}} 
	\le 
	0,
	\)
	then the method fails to produce
	a nonnegative estimate of 
	\(
	\widehat{\sigma}_{\mbox{\tiny T}} \ge 0.
	\)
	
	\begin{algorithm}[h!bt]
		\caption{Trimmed Estimation of \(\left( \widehat{\theta}_{\text{\tiny T}}, \widehat{\sigma}_{\text{\tiny T}} \right)\) in Location-Scale Models}
		\label{alg:LSAlgorithm1}
		\SetAlgoLined
		\LinesNumbered
		\SetNlSty{textbf}{}{:}
		\SetAlgoNlRelativeSize{-0}
		
		\KwIn{Sample data and trimming proportions \((a_1, b_1), (a_2, b_2)\) satisfying \eqref{eqn:abCondition2} or \eqref{eqn:abCondition3}.}
		
		\KwOut{Trimmed estimator vector 
			\(
			\lp 
			\widehat{\theta}_{\text{\tiny T}}, \widehat{\sigma}_{\text{\tiny T}}
			\rp 
			\).}
		
		\textbf{Compute}\;
		\Indp
		\( \widehat{T}_1 \) and \( \widehat{T}_2 \) from Eq.~\eqref{eqn:S2}\;
		\( c_1(a_1, \ol{b}_1), \; c_1(a_2, \ol{b}_2) \),
		and 
		\(c_2(a_2, \ol{b}_2) \) from Eq.~\eqref{eqn:ConstantC1}\;
		\( \eta(a_1, \ol{b}_2), \; \eta(a_2, \ol{b}_2) \), and \( \eta_r \) from Eq.~\eqref{eqn:EtaDefn1}\;
		\( \widehat{\sigma}_{\text{\tiny FT}} \) 
		and 
		\( \widehat{\sigma}_{\text{\tiny ST}} \) 
		from Eq.~\eqref{eqn:SigmaTwoTerms1Hat}
		or Eq.~\eqref{eqn:SigmaTwoTerms2Hat}\;
		\( \widehat{\sigma}_{\mbox{\tiny MLE}} \) from Eq.~\eqref{eqn:LS_CompleteSD}\;
		\Indm
		
		\textbf{Set}
		\( \widehat{\sigma}_+ \leftarrow \widehat{\sigma}_{\text{\tiny FT}} 
		+ \widehat{\sigma}_{\text{\tiny ST}} \)
		and 
		\( \widehat{\sigma}_- \leftarrow -\widehat{\sigma}_{\text{\tiny FT}} + \widehat{\sigma}_{\text{\tiny ST}} \)\;
		\If{\( a_1 = a_2 \) and \( b_1 = b_2 \)}{
			\( \widehat{\sigma}_{\text{\tiny ST}} \leftarrow 0 \),
			resulting in
			\( \widehat{\sigma}_{\text{\tiny T}}
			\leftarrow 
			{\widehat{\sigma}_{\text{\tiny FT}}} \)\;
			\Return \( \widehat{\sigma}_{\text{\tiny T}} \)\;
		}
		\ElseIf{
			\( (a_2 \le a_1 \text{ and } b_1 \le b_2) \)
			\text{ or }
			\( (a_1 \le a_2 \text{ and } b_2 \le b_1) \)
		}
		{
			\If{\( 
				\max 
				\left\{ 
				\widehat{\sigma}_{-}, 
				\widehat{\sigma}_{+}
				\right\}
				\le  
				0 \)}{
				\tcc{Exit and update trimming proportions to ensure 
					\(\widehat{\sigma}_{\text{\tiny T}} > 0.\)}
				\textbf{Ensure} 
				\(
				\max 
				\left\{ 
				\widehat{\sigma}_{-}, 
				\widehat{\sigma}_{+}
				\right\}
				> 
				0
				\)\;
			}
			\ElseIf{\( 
				\widehat{\sigma}_{-}
				\le 
				0
				\mbox{ and }
				\widehat{\sigma}_{+}
				>
				0
				\)}
			{\( \widehat{\sigma}_{\text{\tiny T}} \leftarrow \widehat{\sigma}_+ \)\;}
			\ElseIf{\( 
				\widehat{\sigma}_{-}
				>
				0
				\mbox{ and }
				\widehat{\sigma}_{+}
				\le 
				0
				\)}
			{\( \widehat{\sigma}_{\text{\tiny T}} \leftarrow \widehat{\sigma}_- \)\;}
			\ElseIf{
				\( 
				\min 
				\left\{ 
				\widehat{\sigma}_{-}, 
				\widehat{\sigma}_{+}
				\right\}
				\ge 
				0
				\)
			}
			{
				\If{\( |\widehat{\sigma}_- - \widehat{\sigma}_{\text{\tiny MLE}}| < |\widehat{\sigma}_+ - \widehat{\sigma}_{\text{\tiny MLE}}| \)}
				{
					\( \widehat{\sigma}_{\text{\tiny T}} \leftarrow \widehat{\sigma}_- \)
				}
				\Else{
					\( \widehat{\sigma}_{\text{\tiny T}} \leftarrow \widehat{\sigma}_+ \)\;
				}
			}
		}
		
		\Return \( \widehat{\sigma}_{\text{\tiny T}} \)\;
		
		\( \widehat{\theta}_{\text{\tiny T}} \leftarrow \widehat{T}_1 - c_1(a_1, \ol{b}_1) \, \widehat{\sigma}_{\text{\tiny T}} \)\;
		\Return 
		\( 
		\lp 
		\widehat{\theta}_{\text{\tiny T}}, \; \widehat{\sigma}_{\text{\tiny T}}
		\rp 
		\)\;
	\end{algorithm}
	
	\begin{itemize}
		\item 
		If $a_{1} = a_{2} = a$ and $b_{1} = b_{2} = b$,
		then it follows that 
		\( 
		\eta(a_{1},\ol{b}_2) 
		= 
		\eta(a_{2},\ol{b}_2), 
		\)
		that is, 
		$\eta_{r} = 1$.
		Also, 
		\( 
		c_1(a_{1},\ol{b}_{1})
		- 
		c_1(a_{2}, \ol{b}_{2})
		=
		0.
		\) 
		Thus, 
		\[
		\widehat{\sigma}_{\text{\tiny T}}
		=
		\widehat{\sigma}_{+}
		=
		\left( 
		\dfrac{\widehat{T}_{2} 
			-  
			\widehat{T}_{1}^{2}
		}
		{
			c_{2}(a, \ol{b}) 
			- 
			c_{1}(a, \ol{b})^{2}
		}
		\right)^{1/2},
		\]
		and this result coincides exactly with 
		the solution presented in Eq.~(2.7)
		of \cite{MR2497558}, as expected.
		
		\item 
		If \( a_{1} \ne a_{2} \) or \( b_{1} \ne b_{2} \),  
		selecting the appropriate sign in
		Eq.~\eqref{eqn:MTMSol1} becomes nontrivial.  
		For either trimming inequality \eqref{eqn:abCondition2} 
		or \eqref{eqn:abCondition3},  
		we begin by computing the standard deviation
		of the full sample  
		\( X_{1}, X_{2}, \ldots, X_{n} \) as
		\begin{align}
			\label{eqn:LS_CompleteSD}
			\widehat{\sigma}_{\mbox{\tiny MLE}}
			& = 
			\sqrt{
				\dfrac{1}{n} 
				\sum_{i=1}^{n} 
				\lp 
				X_{i} - \ol{X}
				\rp^{2}.
			}
		\end{align}
		If
		\(
		\max 
		\left\{ 
		\widehat{\sigma}_{-}, 
		\widehat{\sigma}_{+}
		\right\}
		\le 
		0,
		\)
		then there is no positive solution 
		for the scale parameter \(\sigma\). 
		Otherwise, 
		if 
		\(
		\widehat{\sigma}_{-}
		\le 
		0
		\mbox{ and }
		\widehat{\sigma}_{+}
		>
		0
		\),
		then we simply assign 
		\(
		\widehat{\sigma}_{\text{\tiny T}}
		=
		\widehat{\sigma}_+.
		\)
		Similarly,
		if
		\( 
		\widehat{\sigma}_{-}
		>
		0
		\mbox{ and }
		\widehat{\sigma}_{+}
		\le 
		0,
		\)
		then 
		\(
		\widehat{\sigma}_{\text{\tiny T}} 
		=
		\widehat{\sigma}_-.
		\)
		Finally, 
		assuming 
		\(
		\min 
		\left\{ 
		\widehat{\sigma}_{-}, 
		\widehat{\sigma}_{+}
		\right\}
		\ge 
		0,
		\)
		and following a rationale similar to that
		used in selecting a decision threshold 
		in logistic classification, 
		we choose between 
		\(\widehat{\sigma}_-\) and \(\widehat{\sigma}_+\) 
		based on their proximity to the maximum likelihood estimate.
		That is, 
		\begin{eqnarray}
			\label{eqn:SigmaChoice1}
			\widehat{\sigma}_{\text{\tiny T}}
			& = &
			\begin{cases}
				\widehat{\sigma}_{-}, & 
				\mbox{if }
				\left| 
				\widehat{\sigma}_{-} 
				- 
				\widehat{\sigma}_{\mbox{\tiny MLE}}
				\right| 
				< 
				\left| 
				\widehat{\sigma}_{+} 
				- 
				\widehat{\sigma}_{\mbox{\tiny MLE}}
				\right|, \\
				\widehat{\sigma}_{+}, & 
				\mbox{if }
				\left| 
				\widehat{\sigma}_{+} 
				- 
				\widehat{\sigma}_{\mbox{\tiny MLE}}
				\right| 
				\le 
				\left| 
				\widehat{\sigma}_{-} 
				- 
				\widehat{\sigma}_{\mbox{\tiny MLE}}
				\right|.
			\end{cases}
		\end{eqnarray}
	\end{itemize}
	
	A formal procedure for estimating the parameter vector 
	\(
	\left( 
	\widehat{\theta}_{\text{\tiny T}}, 
	\widehat{\sigma}_{\text{\tiny T}} 
	\right)
	\)
	is summarized in Algorithm~\ref{alg:LSAlgorithm1}.
	
	\begin{note}
		For \( (a_{1}, b_{1}) = (a_{2}, b_{2}) = (a, b) \),
		it follows that 
		\( 
		\eta(a_{1},\ol{b}_2) 
		= 
		\eta(a_{2},\ol{b}_2), 
		\)
		that is, 
		$\eta_{r} = 1$,
		and the result given by Eq.~\eqref{eqn:MTMSol1} 
		coincides exactly with the solution presented 
		in Eq.~(2.7) of \cite{MR2497558}, as expected.
		\qed
	\end{note}
	
	From Theorem \ref{thm:MTM_Var1} along with 
	Eqs.~\eqref{eqn:LSH1D} and \eqref{eqn:LSH2D},
	we calculate the entries of the covariance matrix 
	$
	\bm{\Sigma}_{\mbox{\tiny T}}
	=
	\lb 
	\sigma_{ij}^{2}
	\rb_{i,j=1}^{2}
	$
	as below:
	\begin{eqnarray}
		\sigma_{11}^{2} 
		& = & 
		\Gamma(1,1) V(1,1) 
		\nonumber \\ 
		& = & 
		\Gamma(1,1) 
		\int_{a_{1}}^{\ol{b}_{1}}\int_{a_{1}}^{\ol{b}_{1}}
		{K(w,v)} 
		H_{1}'(w) H_{1}'(v) \, dv \, dw 
		\nonumber \\ 
		& = & 
		\sigma^{2} \
		\Gamma(1,1) 
		\int_{a_{1}}^{\ol{b}_{1}}\int_{a_{1}}^{\ol{b}_{1}}
		K(w, v) \, 
		d F_{0}^{-1}(v) \, 
		d F_{0}^{-1}(w)
		\, dv \, dw 
		\nonumber \\
		& = & 
		\sigma^{2} \,
		\Lambda_{111},
		\label{eqn:MTM_Sigma112} \\
		\sigma_{12}^{2}
		& = & 
		\Gamma(1,2) 
		V(1,2) 
		\nonumber \\ 
		& = & 
		\Gamma(1,2) 
		\int_{a_{1}}^{\ol{b}_{1}}\int_{a_{2}}^{\ol{b}_{2}}
		{K(w,v)} 
		H_{1}'(w) H_{2}'(v) \, dv \, dw 
		\nonumber \\ 
		& = & 
		\Gamma(1,2) 
		\scalebox{2.5}{\{}
		2 \, \theta \, \sigma^{2} \
		\int_{a_{1}}^{\ol{b}_{1}}
		\int_{a_{2}}^{\ol{b}_{2}}
		K(w, v) \, 
		d F_{0}^{-1}(v) \, 
		d F_{0}^{-1}(w)
		\, dv \, dw 
		\nonumber \\ 
		& & 
		+
		2 \, \sigma^{3} \,  
		\int_{a_{1}}^{\ol{b}_{1}}
		\int_{a_{2}}^{\ol{b}_{2}}
		K(w, v) \, 
		F_{0}^{-1}(v) \, 
		d F_{0}^{-1}(v) \, 
		d F_{0}^{-1}(w)
		\, dv \, dw 
		\scalebox{2.5}{\}}
		\nonumber \\
		& = & 
		2 \, \theta \, \sigma^{2} \, \Lambda_{121} 
		+ 
		2 \, \sigma^{3} \, \Lambda_{122},
		\label{eqn:MTM_Sigma122} \\
		\sigma_{22}^{2} 
		& = & 
		\Gamma(2,2) \, V(2,2) 
		\nonumber \\ 
		& = & 
		\Gamma(2,2) 
		\int_{a_{2}}^{\ol{b}_{2}}\int_{a_{2}}^{\ol{b}_{2}}
		{K(w,v)} 
		H_{2}'(w) H_{2}'(v) \, dv \, dw 
		\nonumber \\ 
		& = &
		\Gamma(2,2) 
		\scalebox{2.5}{\{}
		4 \, \theta^{2} \, \sigma^{2} \
		\int_{a_{2}}^{\ol{b}_{2}}
		\int_{a_{2}}^{\ol{b}_{2}}
		K(w, v) \, 
		d F_{0}^{-1}(v) \, 
		d F_{0}^{-1}(w)
		\, dv \, dw  
		\nonumber \\ 
		& &
		+ 
		8 \, \theta \, \sigma^{3} \
		\int_{a_{2}}^{\ol{b}_{2}}
		\int_{a_{2}}^{\ol{b}_{2}}
		K(w, v) \, F_{0}^{-1}(w) \, 
		d F_{0}^{-1}(v) \, 
		d F_{0}^{-1}(w)
		\, dv \, dw 
		\nonumber \\ 
		& & 
		+
		4 \, \sigma^{4} \,  
		\int_{a_{2}}^{\ol{b}_{2}}
		\int_{a_{2}}^{\ol{b}_{2}}
		K(w, v) \, 
		F_{0}^{-1}(w) \, F_{0}^{-1}(v) \, 
		d F_{0}^{-1}(v) \, 
		d F_{0}^{-1}(w)
		\, dv \, dw 
		\scalebox{2.5}{\}}
		\nonumber \\
		& = & 
		4 \theta^{2} \sigma^{2}
		\Lambda_{221}
		+ 
		8 \theta \sigma^{3} 
		\Lambda_{222}
		+ 
		4 \sigma^{4} 
		\Lambda_{223},
		\label{eqn:MTM_Sigma222}
	\end{eqnarray}
	where the notations $\Lambda_{ijk}$, 
	for $1 \le i, j \le 2$ and 
	$1 \le k \le 3$ do not depend 
	on the parameters to be 
	estimated and are listed 
	in Appendix \ref{sec:Appendix2}.
	
	\begin{note}
		For the equal trimming proportions
		$(a_1,b_1) = (a_2,b_2) = (a,b)$,
		we have 
		\begin{align*}
			\Lambda_{111}
			& = 
			\Lambda_{121}
			=
			\Lambda_{221}
			= 
			c_{1}^{*},
			\quad 
			\Lambda_{122}
			=
			\Lambda_{222}
			= 
			c_{2}^{*}, 
			\quad 
			\mbox{and}
			\quad 
			\Lambda_{223}
			= 
			c_{3}^{*},
		\end{align*}
		where the notations $c_{i}^{*}$, 
		$i = 1, 2, 3$ can be found in \cite{MR2497558}.
		\qed 
	\end{note}
	
	As defined in Corollary \ref{thm:CGJ3},
	the entries of the matrix 
	\( 
	\bm{D}_{\text{\tiny T}}
	=
	\lb 
	d_{ij}
	\rb_{i,j = 1}^{2},
	\)
	are obtained by differentiating the functions 
	\( g_i \) from Eqs.~\eqref{eqn:MTMSol1}:
	\begin{align*}
		d_{11} 
		& =
		1 
		- 
		c_{1}(a_{1}, \ol{b}_{1}) \, 
		\frac{\partial g_{2}}{\partial \widehat{T}_{1}}, 
		\quad 
		d_{12} 
		= 
		- 
		c_{1}(a_{1}, \ol{b}_{1}) \, 
		\frac{\partial g_{2}}{\partial \widehat{T}_{2}}.
	\end{align*}
	
	The two entries $d_{21}$ and $d_{22}$ 
	depend on the sign of the 
	$\widehat{\sigma}_{\mbox{\tiny T}}$
	as seen in Eq.~\eqref{eqn:SigmaChoice1}.
	That is, if 
	\(
	\widehat{\sigma}_{\text{\tiny T}}
	= 
	\widehat{\sigma}_{-},
	\)
	then 
	\begin{align}
		d_{21}^{-}
		& =
		\frac{\partial g_{2}}{\partial \widehat{T}_{1}} 
		= 
		\frac{\eta_{r} \, \widehat{T}_{1}}
		{\sqrt{\eta(a_1, \ol{b}_2)}} 
		\left( 
		\widehat{T}_{2} 
		- 
		\eta_{r} \,
		\widehat{T}_{1}^{2} 
		\right)^{-\frac{1}{2}}
		+ 
		\frac{c_1(a_1, \ol{b}_{1}) - c_1(a_2, \ol{b}_{2})}
		{\eta(a_1, \ol{b}_{2})}, 
		\label{eqn:D211} \\ 
		d_{22}^{-}
		& = 
		\frac{\partial g_{2}}{\partial \widehat{T}_{2}} 
		= 
		\frac{- 1}
		{2 \, \sqrt{\eta(a_1, \ol{b}_2)}} 
		\left( 
		\widehat{T}_2 
		- 
		\eta_{r} \,
		\widehat{T}_{1}^{2} 
		\right)^{-\frac{1}{2}}.
		\label{eqn:D221}
	\end{align}
	
	And, if 
	\(
	\widehat{\sigma}_{\text{\tiny T}}
	= 
	\widehat{\sigma}_{+},
	\)
	then
	\begin{align}
		d_{21}^{+}
		& =
		\frac{\partial g_{2}}{\partial \widehat{T}_{1}} 
		= 
		\frac{- \eta_{r} \, \widehat{T}_{1}}
		{\sqrt{\eta(a_1, \ol{b}_2)}} 
		\left( 
		\widehat{T}_{2} 
		- 
		\eta_{r} \,
		\widehat{T}_{1}^{2} 
		\right)^{-\frac{1}{2}}
		+ 
		\frac{c_1(a_1, \ol{b}_{1}) - c_1(a_2, \ol{b}_{2})}
		{\eta(a_1, \ol{b}_{2})}, 
		\label{eqn:D212} \\ 
		d_{22}^{+}
		& = 
		\frac{\partial g_{2}}{\partial \widehat{T}_{2}} 
		= 
		\frac{1}
		{2 \, \sqrt{\eta(a_1, \ol{b}_2)}} 
		\left( 
		\widehat{T}_2 
		- 
		\eta_{r} \,
		\widehat{T}_{1}^{2} 
		\right)^{-\frac{1}{2}}.
		\label{eqn:D222}
	\end{align}
	
	\begin{lemma}
		\label{lemma:D_matrix_Signs1}
		Define 
		\[
		\bm{D}_{\mbox{\tiny T}}^{-}
		=
		\begin{bmatrix}
			d_{11} & d_{12} \\[5pt]
			d_{21}^{-} & d_{22}^{-}
		\end{bmatrix}
		\quad 
		\mbox{and} 
		\quad 
		\bm{D}_{\mbox{\tiny T}}^{+}
		=
		\begin{bmatrix}
			d_{11} & d_{12} \\[5pt]
			d_{21}^{+} & d_{22}^{+}
		\end{bmatrix},
		\] 
		then it follows that 
		\(
		\mbox{det}
		\lp 
		\bm{D}_{\mbox{\tiny T}}^{-}
		\rp 
		+
		\mbox{det}
		\lp 
		\bm{D}_{\mbox{\tiny T}}^{+}
		\rp
		=
		0.
		\)
	\end{lemma}
	
	We will be using the Jacobian 
	matrix $\bm{D}_{\text{\tiny T}}$
	to compute the ARE as presented in
	Eq.~\eqref{eqn:ARE1}, where 
	\( 
	\bm{\Sigma}_{\text{\tiny $\mathcal{C}$}}
	=
	\bm{D}_{\text{\tiny T}}
	\bm{\Sigma}_{\text{\tiny T}}
	\bm{D}_{\text{\tiny T}}'.
	\)
	Since both matrices 
	\(
	\bm{D}_{\text{\tiny T}}
	\)
	and 
	\(\bm{\Sigma}_{\text{\tiny T}}\)
	are $2 \times 2$
	square matrices, 
	it follows from
	\citet[][Theorem 1.7]{MR3497549} 
	that
	\(
	\mbox{det}
	\lp 
	\bm{D}_{\text{\tiny T}}
	\bm{\Sigma}_{\text{\tiny T}}
	\bm{D}_{\text{\tiny T}}'
	\rp 
	= 
	\lp 
	\mbox{det}
	\lp 
	\bm{D}_{\text{\tiny T}}
	\rp 
	\rp^{2}
	\mbox{det}
	\lp 
	\bm{\Sigma}_{\text{\tiny T}}
	\rp.
	\)
	Therefore, 
	by the property established in Lemma \ref{lemma:D_matrix_Signs1}, 
	choosing either 
	$\bm{D}_{\text{\tiny T}}^{-}$ or
	$\bm{D}_{\text{\tiny T}}^{+}$
	for the Jacobian matrix $\bm{D}_{\text{\tiny T}}$
	does not affect the 
	final ARE calculation. 
	Thus, without loss of generality, 
	we proceed with 
	$\bm{D}_{\text{\tiny T}} = \bm{D}_{\text{\tiny T}}^{+}$.
	
	Further, 
	define
	\[
	\Omega
	:= 
	\left| 
	\sigma 
	\lp 
	c_{2}(a_{2}, \ol{b}_{2})
	- 
	c_{1}(a_{1}, \ol{b}_{1}) \, 
	c_{1}(a_{2}, \ol{b}_{2})
	\rp 
	+ 
	\theta 
	\lp
	c_{1}(a_{2}, \ol{b}_{2})
	-
	c_{1}(a_{1}, \ol{b}_{1})
	\rp
	\right|,
	\]
	then it follows that 
	\begin{eqnarray*} 
		d_{11} 
		& = & 
		\left. \frac{\partial g_{1}}
		{\partial \widehat{T}_{1}}
		\right\vert_{
			\lp 
			T_{1}, T_{2}
			\rp}
		= 
		\dfrac{\eta_{r}}
		{\Omega}
		\lb 
		\sigma \, 
		c_{1}^{2}(a_{1}, \ol{b}_{1}) 
		+
		\theta \, 
		c_{1}(a_{1}, \ol{b}_{1}) 
		\rb
		+
		\dfrac{c_{2}(a_{2}, \ol{b}_{2})
			- 
			c_{1}(a_{1}, \ol{b}_{1}) \, 
			c_{1}(a_{2}, \ol{b}_{2})}
		{\eta(a_{1}, \ol{b}_{2})}, \\[5pt]
		d_{12} 
		& = & 
		\left. 
		\frac{\partial g_{1}}
		{\partial \widehat{T}_{2}}
		\right\vert_{
			\lp 
			T_{1}, T_{2}
			\rp}
		= 
		-
		\frac{c_{1}(a_{1}, \ol{b}_{1})}
		{2 \, \Omega}, \\[5pt]
		d_{21} 
		& = & 
		\left. \frac{\partial g_{2}}
		{\partial \widehat{T}_{1}}
		\right\vert_{
			\lp 
			T_{1}, T_{2}
			\rp}
		= 
		- 
		\dfrac{1}
		{\Omega \, \eta(a_{1}, \ol{b}_{2})} 
		\lb 
		\Omega 
		\lp 
		c_{1}(a_{2}, \ol{b}_{2}) 
		- 
		c_{1}(a_{1}, \ol{b}_{1}) 
		\rp 
		+ 
		\lp 
		\theta 
		+ 
		\sigma \, c_{1}(a_{1}, \ol{b}_{1})
		\rp 
		\eta(a_{2}, \ol{b}_{2})
		\rb, \\[5pt]
		d_{22} 
		& = & 
		\left. \frac{\partial g_{2}}
		{\partial \widehat{T}_{2}}
		\right\vert_{
			\lp 
			T_{1}, T_{2}
			\rp}
		= 
		\dfrac{1}{2 \, \Omega}.
	\end{eqnarray*}
	
	Therefore,
	we get
	\begin{align}
		\label{eqn:ThetaAsym1}
		\left( 
		\widehat{\theta}_{\mbox{\tiny T}}, 
		\widehat{\sigma}_{\mbox{\tiny T}}
		\right) 
		& \sim  
		\mathcal{AN}
		\left( 
		\left(\theta, \sigma \right), 
		\dfrac{1}{n} \bm{S}_{\mbox{\tiny T}}
		\right),
		\quad 
		\mbox{where}
		\quad 
		\bm{S}_{\mbox{\tiny T}}
		= 
		\bm{D}_{\mbox{\tiny T}}
		\bm{\Sigma}_{\mbox{\tiny T}} 
		\bm{D}_{\mbox{\tiny T}}'. 
	\end{align}
	
	Thus,
	from Eq.~\eqref{eqn:ARE1} 
	and 
	Eq.~\eqref{eqn:ThetaAsym1}, 
	we have 
	\begin{align}
		\label{eqn:LN_LC_K_MLE_ARE}
		\mbox{ARE}
		\left( 
		\left(\widehat{\theta}_{\mbox{\tiny T}},
		\widehat{\sigma}_{\mbox{\tiny T}}\right),
		\left(\widehat{\theta}_{\mbox{\tiny MLE}},\widehat{\sigma}_{\mbox{\tiny MLE}}\right)
		\right)
		& =
		\left(
		\mbox{det}
		\left(
		\bm{S}_{\mbox{\tiny MLE}}
		\right)/
		{
			\mbox{det}
			\left(
			\bm{S}_{\mbox{\tiny T}}
			\right)
		}
		\right)^{0.5}.
	\end{align}
	
	\begin{table}[hbt!]
		\caption{From $N(\theta, \sigma^{2} = 3^{2})$, 
			and we vary the location parameter $\theta$.
			Inequality used \eqref{eqn:abCondition2}.}
		\label{table:ARETable2}
		\centering
		\begin{tabular}{|c|c|c|c|c|c|c|c|c|c|c|}
			\hline 
			\multicolumn{2}{|c|}{Proportions} & 
			\multicolumn{9}{|c|}{$\theta$} \\
			\hline 
			$(a_{1}, b_{1})$ & $(a_{2}, b_{2})$ & 
			$-25$ & $-15$ & $-10$ & $-5$ & 0 & 5 & 10 & 15 & 25 \\
			\hline\hline 
			$(0.02, 0.02)$ & 
			$(0.02, 0.02)$ & 
			0.943 & 0.943 & 0.943 & 0.943 & 0.943 & 0.943 & 0.943 & 0.943 & 0.943 \\
			$(0.02, 0.02)$ & 
			$(0.00, 0.04)$ & 
			0.903 & 0.931 & 0.944 & 0.952 & 0.946 & 0.903 & 0.794 & 0.599 & 0.121 \\
			\hline\hline 
			$(0.05, 0.05)$ & 
			$(0.05, 0.05)$ & 
			0.872 & 0.872 & 0.872 & 0.872 & 0.872 & 0.872 & 0.872 & 0.872 & 0.872 \\
			$(0.05, 0.05)$ & 
			$(0.00, 0.10)$ & 
			0.878 & 0.890 & 0.897 & 0.901 & 0.883 & 0.746 & 0.206 & 0.334 & 0.650 \\
			\hline\hline 
			$(0.10, 0.10)$ & 
			$(0.10, 0.10)$ & 
			0.769 & 0.769 & 0.769 & 0.769 & 0.769 & 0.769 & 0.769 & 0.769 & 0.769 \\
			$(0.10, 0.10)$ & 
			$(0.00, 0.20)$ & 
			0.851 & 0.850 & 0.849 & 0.842 & 0.805 & 0.330 & 0.684 & 0.797 & 0.831 \\
			\hline\hline 
			$(0.15, 0.15)$ & 
			$(0.15, 0.15)$ & 
			0.676 & 0.676 & 0.676 & 0.676 & 0.676 & 0.676 & 0.676 & 0.676 & 0.676 \\
			$(0.15, 0.15)$ & 
			$(0.00, 0.30)$ &
			0.812 & 0.809 & 0.806 & 0.797 & 0.753 & 0.249 & 0.788 & 0.810 & 0.815 \\
			\hline 
		\end{tabular}
	\end{table}
	
\begin{figure}[hbt!]
		\centering
		\includegraphics[width=0.85\linewidth]{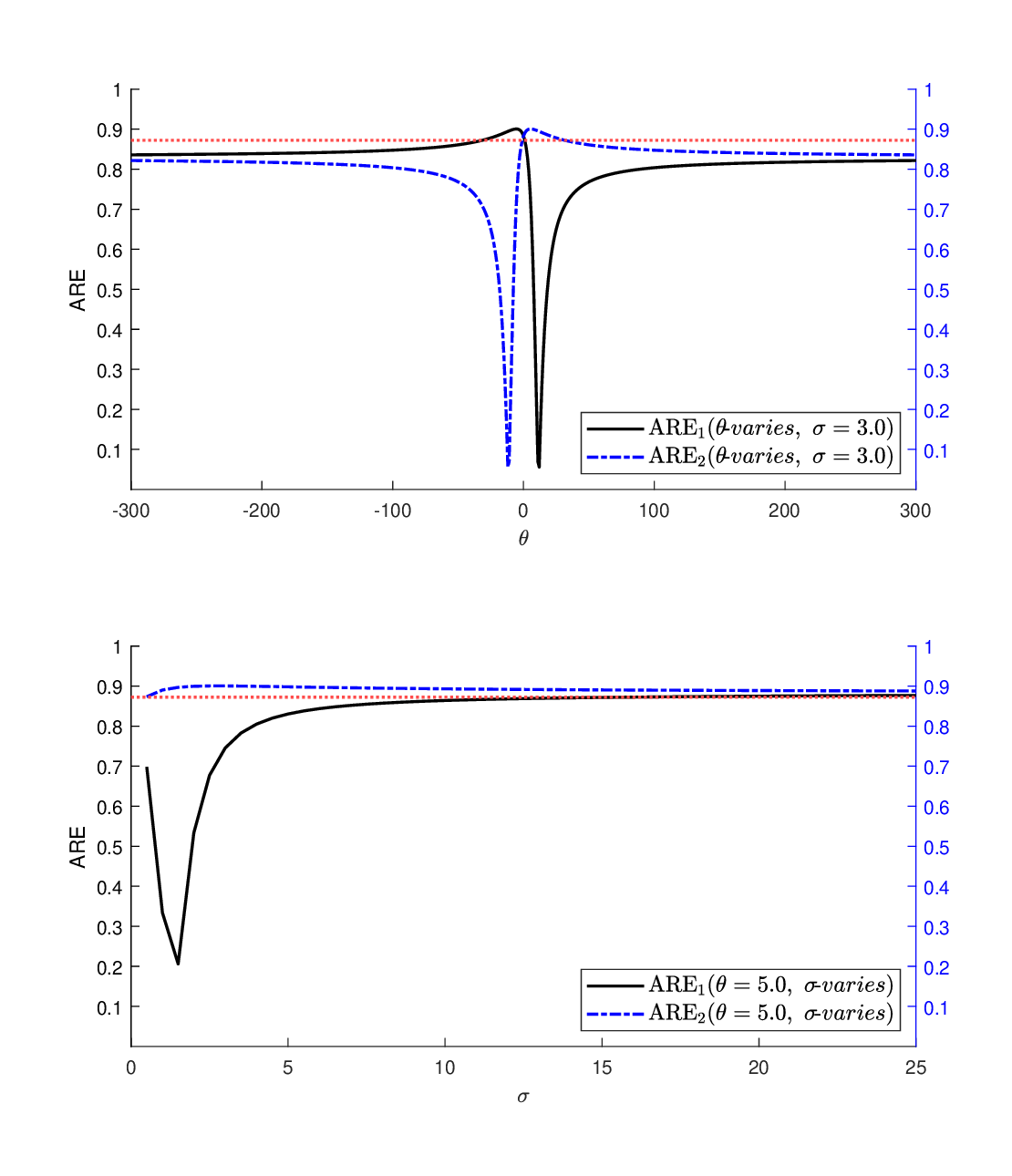}
		\vspace{-1.10cm}
		\begin{table}[H]
			\centering
			{\scriptsize
				\begin{tabular}{lcl}
					$ARE_{1}(\theta \mbox{-varies}, \sigma = 3)$ & $=$ & 
					$ARE(\theta \mbox{-varies}, \sigma = 3, 
					(a_{1},b_{1}) = (0.05, 0.05), 
					(a_{2},b_{2}) = (0.00, 0.10))$ \\
					$ARE_{2}(\theta \mbox{-varies}, \sigma = 3)$ & $=$ &
					$ARE(\theta \mbox{-varies}, \sigma = 3, 
					(a_{1},b_{1}) = (0.05, 0.05), 
					(a_{2},b_{2}) = (0.10, 0.00))$ \\
					$ARE_{1}(\theta = 5, \sigma \mbox{-varies})$ & $=$ &
					$ARE(\theta = 5, \sigma \mbox{-varies}, 
					(a_{1},b_{1}) = (0.05, 0.05), 
					(a_{2},b_{2}) = (0.00, 0.10))$ \\
					$ARE_{2}(\theta = 5, \sigma \mbox{-varies})$ & $=$ &
					$ARE(\theta = 5, \sigma \mbox{-varies}, 
					(a_{1},b_{1}) = (0.05, 0.05), 
					(a_{2},b_{2}) = (0.10, 0.00))$ \\
				\end{tabular}
			}
		\end{table}
		\vspace{-0.5cm}
		\caption{Normal ARE curves under trimming inequalities~\eqref{eqn:abCondition2} 
			or~\eqref{eqn:abCondition3}.
		}
		\label{fig:Normal_ARE_Curves}
	\end{figure}

	Numerical values of the AREs, computed using Eq.~\eqref{eqn:LN_LC_K_MLE_ARE}, are reported in Table~\ref{table:ARETable2} for various trimming proportions satisfying inequality~\eqref{eqn:abCondition2}, under a normal distribution with fixed $\sigma = 3$ and varying location parameter $\theta$. The corresponding ARE curve is shown in the top panel of Figure~\ref{fig:Normal_ARE_Curves} (solid black line) for trimming proportions $(a_1, b_1) = (0.05, 0.05)$ and $(a_2, b_2) = (0.00, 0.10)$.
	
	Due to space constraints, we omit the ARE table for the reverse setting where $\theta = 5$ is fixed and $\sigma$ varies, but the associated curve is displayed in the bottom panel of Figure~\ref{eqn:abCondition3} (solid black line), using the same trimming configuration.
	
	Although similar tables can be generated for other trimming proportions under inequality~\eqref{eqn:abCondition3}, we include only the corresponding ARE curves. For the case with $\sigma = 3$ fixed and varying $\theta$, the top panel of Figure~\ref{eqn:abCondition3} shows the result as a dash-dot blue curve. 
	Likewise, when $\theta = 5$ is fixed and $\sigma$ varies, the bottom panel presents the corresponding ARE curve using the same line style.
	
	Several key findings emerge from Table~\ref{table:ARETable2} and Figure~\ref{fig:Normal_ARE_Curves}. 
	In Figure~\ref{fig:Normal_ARE_Curves}, 
	the horizontal red dotted line represents the ARE value obtained from a normal distribution using identical trimming proportions for both moments, i.e., 
	\((a_1, b_1) = (a_2, b_2) = (0.05, 0.05)\).
	When the location parameter \(\theta\) is close to zero, the AREs corresponding to trimming inequalities~\eqref{eqn:abCondition2} or~\eqref{eqn:abCondition3} exceed those obtained under equal trimming for both moments. This observation supports the primary motivation of this study: when the data and underlying model are approximately symmetric about the origin, applying the same trimming to both moments can unintentionally exclude different sets of observations, thereby reducing efficiency (see Note~\ref{note:DiffRel2}). In contrast, under inequalities~\eqref{eqn:abCondition2} or~\eqref{eqn:abCondition3}, asymmetric trimming may preserve informative observations for specific moments, leading to higher efficiency.
	
	However, when \(\left| \theta \right|\) becomes large, 
	trimming inequalities~\eqref{eqn:abCondition2} 
	and~\eqref{eqn:abCondition3} do not trim the same 
	data points across moments, unlike equal trimming,
	which can lead to reduced efficiency and lower AREs.
	
	Table~\ref{table:ARETable2} also illustrates that the ARE generally decreases with increasing trimming proportions. Nonetheless, the decline in ARE is slower under asymmetric trimming (inequalities~\eqref{eqn:abCondition2} and~\eqref{eqn:abCondition3}) compared to equal trimming. In other words, the ARE curves in Figure~\ref{fig:Normal_ARE_Curves} exhibit similar shapes across trimming levels, but their height decreases more gradually under asymmetric configurations, suggesting a more stable robustness-efficiency trade-off.
	
	Importantly, the lowest ARE values occur when  
	\(
	T_2 - \eta_r T_1^2 \to 0^+,
	\)
	indicating near-singularity in the scale estimation formula. If the true values of \(\theta\) and \(\sigma\) result in  
	\(
	T_2 - \eta_r T_1^2 \to 0^+,
	\)
	this may lead to  
	\(
	\widehat{T}_{2} 
	-  
	\eta_{r} \,
	\widehat{T}_{1}^{2}
	< 
	0,
	\)
	or equivalently,  
	\(
	\widehat{T}_{2} / \widehat{T}_{1}^{2}
	<
	\eta_{r},
	\)
	making the estimation unstable. 
	For instance, with fixed \(\theta = 5\) and varying \(\sigma\) (Figure~\ref{fig:Normal_ARE_Curves}, bottom panel), 
	the ARE drops sharply as  
	\(
	T_2 - \eta_r T_1^2 \to 0^+,
	\)
	with the trimming inequality~\eqref{eqn:abCondition2}.
	Otherwise, for both trimming inequalities, 
	the ARE stabilizes as \(\sigma\) increases, 
	showing that the MTM method performs reliably 
	even with widely dispersed or contaminated data.
	
	For a fixed negative value, such as \(\theta = -5\), 
	the ARE curves in Figure~\ref{fig:Normal_ARE_Curves}
	(bottom panel) retain their shape,
	but the roles of the black and blue curves are reversed.
	
	\subsection{Fr\'{e}chet Model}
	
	As a member of the location-scale family 
	of distributions, the pdf and cdf are,
	respectively, given by 
	\begin{eqnarray*}
		f(x)
		&=& 
		\dfrac{\alpha}{\sigma}
		\left( 
		\dfrac{x - \theta}{\sigma}
		\right)^{-(\alpha + 1)}
		\exp 
		\left[ 
		- 
		\left( 
		\dfrac{\sigma}{x - \theta}
		\right)^{\alpha}
		\right]; 
		\quad 
		x > \theta, \\[5pt]
		F(x) 
		&=&
		\exp 
		\left[
		-\left( 
		\frac{\sigma }{x - \theta}
		\right)^{^{\alpha }}
		\right],
	\end{eqnarray*}
	where 
	$-\infty < \theta < \infty$, 
	$\sigma > 0$, and $\alpha > 0$
	are, respectively, 
	the location, scale, and shape parameters. 
	But to investigate the performance of 
	general MTM for strictly positive 
	distribution,  
	we set the location parameter 
	$\theta = 0$.
	Thus, 
	we are left to estimate the scale parameter, $\sigma$, 
	and the shape parameter, $\alpha$,
	of the distribution.
	Furthermore, for ease of estimation, instead of directly estimating the shape 
	parameter \( \alpha \), we estimate the \textit{tail index}, defined as its 
	reciprocal, \( \beta := 1/\alpha \).
	The moments and quantile functions
	are given by
	\begin{eqnarray*}
		\E 
		\left[ 
		X^k
		\right] 
		&=&
		\sigma^{k} \,
		\Gamma
		\left(
		1 - k \, \beta
		\right);
		\quad 
		k \, \beta < 1, 
		\quad 
		\mbox{and} 
		\quad 
		F^{-1}(u) 
		= 
		\sigma 
		\left( - \log (u) \right)^{-\beta}.
	\end{eqnarray*}

	\begin{figure}[hbt!]
		\centering
		\includegraphics[width=0.95\linewidth]{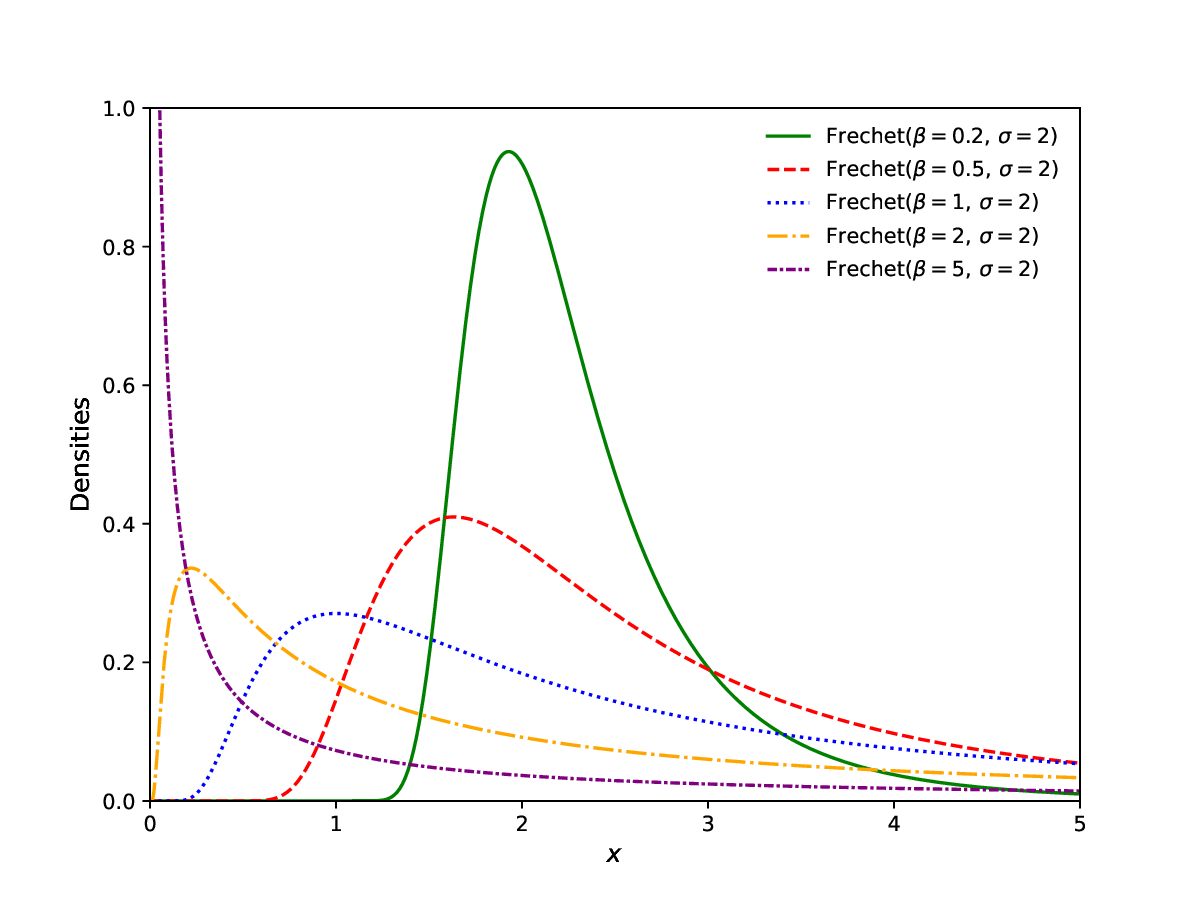}
		\caption{Frechet Densities. 
			A larger $\beta$ (i.e., a smaller shape parameter $\alpha$) implies a heavier tail. This is because
			$
			S(x; \sigma, \beta_{1}) 
			\le 
			S(x; \sigma, \beta_{2})
			$
			for all $x \ge 0$ and $\beta_{1} \le \beta_{2}$, where $S$ denotes the survival function.
		}
		\label{fig:FrechetDensities}
	\end{figure}
	
	Unlike for location-scale member, 
	we take the two functions
	\begin{align}
		\label{eqn:FHFun1}
		h_{1}(x)
		& = 
		\log{(x)}
		\quad 
		\mbox{and} 
		\quad 
		h_{2}(x) 
		= 
		\left( 
		\log{(x)}
		\right)^{2}.
	\end{align}
	
	Thus, with $H_{j} := h_{j} \circ F^{-1}$,
	it follows that 
	\begin{eqnarray}
		& & 
		H_{1}(u) 
		=
		h_{1}
		\left(
		F^{-1}(u)
		\right) 
		= 
		\log{(\sigma)} 
		- 
		\beta \,\log{(-\log{(u)})}, 
		\label{eqn:FH1} \\
		\implies 
		& & 
		H_{1}^{'}(u) 
		= 
		- 
		\dfrac{\beta}{u \log{(u)}},
		\label{eqn:FDerH1} \\ 
		& & 
		H_{2}(u) 
		=
		h_{2}
		\left(
		F^{-1}(u)
		\right) 
		= 
		\left( 
		\log{(\sigma)}
		\right)^{2}
		- 
		2 \beta
		\log{(\sigma)}
		\log{( -\log{(u)})} 
		+ 
		\beta^{2} 
		\left( 
		\log{( -\log{(u)})}
		\right)^{2}, 
		\label{eqn:FH2} \\
		\implies & & 
		H_{2}^{'}(u) 
		= 
		\frac{2 \beta^{2} \log{( -\log(u))}}{u \log(u)}
		-
		\frac{2 \beta \log(\sigma)}{u \log(u)}
		=
		\frac{2 \beta}{u \log{(u)}}
		\left( 
		\beta 
		\log{( -\log(u))}
		- 
		\log{(\sigma)}
		\right).
		\label{eqn:FDerH2}
	\end{eqnarray}
	
	With the trimming proportions 
	$(a_{1}, b_{1})$ and $(a_{2},b_{2})$
	as given in \eqref{eqn:abCondition2},
	then from Eq.~\eqref{eq:sample_mtmG}, 
	the first two sample trimmed-moments 
	are given by:
	\begin{align}
		\label{eqn:FS1}
		\begin{cases}
			\displaystyle 
			\widehat{T}_{1} 
			=
			\frac{1}{n-\lfloor na_{1} \rfloor - \lfloor nb_{1} \rfloor}
			\sum_{i=\lfloor na_{1} \rfloor+1}^{n-\lfloor nb_{1} \rfloor}{h_{1}(X_{i:n})}
			= 
			\frac{1}{n-\lfloor na_{1} \rfloor - \lfloor nb_{1} \rfloor}
			\sum_{i=\lfloor na_{1} \rfloor+1}^{n-\lfloor nb_{1} \rfloor}
			\log {\lp X_{i:n} \rp }, \\[15pt]
			\displaystyle 
			\widehat{T}_{2} 
			=
			\frac{1}{n-\lfloor na_{2} \rfloor - \lfloor nb_{2} \rfloor}
			\sum_{i=\lfloor na_{2} \rfloor+1}^{n-\lfloor nb_{2} \rfloor}{h_{2}(X_{i:n})}
			= 
			\frac{1}{n-\lfloor na_{2} \rfloor - \lfloor nb_{2} \rfloor}
			\sum_{i=\lfloor na_{2} \rfloor+1}^{n-\lfloor nb_{2} \rfloor}
			\lp 
			\log{ \lp X_{i:n} \rp }
			\rp^{2}.
		\end{cases}
	\end{align}
	
	The corresponding first two population 
	trimmed-moments using Eq.~\eqref{eq:pop_mtmG}
	takes the form:
	\begin{align}
		\label{eqn:FP1}
		\begin{cases}
			T_{1}
			= 
			\displaystyle 
			\frac{1}{1-a_{1}-b_{1}}
			\int_{a_{1}}^{\ol{b}_{1}}
			H_{1}
			\lp 
			u
			\rp \, du
			= 
			\log{(\sigma)} 
			-
			\beta \kappa_{1}(a_{1},\ol{b}_{1}), \\[10pt]
			T_{2}
			= 
			\displaystyle 
			\frac{1}{1-a_{2}-b_{2}}
			\int_{a_{2}}^{\ol{b}_{2}}
			H_{2}
			\lp 
			u
			\rp \, du
			=
			\left(\log{(\sigma)}\right)^{2}
			- 
			2 \beta \log{(\sigma)}
			\kappa_{1}(a_{2},\ol{b}_{2})
			+ 
			\beta^{2} 
			\kappa_{2}(a_{2},\ol{b}_{2}),
		\end{cases}
	\end{align}
	where
	\begin{align}
		\label{eqn:IntDefn1}
		\kappa_{k}(a,b) 
		& : = 
		\dfrac{1}{b - a}
		\int_{a}^{b} 
		\lb 
		\Delta (u)
		\rb^{k} \, du, 
		\quad 
		\Delta(u) 
		:= 
		\log{\left( - \log{(u)} \right)}, 
		\quad 
		k \ge 1,
	\end{align}
	do not depend on the parameters 
	$\beta$ and $\sigma$ to be estimated.
	
	Setting 
	$
	\left(
	T_{1}, T_{2} 
	\right) 
	= 
	\left( 
	\widehat{T}_{1}, \widehat{T}_{2}
	\right)
	$
	from Eqs.~\eqref{eqn:FS1} and 
	\eqref{eqn:FP1}, 
	and solving for $\beta$ and $\sigma$, 
	we get the explicit trimmed $L$-estimators as 
	\begin{eqnarray}
		\label{eqn:FEst1}
		\begin{cases}
			\widehat{\beta}_{\mbox{\tiny T}}
			=
			\dfrac{\pm \, 1}
			{
				\sqrt{
					\zeta(a_{1},\ol{b}_2)}}
			\lp 
			\widehat{T}_{2} 
			-  
			\zeta_{r} \,
			\widehat{T}_{1}^{2}
			\rp^{1/2}
			+ 
			\dfrac{
				\widehat{T}_{1}
				\lp 
				\kappa_1(a_{2},\ol{b}_{2})
				- 
				\kappa_1(a_{1}, \ol{b}_{1})
				\rp 
			}
			{\zeta(a_{1},\ol{b}_{2})}
			=:
			g_{1}
			\left( 
			\widehat{T}_{1}, 
			\widehat{T}_{2}
			\right), \\[15pt]
			\widehat{\sigma}_{\mbox{\tiny T}}
			= 
			\exp 
			\left\{ 
			\widehat{T}_{1}
			+ 
			\widehat{\beta}_{\mbox{\tiny T}} \, 
			\kappa_{1}(a_{1}, \ol{b}_{1})
			\right\}
			=:
			g_{2}
			\left( 
			\widehat{T}_{1}, 
			\widehat{T}_{2}
			\right),
		\end{cases}
	\end{eqnarray}
	where, 
	as in Eq.~\eqref{eqn:EtaDefn1}, 
	and for \( 1 \le i, j \le 2 \), 
	we define
	\begin{align}
		\label{eqn:ZetaDefn1}
		\zeta(a_{i},\ol{b}_{j}) 
		& := 
		\kappa_{1}^{2}(a_{i},\ol{b}_{i})
		- 
		2 \kappa_{1}(a_{i},\ol{b}_{i}) \kappa_1(a_{j},\ol{b}_{j}) 
		+ 
		\kappa_{2}(a_{j},\ol{b}_j), \
		1 \le i,j \le 2, 
		\quad 
		\zeta_{r} 
		: = 
		\dfrac{\zeta(a_j,\ol{b}_{j})}
		{\zeta(a_{i},\ol{b}_j)}.
	\end{align}
	
	We now summarize some results, 
	similar to those in Proposition \ref{prop:EtaRealtion1},
	in Corollary \ref{cor:ZetaRealtion1}.
	
	\begin{cor}
		\label{cor:ZetaRealtion1}
		With the trimming proportions
		$\lp a_{i}, b_{i} \rp$ and 
		$\lp a_{j}, b_{j} \rp$
		satisfying the inequality \eqref{eqn:abCondition1}, 
		it follows that 
		\begin{itemize}
			\item[(i)]
			\(
			\zeta 
			\lp 
			a_{i},\ol{b}_{j}
			\rp 
			= 
			\kappa_{1}^{2} 
			\lp 
			a_{i}, \ol{b}_{i}
			\rp 
			-
			2 
			\kappa_{1} 
			\lp 
			a_{i}, \ol{b}_{i}
			\rp 
			\kappa_{1} 
			\lp 
			a_{j}, \ol{b}_{j}
			\rp 
			+ 
			\kappa_{2} 
			\lp 
			a_{j}, \ol{b}_{j}
			\rp 
			> 
			0.
			\)
			
			\item[(ii)]
			\(
			\kappa_{k}
			\lp 
			a_{i}, \ol{b}_{i}
			\rp 
			\le 
			\kappa_{k}
			\lp 
			a_{j}, \ol{b}_{j}
			\rp,
			\)
			for any odd positive integer $k$.
			
			\item[(iii)]
			\(
			0
			< 
			\zeta_{r} 
			= 
			\dfrac{\zeta(a_j,\ol{b}_{j})}
			{\zeta(a_{i},\ol{b}_j)}
			\le 
			1.
			\)
			
			\item[(iv)]
			\(
			\kappa_2
			\lp 
			a_j, \overline{b}_{j}
			\rp 
			\ge  
			\zeta_r \, 
			\kappa_1^2
			\lp 
			a_i, \overline{b}_i
			\rp.
			\)
			
			\item[(v)]
			\(
			T_{2} - \zeta_{r} T_{1}^{2} \ge 0.
			\)
		\end{itemize}
		
		\begin{proof}
			See Appendix \ref{sec:Appendix1}.
		\end{proof}
	\end{cor}
	
	Similar to Corollary \ref{cor:InequalityFlip1},
	we have the following result. 
	
	\begin{cor}
		\label{cor:InequalityFlip2}
		Under the trimming inequality 
		\eqref{eqn:abCondition1Flip},
		that is, 
		\eqref{eqn:abCondition3} for Fr{\'e}chet model,
		all results of Corollary~\ref{cor:ZetaRealtion1}
		remain valid except for Part~\((ii)\),
		which instead takes the form
		\(
		\kappa_{k}
		\lp 
		a_{i}, \ol{b}_{i}
		\rp 
		\ge  
		\kappa_{k}
		\lp 
		a_{j}, \ol{b}_{j}
		\rp,
		\)
		for any odd positive integer $k$.
	\end{cor}
	
	\begin{note}
		\label{note:Frechet_Beta_Terms1}
		Define 
		\begin{align}
			\beta_{\mbox{\tiny FT}}
			& = 
			\dfrac{1}
			{
				\sqrt{
					\zeta(a_{1},\ol{b}_2)}}
			\lp 
			{T}_{2} 
			-  
			\zeta_{r} \,
			{T}_{1}^{2}
			\rp^{1/2}
			\quad 
			\mbox{and} 
			\quad 
			\beta_{\mbox{\tiny ST}}
			=
			\dfrac{
				{T}_{1}
				\lp 
				\kappa_1(a_{2},\ol{b}_{2})
				- 
				\kappa_1(a_{1}, \ol{b}_{1})
				\rp 
			}
			{\zeta(a_{1},\ol{b}_{2})}.
		\end{align}
		Similarly as mentioned in 
		Note \ref{note:LS_Sigma_Terms1},
		under the trimming inequality 
		\eqref{eqn:abCondition2}, 
		a consistent directional relationship between
		\(\beta_{\text{\tiny FT}}\) and 
		\(\beta_{\text{\tiny ST}}\) is not guaranteed.
		
		For illustration, consider 
		\( X \sim \mbox{Fr{\'e}chet}
		(\beta = 2, \sigma = 3) \). 
		Then, for
		\[
		(a_{1}, b_{1}) = (0.02, 0.02) 
		\quad \text{and} \quad 
		(a_{2}, b_{2}) = (0.00, 0.03),
		\]
		we have
		\[
		\beta_{\mbox{\tiny FT}}
		= 
		1.860
		> 
		0.139
		=
		\beta_{\mbox{\tiny ST}}.
		\]
		
		However, for
		\[
		(a_{1}, b_{1}) = (0.02, 0.02) 
		\quad \text{and} \quad 
		(a_{2}, b_{2}) = (0.00, 0.20),
		\]
		we obtain
		\[
		\beta_{\mbox{\tiny FT}}
		= 
		0.738
		< 
		1.262
		=
		\beta_{\mbox{\tiny ST}}.
		\]
		\qed 
	\end{note}
	
	As in Section~\ref{sec:LS1}, 
	the trimmed estimators 
	\( 
	\lp 
	\widehat{\beta}_{\mbox{\tiny T}},
	\widehat{\sigma}_{\mbox{\tiny T}}
	\rp 
	\)
	in Eq.~\eqref{eqn:FEst1} are derived 
	from Eq.~\eqref{eqn:FP1}, 
	ensuring that one of the solutions 
	from the \(\pm\) branch for \(\beta\) 
	is theoretically positive. 
	However, 
	for finite samples, it is possible that 
	\(
	\widehat{\beta}_{\mbox{\tiny T}}
	< 
	0,
	\)
	in which case the proposed trimmed \(L\)-moment estimation is invalid.
	To determine the correct sign of 
	\(\widehat{\beta}_{\text{\tiny T}}\), 
	in Eq.~\eqref{eqn:FEst1}, 
	we adopt the following strategy. 
	As established in Corollary~\ref{cor:ZetaRealtion1}, 
	it holds that 
	\(
	T_{2} - \zeta_{r} T_{1}^{2} \ge 0.
	\)
	Therefore, assuming 
	\(
	\lp 
	\widehat{T}_{2} 
	-  
	\eta_{r} \,
	\widehat{T}_{1}^{2}
	\rp^{1/2} 
	\ge 
	0,
	\)
	we consider  
	\begin{align} 
		\label{eqn:BetaTwoTerms1Hat}
		\widehat{\beta}_{\mbox{\tiny FT}}
		& = 
		\dfrac{1}
		{
			\sqrt{
				\zeta(a_{1},\ol{b}_2)}}
		\lp 
		\widehat{T}_{2} 
		-  
		\zeta_{r} \,
		\widehat{T}_{1}^{2}
		\rp^{1/2} 
		\quad 
		\mbox{and} 
		\quad 
		\widehat{\beta}_{\mbox{\tiny ST}}
		=
		\dfrac{
			\widehat{T}_{1}
			\lp 
			\kappa_1(a_{2},\ol{b}_{2})
			- 
			\kappa_1(a_{1}, \ol{b}_{1})
			\rp 
		}
		{\zeta(a_{1},\ol{b}_{2})}. 
	\end{align}
	
	But 
	for $a_{1} \ne a_{2}$ or $b_{1} \ne b_{2}$,
	it is possible to have
	\(
	\widehat{T}_{2} 
	-  
	\zeta_{r} \,
	\widehat{T}_{1}^{2}
	< 
	0,
	\)
	i.e., 
	\(
	\widehat{T}_{2} / \widehat{T}_{1}^{2}
	<
	\zeta_{r},
	\)
	and in that case we consider 
	\begin{align}
		\label{eqn:BetaTwoTerms2Hat}
		\widehat{\beta}_{\mbox{\tiny FT}}
		& = 
		\dfrac{\pm \, 1}
		{
			\sqrt{
				\zeta(a_{1},\ol{b}_2)}}
		\lp 
		\left| 
		\widehat{T}_{2} 
		-  
		\zeta_{r} \,
		\widehat{T}_{1}^{2}
		\right|
		\rp^{1/2}
		\quad 
		\mbox{and} 
		\quad 
		\widehat{\beta}_{\mbox{\tiny ST}}
		= 
		\dfrac{
			\widehat{T}_{1}
			\lp 
			\kappa_1(a_{2},\ol{b}_{21})
			- 
			\kappa_1(a_{1}, \ol{b}_{1})
			\rp 
		}
		{\zeta(a_{1},\ol{b}_{2})}.
	\end{align}
	
	Therefore, for Fr{\'e}chet models, we have 
	\(
	\widehat{\beta}_{\text{\tiny FT}} \ge 0,
	\)
	by construction. In contrast, 
	\(
	\widehat{\beta}_{\text{\tiny ST}}
	\)
	may take any real value, i.e., 
	\(
	\widehat{\beta}_{\text{\tiny ST}} \in \mathbb{R}.
	\)
	
	If 
	\(
	\widehat{\beta}_{\text{\tiny FT}} 
	< 
	\widehat{\beta}_{\text{\tiny ST}},
	\)
	we define 
	\(
	\widehat{\beta}_{-}
	:= 
	- 
	\widehat{\beta}_{\text{\tiny FT}} 
	+ 
	\widehat{\beta}_{\text{\tiny ST}} > 0.
	\) 
	Similarly, if 
	\(
	\widehat{\beta}_{\text{\tiny FT}} 
	> 
	\left| 
	\widehat{\beta}_{\text{\tiny ST}} 
	\right|,
	\)
	we define 
	\(
	\widehat{\beta}_{+}
	:= 
	\widehat{\beta}_{\text{\tiny FT}} 
	+ 
	\widehat{\beta}_{\text{\tiny ST}} > 0.
	\)
	Finally, if 
	\(
	\widehat{\beta}_{\text{\tiny FT}} 
	\le 
	\left| 
	\widehat{\beta}_{\text{\tiny ST}} 
	\right|
	\)
	and 
	\(
	\widehat{\beta}_{\text{\tiny ST}} 
	\le 
	0,
	\)
	then the method fails to yield
	a nonnegative estimate of 
	\(
	\widehat{\beta}_{\text{\tiny T}} \ge 0.
	\)
	
	\begin{algorithm}[h!bt]
		\caption{Trimmed Estimation of \(\left( \widehat{\beta}_{\text{\tiny T}}, \widehat{\sigma}_{\text{\tiny T}} \right)\) in Fr{\'e}chet Model}
		\label{alg:FrechetAlgorithm1}
		\SetAlgoLined
		\LinesNumbered
		\SetNlSty{textbf}{}{:}
		\SetAlgoNlRelativeSize{-0}
		
		\KwIn{Sample data and trimming proportions \((a_1, b_1), (a_2, b_2)\) satisfying \eqref{eqn:abCondition2} or \eqref{eqn:abCondition3}.}
		
		\KwOut{Trimmed estimator vector 
			\(
			\lp 
			\widehat{\beta}_{\text{\tiny T}}, 
			\widehat{\sigma}_{\text{\tiny T}}
			\rp 
			\).}
		
		\textbf{Compute}\;
		\Indp
		\( \widehat{T}_1 \) and \( \widehat{T}_2 \) from Eq.~\eqref{eqn:FS1}\;
		\( \kappa_1(a_1, \ol{b}_1), \; \kappa_1(a_2, \ol{b}_2) \),
		and 
		\(\kappa_2(a_2, \ol{b}_2) \) from Eq.~\eqref{eqn:IntDefn1}\;
		\( \zeta(a_1, \ol{b}_2), \; \zeta(a_2, \ol{b}_2) \), and \( \zeta_r \) 
		from Eq.~\eqref{eqn:ZetaDefn1}\;
		\( \widehat{\beta}_{\text{\tiny FT}} \) 
		and 
		\( \widehat{\beta}_{\text{\tiny ST}} \) from Eq.~\eqref{eqn:BetaTwoTerms1Hat}
		or 
		Eq.~\eqref{eqn:BetaTwoTerms2Hat}\;
		\( \widehat{\beta}_{\mbox{\tiny MLE}} \) from Eq.~\eqref{eqn:FrechetBetaMLE}\;
		\Indm
		
		\textbf{Set}
		\( \widehat{\beta}_+ \leftarrow \widehat{\beta}_{\text{\tiny FT}} 
		+ \widehat{\beta}_{\text{\tiny ST}} \)
		and 
		\( \widehat{\beta}_- \leftarrow -\widehat{\beta}_{\text{\tiny FT}} + \widehat{\beta}_{\text{\tiny ST}} \)\;
		\If{\( a_1 = a_2 \) and \( b_1 = b_2 \)}{
			\( \widehat{\beta}_{\text{\tiny ST}} \leftarrow 0 \),
			resulting in
			\( \widehat{\beta}_{\text{\tiny T}} 
			\leftarrow
			{\widehat{\beta}_{\text{\tiny FT}}} \)\;
			\Return \( \widehat{\beta}_{\text{\tiny T}} \)\;
		}
		\ElseIf{
			\( (a_2 \le a_1 \text{ and } b_1 \le b_2) \)
			\text{ or }
			\( (a_1 \le a_2 \text{ and } b_2 \le b_1) \)
		}
		{
			\If{\( 
				\max 
				\left\{ 
				\widehat{\beta}_{-}, 
				\widehat{\beta}_{+}
				\right\}
				\le  
				0 \)}{
				\tcc{Exit and update trimming proportions to ensure 
					\(\widehat{\beta}_{\text{\tiny T}} > 0.\)}
				\textbf{Ensure} 
				\(
				\max 
				\left\{ 
				\widehat{\beta}_{-}, 
				\widehat{\beta}_{+}
				\right\}
				> 
				0
				\)\;
			}
			\ElseIf{\( 
				\widehat{\beta}_{-}
				\le 
				0
				\mbox{ and }
				\widehat{\beta}_{+}
				>
				0
				\)}
			{\( \widehat{\beta}_{\text{\tiny T}} \leftarrow \widehat{\beta}_+ \)\;}
			\ElseIf{\( 
				\widehat{\beta}_{-}
				>
				0
				\mbox{ and }
				\widehat{\beta}_{+}
				\le 
				0
				\)}
			{\( \widehat{\beta}_{\text{\tiny T}} \leftarrow \widehat{\beta}_- \)\;}
			\ElseIf{
				\( 
				\min 
				\left\{ 
				\widehat{\beta}_{-}, 
				\widehat{\beta}_{+}
				\right\}
				\ge 
				0
				\)
			}
			{
				\If{\( |\widehat{\beta}_- - \widehat{\beta}_{\text{\tiny MLE}}| < |\widehat{\beta}_+ - \widehat{\beta}_{\text{\tiny MLE}}| \)}
				{
					\( \widehat{\beta}_{\text{\tiny T}} \leftarrow \widehat{\beta}_- \)
				}
				\Else{
					\( \widehat{\beta}_{\text{\tiny T}} \leftarrow \widehat{\beta}_+ \)\;
				}
			}
		}
		
		\Return \( \widehat{\beta}_{\text{\tiny T}} \)\;
		
		\( \widehat{\sigma}_{\text{\tiny T}} 
		\leftarrow 
		\exp 
		\left\{ 
		\widehat{T}_{1}
		+ 
		\widehat{\beta}_{\mbox{\tiny T}} \, 
		\kappa_{1}(a_{1}, \ol{b}_{1})
		\right\}
		\)\;
		\Return 
		\( 
		\lp 
		\widehat{\beta}_{\text{\tiny T}}, 
		\;
		\widehat{\sigma}_{\text{\tiny T}}
		\rp 
		\)\;
	\end{algorithm}
	
	\begin{itemize}
		\item 
		If $a_{1} = a_{2} = a$ and $b_{1} = b_{2} = b$,
		then it follows that 
		\( 
		\zeta(a_{1},\ol{b}_2) 
		= 
		\eta(a_{2},\ol{b}_2), 
		\)
		that is, 
		$\zeta_{r} = 1$.
		Also, 
		\( 
		\kappa_1(a_{1},\ol{b}_{1})
		- 
		\kappa_1(a_{2}, \ol{b}_{2})
		=
		0.
		\) 
		Thus, 
		\[
		\widehat{\beta}_{\text{\tiny T}}
		=
		\widehat{\beta}_{+}
		=
		\left( 
		\dfrac{\widehat{T}_{2} 
			-  
			\widehat{T}_{1}^{2}
		}
		{
			\kappa_{2}(a, \ol{b}) 
			- 
			\kappa_{1}(a, \ol{b})^{2}
		}
		\right)^{1/2}.
		\]
		
		\item 
		If \( a_{1} \ne a_{2} \) 
		or \( b_{1} \ne b_{2} \),
		selecting the appropriate sign in
		Eq.~\eqref{eqn:FEst1} becomes nontrivial. 
		To proceed, 
		and following \citet[][Eq.~(6)]{MR4801492}, 
		the \( L \)-moment estimate of \( \beta \) is given by
		
		\[
		\widehat{\beta}_{\mbox{\tiny LM}}
		= 
		\dfrac{
			\log{(2)} 
			+
			\log \left( \sum_{i=1}^n (i-1) X_{i:n} \right) 
			- 
			\log \left( n(n-1) \overline{X} \right)
		}
		{\log{(2)}},
		\]
		but this closed-form expression is valid only if $\beta < 1$.
		Thus, 
		instead of relying on this restricted formula, 
		we treat $\beta$ as a tunable parameter and estimate
		it using the method of maximum likelihood.
		Following \citet{MR3706798}, 
		$\widehat{\beta}_{\mbox{\tiny MLE}}$ 
		is the unique root of the strictly increasing function
		\begin{align}
			\label{eqn:FrechetBetaMLE}
			\xi(\beta) 
			& =
			\beta 
			+ 
			\sum_{i=1}^{n} x_{i}^{-1/\beta}
			\log{(x_{i})}
			\left(
			\sum_{i=1}^{n} x_{i}^{-1/\beta }
			\right)^{-1}
			-
			n^{-1} 
			\sum_{i=1}^{n}
			\log{(x_{i})}, 
			\quad 
			\mbox{i.e.,}
			\quad 
			\xi
			\left( 
			\widehat{\beta}_{\mbox{\tiny MLE}}
			\right) 
			=
			0.
		\end{align}
		The maximum likelihood estimator of 
		$\sigma$ is given by 
		\begin{align}
			\widehat{\sigma}_{\mbox{\tiny MLE}}
			& = 
			\left(
			\dfrac{1}{n} 
			\sum_{i=1}^{n} 
			x_{i}^{-1/\widehat{\beta}_{\mbox{\tiny MLE}}}
			\right)^{-\widehat{\beta}_{\mbox{\tiny MLE}}}.
		\end{align}
		
		If
		\(
		\max 
		\left\{ 
		\widehat{\beta}_{-}, 
		\widehat{\beta}_{+}
		\right\}
		\le 
		0,
		\)
		there is no valid positive solution for the tail index parameter \(\beta\). 
		Otherwise, if 
		\(
		\widehat{\beta}_{-} \le 0
		\) 
		and 
		\(
		\widehat{\beta}_{+} > 0,
		\) 
		we set 
		\(
		\widehat{\beta}_{\text{\tiny T}} = \widehat{\beta}_+.
		\) 
		Similarly, if 
		\(
		\widehat{\beta}_{-} > 0
		\) 
		and 
		\(
		\widehat{\beta}_{+} \le 0,
		\)
		we set 
		\(
		\widehat{\beta}_{\text{\tiny T}} = \widehat{\beta}_-.
		\) 
		Finally, 
		if 
		\(
		\min 
		\left\{ 
		\widehat{\beta}_{-}, 
		\widehat{\beta}_{+}
		\right\}
		\ge 
		0,
		\) 
		we select the estimate closer 
		to the maximum likelihood estimate. 
		That is,
		\begin{eqnarray}
			\label{eqn:BetaChoice1}
			\widehat{\beta}_{\text{\tiny T}}
			& = &
			\begin{cases}
				\widehat{\beta}_{-}, & 
				\mbox{if }
				\left| 
				\widehat{\beta}_{-} 
				- 
				\widehat{\beta}_{\mbox{\tiny MLE}}
				\right| 
				< 
				\left| 
				\widehat{\beta}_{+} 
				- 
				\widehat{\beta}_{\mbox{\tiny MLE}}
				\right|, \\[10pt]
				\widehat{\beta}_{+}, & 
				\mbox{if }
				\left| 
				\widehat{\beta}_{+} 
				- 
				\widehat{\beta}_{\mbox{\tiny MLE}}
				\right| 
				\le 
				\left| 
				\widehat{\beta}_{-} 
				- 
				\widehat{\beta}_{\mbox{\tiny MLE}}
				\right|.
			\end{cases}
		\end{eqnarray}
	\end{itemize}
	
	\begin{note}
		Heuristically, 
		the shape parameter $\alpha$ of a distribution 
		is often approximately inversely proportional 
		to the coefficient of variation (CV). 
		Thus, defining $\beta = \alpha^{-1}$, 
		we have that $\beta$ is approximately 
		proportional to the CV. 
		Therefore, 
		to solve Eq.~\eqref{eqn:FrechetBetaMLE} numerically, 
		we can initialize the iterative algorithm as
		\(
		\beta_{\text{\tiny start}} = \widehat{\text{CV}}.
		\)
		\qed 
	\end{note}
	
	A formal procedure for estimating the parameter vector 
	\(
	\left( 
	\widehat{\beta}_{\text{\tiny T}}, 
	\widehat{\sigma}_{\text{\tiny T}} 
	\right)
	\)
	is summarized in Algorithm~\ref{alg:FrechetAlgorithm1}.
	
	We now calculate $\mathbf{\Sigma}_{\text{\tiny T}}$ by 
	using Theorem \ref{thm:MTM_Var1}.
	That is,
	\begin{eqnarray}
		\sigma_{11}^{2} 
		& = & 
		\Gamma(1,1) V(1,1) 
		\nonumber \\ 
		& = & 
		\Gamma(1,1) 
		\int_{a_{1}}^{\ol{b}_{1}}\int_{a_{1}}^{\ol{b}_{1}}
		{K(w,v)} 
		H_{1}'(w) H_{1}'(v) \, dv \, dw 
		\nonumber \\
		& = & 
		\beta^{2} \,
		\Gamma(1,1) 
		\int_{a_{1}}^{\ol{b}_{1}}\int_{a_{1}}^{\ol{b}_{1}}
		\dfrac{K(w,v)}
		{v w \log(v) \, \log(w)}
		\, dv \, dw 
		\nonumber \\
		& = & 
		\beta^{2} \,
		\Psi_{111}
		\label{eqn:MTM_FSigma112}, 
		\nonumber \\
		\sigma_{12}^{2}
		& = & 
		\Gamma(1,2) 
		V(1,2) 
		\nonumber \\ 
		& = & 
		\Gamma(1,2) 
		\int_{a_{1}}^{\ol{b}_{1}}\int_{a_{2}}^{\ol{b}_{2}}
		{K(w,v)} 
		H_{1}'(w) H_{2}'(v) \, dv \, dw 
		\nonumber \\ 
		& = & 
		2 \beta^{2} \Gamma(1,2)
		\int_{a_{1}}^{\ol{b}_{1}} 
		\int_{a_{2}}^{\ol{b}_{2}} 
		\frac{K(w,v)}
		{v w \log{(v)} \log{(w)}}
		\left( 
		\log{(\sigma)}
		-
		\beta 
		\log{( -\log(v))}
		\right)
		dv \, dw 
		\nonumber \\
		& = & 
		2 \beta^{2} \log{(\sigma)} 
		\Psi_{121}
		-
		2 \beta^{3}
		\Psi_{122}
		\label{eqn:MTM_FSigma122}, \\
		\sigma_{22}^{2} 
		& = & 
		\Gamma(2,2) \, V(2,2) 
		\nonumber \\ 
		& = & 
		\Gamma(2,2) 
		\int_{a_{2}}^{\ol{b}_{2}}
		\int_{a_{2}}^{\ol{b}_{2}}
		{K(w,v)} 
		H_{2}'(w) H_{2}'(v) \, dv \, dw 
		\nonumber \\
		& = & 
		4 \beta^{2} \Gamma(2,2)
		\int_{a_{2}}^{\ol{b}_{2}}
		\int_{a_{2}}^{\ol{b}_{2}}
		\dfrac{K(w,v)} 
		{{v w \log{(v)} \log{(w)}}}
		\left( 
		\beta \log{( -\log(w))}
		- 
		\log{(\sigma)}
		\right)
		\left( 
		\beta \log{( -\log(v))}
		- 
		\log{(\sigma)}
		\right)
		dv \, dw 
		\nonumber \\ 
		& = & 
		4 \beta^{2} \left( \log{(\sigma)} \right)^{2}
		\Psi_{221}
		- 
		8 \beta^{3} \log{(\sigma)} 
		\Psi_{222}
		+ 
		4 \beta^{4}
		\Psi_{223},
		\label{eqn:MTM_FSigma222}
	\end{eqnarray}
	where the double integral terms 
	\(\Psi_{ijk}\), for \(1 \le i, j \le 2\) 
	and \(1 \le k \le 3\), 
	do not depend on the parameters to be estimated.
	The corresponding simplified single-integral expressions, 
	derived using Theorem~\ref{thm:MTM_Var1}, 
	are provided in Appendix~\ref{sec:Appendix1}.
	
	The Jacobian matrix $\bm{D}_{\mbox{\tiny T}}$ 
	is found by differentiating the functions 
	$g_{1}$ and $g_{2}$ from Eq.~\eqref{eqn:FEst1}
	and then evaluating its derivatives at 
	$\lp T_{1}, T_{2} \rp$.
	That is, 
	\begin{align}
		d_{11}
		& =
		\left. \frac{\partial g_{1}}
		{\partial \widehat{T}_{1}}
		\right\vert_{
			\lp 
			T_{1}, T_{2}
			\rp}
		= 
		\frac{- \zeta_{r} \, {T}_{1}}
		{\sqrt{\zeta(a_1, \ol{b}_2)}} 
		\left( 
		{T}_{2} 
		- 
		\zeta_{r} \,
		{T}_{1}^{2} 
		\right)^{-\frac{1}{2}}
		+ 
		\frac{\kappa_1(a_2, \ol{b}_{2}) 
			- 
			\kappa_1(a_1, \ol{b}_{1})}
		{\zeta(a_1, \ol{b}_{2})}, 
		\label{eqn:FD11} \\ 
		d_{12}
		& = 
		\left. \frac{\partial g_{1}}
		{\partial \widehat{T}_{2}}
		\right\vert_{
			\lp 
			T_{1}, T_{2}
			\rp}
		= 
		\frac{1}
		{2 \, \sqrt{\zeta(a_1, \ol{b}_2)}} 
		\left( 
		{T}_2 
		- 
		\zeta_{r} \,
		{T}_{1}^{2} 
		\right)^{-\frac{1}{2}},
		\label{eqn:FD12} \\
		d_{21}
		& =
		\left. \frac{\partial g_{2}}
		{\partial \widehat{T}_{1}}
		\right\vert_{
			\lp 
			T_{1}, T_{2}
			\rp}
		= 
		\lp 
		1 + d_{11} \, \kappa_{1} \lp a_{1}, \ol{b}_{1} \rp
		\rp 
		\exp 
		\left\{ 
		{T}_{1}
		+ 
		{\beta} \, 
		\kappa_{1}(a_{1}, \ol{b}_{1})
		\right\}
		= 
		\sigma
		\lp 
		1 + d_{11} \, \kappa_{1} \lp a_{1}, \ol{b}_{1} \rp
		\rp, \\ 
		d_{22}
		& =
		\left. \frac{\partial g_{2}}
		{\partial \widehat{T}_{2}}
		\right\vert_{
			\lp 
			T_{1}, T_{2}
			\rp}
		=
		d_{12} \, \kappa_{1} \lp a_{1}, \ol{b}_{1} \rp
		\exp 
		\left\{ 
		{T}_{1}
		+ 
		{\beta} \, 
		\kappa_{1}(a_{1}, \ol{b}_{1})
		\right\}
		=
		\sigma \, 
		d_{12} \, \kappa_{1} \lp a_{1}, \ol{b}_{1} \rp.
	\end{align}
	
	Thus, we get
	\begin{align}
		\label{eqn:FAsym1}
		\left( 
		\widehat{\beta}_{\mbox{\tiny T}}, 
		\widehat{\sigma}_{\mbox{\tiny T}}
		\right) 
		& \sim  
		\mathcal{AN}
		\left( 
		\left(\beta, \sigma \right), 
		\dfrac{1}{n} \bm{S}_{\mbox{\tiny T}}
		\right),
		\quad 
		\mbox{where}
		\quad 
		\bm{S}_{\mbox{\tiny T}}
		= 
		\bm{D}_{\mbox{\tiny T}}
		\bm{\Sigma}_{\mbox{\tiny T}} 
		\bm{D}_{\mbox{\tiny T}}'. 
	\end{align}
	
	Again from \cite{MR3706798}, 
	we can derive 
	\begin{align}
		\label{eqn:FThetaAsym1}
		\left( 
		\widehat{\beta}_{\mbox{\tiny MLE}}, 
		\widehat{\sigma}_{\mbox{\tiny MLE}}
		\right) 
		& \sim  
		\mathcal{AN}
		\left( 
		\left(\beta, \sigma \right), 
		\dfrac{1}{n} 
		\bm{S}_{\mbox{\tiny MLE}}
		\right),
	\end{align}
	where 
	\begin{align}
		\label{eqn:FThetaAsym2}
		\bm{S}_{\mbox{\tiny MLE}}
		& = 
		\dfrac{6}{\pi^{2}} 
		\begin{bmatrix}
			\beta^{2} & 
			(1 - \gamma) \sigma \beta^{2} \\ 
			(1 - \gamma ) \sigma \beta^{2} & 
			(\sigma \, \beta)^{2} 
			\left( 
			(\gamma - 1)^{2} + \pi^{2}/6
			\right) 
		\end{bmatrix}
		\quad 
		\mbox{giving}
		\quad 
		\mbox{det}
		\left( 
		\bm{S}_{\mbox{\tiny MLE}}
		\right) 
		=
		\dfrac{6 \beta^{4} \sigma^{2}}{\pi^{2}},
	\end{align}
	and 
	$\gamma := - \Gamma'(1) = 0.57721566490$
	is the Euler–Mascheroni constant.
	
	Finally, 
	from Eq.~\eqref{eqn:ARE1} and 
	Eq.~\eqref{eqn:FAsym1}, 
	we have 
	\begin{align}
		\label{eqn:F_ARE}
		\mbox{ARE}
		\left( 
		\left(\widehat{\beta}_{\mbox{\tiny T}},
		\widehat{\sigma}_{\mbox{\tiny T}}\right),
		\left(\widehat{\beta}_{\mbox{\tiny MLE}},\widehat{\sigma}_{\mbox{\tiny MLE}}\right)
		\right)
		& =
		\left(
		\mbox{det}
		\left(
		\bm{S}_{\mbox{\tiny MLE}}
		\right)/
		{
			\mbox{det}
			\left(
			\bm{S}_{\mbox{\tiny T}}
			\right)
		}
		\right)^{0.5}.
	\end{align}
	
	\begin{table}[hbt!]
		\caption{From $\mbox{Frechet}(\beta, \sigma = 2)$, 
			and we vary the tail index $\beta$, 
			reciprocal of the shape parameter.
			A larger $\beta$ 
			(i.e., a smaller shape parameter $\alpha$)
			implies a heavier tail.
			Inequality used \eqref{eqn:abCondition2}.}
		\label{table:ARETable4}
		\centering
		\begin{tabular}{|c|c|c|c|c|c|c|c|c|c|c|}
			\hline 
			\multicolumn{2}{|c|}{Proportions} & 
			\multicolumn{9}{|c|}{$\beta$} \\
			\hline 
			$(a_{1}, b_{1})$ & $(a_{2}, b_{2})$ & 
			0.1 & 0.2 & 0.5 & 1 & 2 & 5 & 10 & 15 & 25 \\
			\hline\hline 
			$(0.02, 0.02)$ & 
			$(0.02, 0.02)$ & 
			0.771 & 0.771 & 0.771 & 0.771 & 0.771 & 0.771 & 0.771 & 0.771 & 0.771 \\ 
			$(0.02, 0.02)$ & 
			$(0.00, 0.04)$ & 
			0.259 & 0.633 & 0.786 & 0.815 & 0.827 & 0.833 & 0.834 & 0.835 & 0.835 \\
			\hline\hline 
			$(0.05, 0.05)$ & 
			$(0.05, 0.05)$ & 
			0.754 & 0.754 & 0.754 & 0.754 & 0.754 & 0.754 & 0.754 & 0.754 & 0.754 \\ 
			$(0.05, 0.05)$ & 
			$(0.00, 0.10)$ & 
			0.458 & 0.004 & 0.610 & 0.759 & 0.809 & 0.833 & 0.840 & 0.842 & 0.844 \\
			\hline\hline 
			$(0.10, 0.10)$ & 
			$(0.10, 0.10)$ & 
			0.693 & 0.693 & 0.693 & 0.693 & 0.693 & 0.693 & 0.693 & 0.693 & 0.693 \\ 
			$(0.10, 0.10)$ & 
			$(0.00, 0.20)$ & 
			0.760 & 0.624 & 0.036 & 0.560 & 0.736 & 0.802 & 0.819 & 0.824 & 0.828 \\
			\hline\hline 
			$(0.15, 0.15)$ & 
			$(0.15, 0.15)$ & 
			0.623 & 0.623 & 0.623 & 0.623 & 0.623 & 0.623 & 0.623 & 0.623 & 0.623 \\
			$(0.15, 0.15)$ & 
			$(0.00, 0.30)$ &
			0.812 & 0.762 & 0.439 & 0.296 & 0.674 & 0.786 & 0.810 & 0.817 & 0.822 \\
			\hline 
		\end{tabular}
	\end{table}
	
	\begin{figure}[hbt!]
		\centering
		\includegraphics[width=0.85\linewidth]{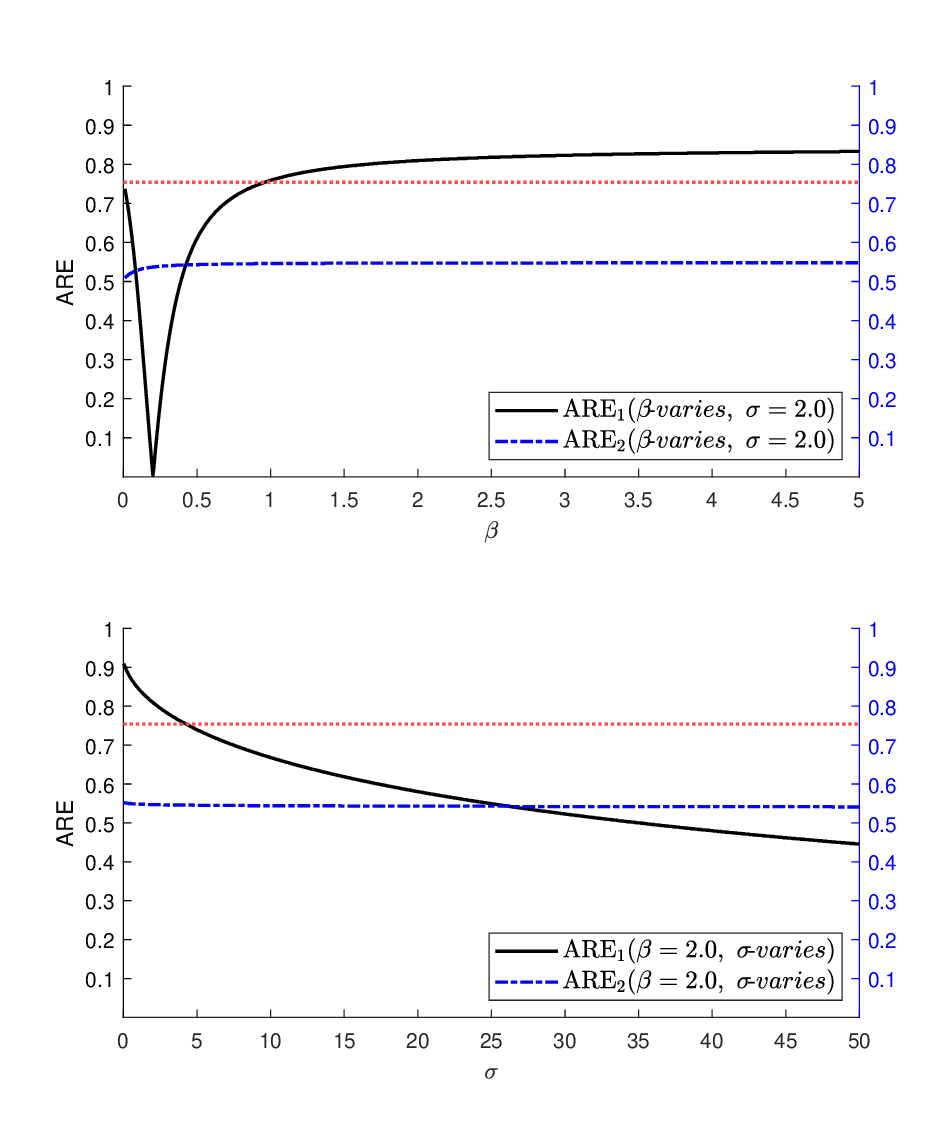}
		\vspace{-1.10cm}
		\begin{table}[H]
			\centering
			{\scriptsize
				\begin{tabular}{lcl}
					$ARE_{1}(\beta \mbox{-varies}, \sigma = 2)$ & $=$ & 
					$ARE(\beta \mbox{-varies}, \sigma = 2, 
					(a_{1},b_{1}) = (0.05, 0.05), 
					(a_{2},b_{2}) = (0.00, 0.10))$ \\
					$ARE_{2}(\beta \mbox{-varies}, \sigma = 2)$ & $=$ &
					$ARE(\beta \mbox{-varies}, \sigma = 2, 
					(a_{1},b_{1}) = (0.05, 0.05), 
					(a_{2},b_{2}) = (0.10, 0.00))$ \\
					$ARE_{1}(\beta = 2, \sigma \mbox{-varies})$ & $=$ &
					$ARE(\beta = 2, \sigma \mbox{-varies}, 
					(a_{1},b_{1}) = (0.05, 0.05), 
					(a_{2},b_{2}) = (0.00, 0.10))$ \\
					$ARE_{2}(\beta = 2, \sigma \mbox{-varies})$ & $=$ &
					$ARE(\beta = 2, \sigma \mbox{-varies}, 
					(a_{1},b_{1}) = (0.05, 0.05), 
					(a_{2},b_{2}) = (0.10, 0.00))$ \\
				\end{tabular}
			}
		\end{table}
		\vspace{-0.50cm}
		\caption{Fr{\'e}chet ARE curves under trimming inequalities~\eqref{eqn:abCondition2} 
			or~\eqref{eqn:abCondition3}.
		}
		\label{fig:Frechet_ARE_Curves}
	\end{figure}
	
	Like in Section~\ref{sec:LS1}, several key findings emerge from Table~\ref{table:ARETable4} and Figure~\ref{fig:Frechet_ARE_Curves}. In Figure~\ref{fig:Frechet_ARE_Curves}, the horizontal red dotted line represents the ARE value obtained using identical trimming proportions for both moments, i.e., \((a_1, b_1) = (a_2, b_2) = (0.05, 0.05)\).
	
	Unlike the location-scale model, where the distribution may be symmetric about the origin, the Fr{\'e}chet model considered here is strictly positive. Therefore, trimming inequalities~\eqref{eqn:abCondition2} and~\eqref{eqn:abCondition3} lead to different subsets of data being excluded for different moments. As shown in the top panel of Figure~\ref{fig:Frechet_ARE_Curves}, and under trimming inequality~\eqref{eqn:abCondition2}, the ARE values remain higher than those from equal trimming when 
	\(
	T_2 - \zeta_r T_1^2 \gg 0.
	\)
	For example, the ARE drops sharply at approximately \(\beta = 0.2\) and \(\sigma = 2\), where 
	\(
	T_2 - \zeta_r T_1^2 = 5.2052 \times 10^{-7}.
	\)
	
	For \(\beta \ge 1\) and \(\sigma = 1\), the ARE values under trimming inequality~\eqref{eqn:abCondition2} exceed those under equal trimming, as seen by the black curve lying above the red dotted line. This indicates that for heavier-tailed Fr{\'e}chet distributions, using distinct trimming proportions for each moment can yield higher efficiency than applying the same trimming to both.
	
	In contrast, the ARE values under trimming inequality~\eqref{eqn:abCondition3} remain consistently flat. This is because the scenario 
	\(
	T_2 - \zeta_r T_1^2 \to 0^+
	\)
	rarely occurs under this configuration.
	
	The interpretation of the bottom panel in Figure~\ref{fig:Frechet_ARE_Curves} is similar. The ARE curve corresponding to trimming inequality~\eqref{eqn:abCondition2} decreases with increasing \(\sigma\) until 
	\(
	T_2 - \zeta_r T_1^2 \to 0^+,
	\)
	after which it begins to rise again. The ARE curve under trimming inequality~\eqref{eqn:abCondition3} remains nearly constant throughout.
	
	\section{Simulation Study}
	\label{sec:Sim}
	
	This section complements the theoretical results developed in previous sections with simulation studies. 
	The primary objectives are to determine the sample size required for the estimators to become effectively unbiased
	(given their asymptotic unbiasedness), validate their asymptotic normality, 
	and evaluate their finite-sample relative efficiencies (REs) in relation to their corresponding asymptotic relative efficiencies (AREs).
	To compute the RE of MTM estimators, we use the MLE as the benchmark. Accordingly, the definition of ARE in 
	Eq.~\eqref{eqn:ARE1} is adapted for finite-sample performance 
	as follows: 
	\begin{equation} 
		\label{eq:finite_relative_efficiency_benchmark_MLE}
		RE(\mbox{MTM, MLE}) 
		=
		\frac{\text{asymptotic variance of MLE estimator}}
		{\text{small-sample variance of a competing MTM estimator}},
	\end{equation}
	where the numerator is defined in Eq.~\eqref{eqn:ARE1}, 
	and the denominator is expressed as:
	\[
	\left(
	\mbox{det}
	\begin{bmatrix}
		E 
		\left[ 
		\left(\widehat{\theta} - \theta \right)^{2}
		\right] & 
		E 
		\left[ 
		\left(\widehat{\theta} - \theta \right)
		\left(\widehat{\sigma} - \sigma \right)
		\right] \\[15pt]
		E 
		\left[ 
		\left(\widehat{\theta} - \theta \right)
		\left(\widehat{\sigma} - \sigma \right)
		\right] & 
		E 
		\left[ 
		\left(\widehat{\sigma} - \sigma \right)^{2}
		\right] 
	\end{bmatrix}
	\right)^{1/2}.
	\]
	
	From a specified distribution \(F\), 
	we generate 10,000 samples of a given length
	\(n\) using Monte Carlo simulations.
	For each sample, 
	we estimate the parameters of \(F\) using various
	\(T\)-estimators under trimming 
	inequality~\eqref{eqn:abCondition2} 
	or~\eqref{eqn:abCondition3}, 
	and compute the sample mean and relative efficiency (RE) 
	from 10{,}000 replicates.
	This procedure is repeated 10 times,
	and the averages of the resulting 10 means and 10 REs, 
	along with their standard deviations, are reported.
	
	We start the simulation study with the
	normal distribution
	$N \lp \theta = 0.1, \sigma^{2} = 5^{2} \rp$,
	using the following specifications. 
	\begin{itemize}
		\item 
		Sample size: 
		$n = 100, 500, 1000.$
		
		\item 
		Estimators of $\theta$ and $\sigma$:
		\begin{itemize}
			\item 
			MLE, 
			
			\item 
			MTM using trimming proportions specified by either inequality~\eqref{eqn:abCondition2} or~\eqref{eqn:abCondition3}
		\end{itemize}
	\end{itemize}
	
	\begin{table}[hbt!]
		\caption{Normal model with parameters 
			$\theta = 0.1$ and $\sigma = 5$.}
		\label{table:NormalSim1}
		\centering
		\begin{tabular}{|c|c|cc|cc|cc|cc|}
			\hline 
			\multicolumn{2}{|c|}{Estimator} & 
			\multicolumn{2}{|c|}{$n = 100$} & 
			\multicolumn{2}{|c|}{$n = 500$} & 
			\multicolumn{2}{|c|}{$n = 1000$} & 
			\multicolumn{2}{|c|}{$n \to \infty$} \\
			\cline{1-2}
			$(a_{1},b_{1})$ & $(a_{2}, b_{2})$ & 
			\multicolumn{1}{c}{$\widehat{\theta}/\theta$} & 
			\multicolumn{1}{c|}{$\widehat{\sigma}/\sigma$} &
			\multicolumn{1}{c}{$\widehat{\theta}/\theta$} & 
			\multicolumn{1}{c|}{$\widehat{\sigma}/\sigma$} &
			\multicolumn{1}{c}{$\widehat{\theta}/\theta$} & 
			\multicolumn{1}{c|}{$\widehat{\sigma}/\sigma$} &  
			\multicolumn{1}{c}{$\widehat{\theta}/\theta$} & 
			\multicolumn{1}{c|}{$\widehat{\sigma}/\sigma$} \\
			\hline\hline 
			\multicolumn{10}{|c|}{} \\[-2.50ex] 
			\multicolumn{10}{|c|}
			{Mean values of
				$\widehat{\theta}/\theta$ 
				and
				$\widehat{\sigma}/\sigma$.} \\
			\hline 
			\multicolumn{2}{|c|}{MLE} & 0.98 & 0.99 & 1.01 & 1.00 & 1.00 & 1.00 & 1 & 1 \\
			\hline 
			$(0.00,0.00)$ & $(0.00,0.00)$ & 0.98 &0.99 &1.01 &1.00 & 1.00&1.00 & 1 & 1 \\
			$(0.00,0.05)$ & $(0.00,0.05)$ & 1.10 &1.00 &1.01 &1.00 &1.00 &1.00 & 1 & 1 \\
			$(0.00,0.10)$ & $(0.00,0.10)$ & 1.10 &1.00 &1.02 & 1.00&1.01 & 1.00& 1 & 1\\
			\hline
			\multicolumn{10}{|c|}{With strict trimming 
				inequality~\eqref{eqn:abCondition2}.} \\
			\hline 
			$(0.10,0.00)$ & $(0.05,0.05)$ & 0.84 & 1.00& 0.98&1.00 &0.98 &1.00 & 1 & 1 \\ 
			$(0.05,0.05)$ & $(0.00,0.10)$ & 1.00 &0.99 &1.00 &1.00 & 1.00&1.00 & 1 & 1 \\
			$(0.10,0.10)$ & $(0.00,0.20)$ & 1.01 & 0.99&1.00 &1.00 &1.00 & 1.00& 1 & 1 \\
			$(0.15,0.15)$ & $(0.00,0.30)$ & 0.97 &0.99 &1.01 &1.00 &1.01 &1.00 & 1 & 1 \\
			\hline
			\multicolumn{10}{|c|}{With strict trimming 
				inequality~\eqref{eqn:abCondition3}.} \\
			\hline 
			$(0.00,0.10)$ & $(0.05,0.05)$ &1.07
			&1.00 &1.02 &1.00 & 1.01&1.00 & 1 & 1 \\
			$(0.05,0.05)$ & $(0.10,0.00)$ &1.00
			&0.99 &0.99 &1.00 &1.00 &1.00 & 1 & 1 \\
			$(0.10,0.10)$ & $(0.20,0.00)$ & 0.99
			& 0.99& 1.00&1.00 &1.00 & 1.00& 1 & 1 \\
			$(0.15,0.15)$ & $(0.30,0.00)$ & 1.00
			&0.99 &1.00 &1.00 &1.00 &1.00 & 1 & 1 \\
			$(0.25,0.50)$ & $(0.50,0.25)$ & 1.03
			&1.00 &1.00 &1.00 & 1.00&1.00 & 1 & 1 \\
			\hline 
			\multicolumn{10}{|c|}
			{\small Finite-sample efficiencies (RE) of MTMs relative to MLEs.} \\
			\hline 
			\multicolumn{2}{|c|}{MLE} & 
			\multicolumn{2}{|c|}{0.999} & 
			\multicolumn{2}{|c|}{0.996} & 
			\multicolumn{2}{|c|}{0.994} & 
			\multicolumn{2}{|c|}{1} \\
			\hline 
			$(0.00,0.00)$ & $(0.00,0.00)$ & 
			\multicolumn{2}{|c|}{0.999} &
			\multicolumn{2}{|c|}{0.996} & 
			\multicolumn{2}{|c|}{0.994} &
			
			\multicolumn{2}{|c|}{1} \\
			$(0.00,0.05)$ & $(0.00,0.05)$ & 
			\multicolumn{2}{|c|}{0.930} &
			\multicolumn{2}{|c|}{0.934} & 
			\multicolumn{2}{|c|}{0.929} &
			
			\multicolumn{2}{|c|}{0.932} \\
			$(0.00,0.10)$ & $(0.00,0.10)$ & 
			\multicolumn{2}{|c|}{0.877} &
			\multicolumn{2}{|c|}{0.876 } & 
			\multicolumn{2}{|c|}{0.876 } &
			\multicolumn{2}{|c|}{0.872} \\
			\hline
			\multicolumn{10}{|c|}{With strict trimming 
				inequality~\eqref{eqn:abCondition2}.} \\
			\hline
			$(0.10,0.00)$ & $(0.05,0.05)$ & 
			\multicolumn{2}{|c|}{0.874} &
			\multicolumn{2}{|c|}{0.870} & 
			\multicolumn{2}{|c|}{0.872} &
			\multicolumn{2}{|c|}{0.872} \\ 
			$(0.05,0.05)$ & $(0.00,0.10)$ & 
			\multicolumn{2}{|c|}{0.884} &
			\multicolumn{2}{|c|}{0.882} & 
			\multicolumn{2}{|c|}{0.881} &
			\multicolumn{2}{|c|}{0.883} \\
			$(0.10,0.10)$ & $(0.00,0.20)$ & 
			\multicolumn{2}{|c|}{0.808} &
			\multicolumn{2}{|c|}{0.801} & 
			\multicolumn{2}{|c|}{0.805} &
			\multicolumn{2}{|c|}{0.805} \\
			$(0.15,0.15)$ & $(0.00,0.30)$ & 
			\multicolumn{2}{|c|}{0.753} &
			\multicolumn{2}{|c|}{0.748} & 
			\multicolumn{2}{|c|}{0.752} &
			\multicolumn{2}{|c|}{0.752} \\
			\hline 
			\multicolumn{10}{|c|}{With strict trimming 
				inequality~\eqref{eqn:abCondition3}.} \\
			\hline 
			$(0.00,0.10)$ & $(0.05,0.05)$ & 
			\multicolumn{2}{|c|}{0.874} &
			\multicolumn{2}{|c|}{0.880} & 
			\multicolumn{2}{|c|}{0.872} &
			\multicolumn{2}{|c|}{0.876} \\ 
			$(0.05,0.05)$ & $(0.10,0.00)$ & 
			\multicolumn{2}{|c|}{0.888} &
			\multicolumn{2}{|c|}{0.889} & 
			\multicolumn{2}{|c|}{0.881} &
			\multicolumn{2}{|c|}{0.884} \\
			$(0.10,0.10)$ & $(0.20,0.00)$ & 
			\multicolumn{2}{|c|}{0.807} &
			\multicolumn{2}{|c|}{0.805} & 
			\multicolumn{2}{|c|}{0.810} &
			\multicolumn{2}{|c|}{0.807} \\
			$(0.15,0.15)$ & $(0.30,0.00)$ & 
			\multicolumn{2}{|c|}{0.758} &
			\multicolumn{2}{|c|}{0.754} & 
			\multicolumn{2}{|c|}{0.760} &
			\multicolumn{2}{|c|}{0.754} \\
			$(0.25,0.50)$ & $(0.50,0.25)$ & 
			\multicolumn{2}{|c|}{0.493} &
			\multicolumn{2}{|c|}{0.491} & 
			\multicolumn{2}{|c|}{0.488} &
			\multicolumn{2}{|c|}{0.491} \\
			\hline 
		\end{tabular}
	\end{table}
	
	The simulation results are summarized in Table~\ref{table:NormalSim1}. 
	All estimators in the normal case accurately recover both the location parameter \(\theta\) and the scale parameter
	\(\sigma\), becoming nearly unbiased for sample sizes as small as \(n = 100\).
	The relative bias of \(\theta\) estimators under trimming inequality~\eqref{eqn:abCondition2} is generally negative, while that under inequality~\eqref{eqn:abCondition3} tends to be positive. 
	This pattern aligns with the explanation in Note~\ref{note:DiffRel2}: when \(\theta = 0.1\), trimming under inequality~\eqref{eqn:abCondition2} is more likely to remove the same observations from both moments, preserving balance. 
	In contrast, under inequality~\eqref{eqn:abCondition3}, where \(\ol{b}_1 \le \ol{b}_2\), fewer large values are excluded from the second moment, potentially inflating the location estimate. 
	The finite relative efficiencies (FREs) also approach their asymptotic limits, with some converging from above.
	
	\begin{table}[t!b!h!]
		\caption{Fr\'{e}chet model with parameters 
			$\beta = 5$ and $\sigma = 2$.}
		\label{table:FrechetSim2}
		\centering
		\begin{tabular}{|c|c|cc|cc|cc|cc|}
			\hline 
			\multicolumn{2}{|c|}{Estimator} & 
			\multicolumn{2}{|c|}{$n = 100$} & 
			\multicolumn{2}{|c|}{$n = 500$} & 
			\multicolumn{2}{|c|}{$n = 1000$} & 
			\multicolumn{2}{|c|}{$n \to \infty$} \\
			\cline{1-2}
			$(a_{1},b_{1})$ & $(a_{2}, b_{2})$ & 
			\multicolumn{1}{c}{$\widehat{\beta}/\beta$} & 
			\multicolumn{1}{c|}{$\widehat{\sigma}/\sigma$} &
			\multicolumn{1}{c}{$\widehat{\beta}/\beta$} & 
			\multicolumn{1}{c|}{$\widehat{\sigma}/\sigma$} &
			\multicolumn{1}{c}{$\widehat{\beta}/\beta$} & 
			\multicolumn{1}{c|}{$\widehat{\sigma}/\sigma$} &  
			\multicolumn{1}{c}{$\widehat{\beta}/\beta$} & 
			\multicolumn{1}{c|}{$\widehat{\sigma}/\sigma$} \\
			\hline\hline 
			\multicolumn{10}{|c|}{} \\[-2.50ex] 
			\multicolumn{10}{|c|}
			{Mean values of
				$\widehat{\theta}/\theta$ 
				and
				$\widehat{\sigma}/\sigma$.} \\
			\hline 
			\multicolumn{2}{|c|}{MLE} & 0.99 & 1.17 & 1.00 & 1.03 & 1.00 & 1.01 & 1 & 1 \\
			\hline 
			$(0.00,0.00)$ & $(0.00,0.00)$ & 0.99 & 1.19&1.00 &1.03 &1.00 &1.02 & 1 & 1 \\
			$(0.00,0.05)$ & $(0.00,0.05)$ &1.00 &1.18 &1.00 &1.03 &1.00 &1.02 & 1 & 1 \\
			$(0.00,0.10)$ & $(0.00,0.10)$ &1.00 & 1.19&1.00 &1.03 &1.00 &1.02 & 1 & 1\\
			\hline
			\multicolumn{10}{|c|}{With strict trimming 
				inequality~\eqref{eqn:abCondition2}.} \\
			\hline 
			$(0.10,0.00)$ & $(0.05,0.05)$ &1.01 &1.15 &1.00 & 1.03&1.00 &1.01 & 1 & 1 \\ 
			$(0.05,0.05)$ & $(0.00,0.10)$ & 1.00&1.17 &1.00 &1.03 &1.00 &1.02 & 1 & 1 \\
			$(0.10,0.10)$ & $(0.00,0.20)$ &1.00 &1.18 &1.00 &1.03 &1.00 &1.02 & 1 & 1 \\
			$(0.15,0.15)$ & $(0.00,0.30)$ &1.00 &1.18 &1.00 & 1.03& 1.00&1.02 & 1 & 1 \\
			\hline 
			\multicolumn{10}{|c|}{With strict trimming 
				inequality~\eqref{eqn:abCondition3}.} \\
			\hline 
			$(0.00,0.10)$ & $(0.05,0.05)$ & 1.00
			& 1.19 & 1.00 & 1.03 & 1.00 & 1.02 & 1 & 1 \\
			$(0.05,0.05)$ & $(0.10,0.00)$ & 0.99
			&1.26 &1.00 &1.05 &1.00 &1.02 & 1 & 1 \\
			$(0.10,0.10)$ & $(0.20,0.00)$ & 0.99
			& 1.27 & 1.00&1.05 &1.00 & 1.03& 1 & 1 \\
			$(0.15,0.15)$ & $(0.30,0.00)$ & 
			0.99 &1.29 &1.00 &1.05 &1.00 &1.03 & 1 & 1 \\
			$(0.25,0.50)$ & $(0.50,0.25)$ & 
			1.00 &1.23 &1.00 &1.04 & 1.00&1.02 & 1 & 1 \\
			\hline 
			\multicolumn{10}{|c|}
			{\small Finite-sample efficiencies (RE) of MTMs relative to MLEs.} \\
			\hline 
			\multicolumn{2}{|c|}{MLE} & 
			\multicolumn{2}{|c|}{0.729} & 
			\multicolumn{2}{|c|}{0.934} & 
			\multicolumn{2}{|c|}{0.971}& 
			\multicolumn{2}{|c|}{1} \\
			\hline 
			$(0.00,0.00)$ & $(0.00,0.00)$ & 
			\multicolumn{2}{|c|}{0.509} &
			\multicolumn{2}{|c|}{0.651} & 
			\multicolumn{2}{|c|}{0.671} &
			\multicolumn{2}{|c|}{0.690} \\
			
			$(0.00,0.05)$ & $(0.00,0.05)$ & 
			\multicolumn{2}{|c|}{0.626} &
			\multicolumn{2}{|c|}{0.803} & 
			\multicolumn{2}{|c|}{0.831} &
			\multicolumn{2}{|c|}{0.856} \\
			
			$(0.00,0.10)$ & $(0.00,0.10)$ & 
			\multicolumn{2}{|c|}{0.629} &
			\multicolumn{2}{|c|}{0.817} & 
			\multicolumn{2}{|c|}{0.849} &
			\multicolumn{2}{|c|}{0.875} \\
			
			\hline
			\multicolumn{10}{|c|}{With strict trimming 
				inequality~\eqref{eqn:abCondition2}.} \\
			\hline 
			$(0.10,0.00)$ & $(0.05,0.05)$ & 
			\multicolumn{2}{|c|}{0.448} &
			\multicolumn{2}{|c|}{0.588} & 
			\multicolumn{2}{|c|}{0.606} &
			\multicolumn{2}{|c|}{0.627} \\ 
			
			$(0.05,0.05)$ & $(0.00,0.10)$ & 
			\multicolumn{2}{|c|}{0.614} &
			\multicolumn{2}{|c|}{0.784} & 
			\multicolumn{2}{|c|}{0.809} &
			\multicolumn{2}{|c|}{0.833} \\
			
			$(0.10,0.10)$ & $(0.00,0.20)$ & 
			\multicolumn{2}{|c|}{0.583} &
			\multicolumn{2}{|c|}{0.753} & 
			\multicolumn{2}{|c|}{0.773} &
			\multicolumn{2}{|c|}{0.802} \\
			
			$(0.15,0.15)$ & $(0.00,0.30)$ & 
			\multicolumn{2}{|c|}{0.549} &
			\multicolumn{2}{|c|}{0.737} & 
			\multicolumn{2}{|c|}{0.759} &
			\multicolumn{2}{|c|}{0.786} \\
			\hline 
			\multicolumn{10}{|c|}{With strict trimming 
				inequality~\eqref{eqn:abCondition3}.} \\
			\hline 
			$(0.00,0.10)$ & $(0.05,0.05)$ & 
			\multicolumn{2}{|c|}{0.563} &
			\multicolumn{2}{|c|}{0.723} & 
			\multicolumn{2}{|c|}{0.753} &
			\multicolumn{2}{|c|}{0.774} \\ 
			$(0.05,0.05)$ & $(0.10,0.00)$ & 
			\multicolumn{2}{|c|}{0.378} &
			\multicolumn{2}{|c|}{0.512} & 
			\multicolumn{2}{|c|}{0.526} &
			\multicolumn{2}{|c|}{0.548} \\
			$(0.10,0.10)$ & $(0.20,0.00)$ & 
			\multicolumn{2}{|c|}{0.348} &
			\multicolumn{2}{|c|}{0.467} & 
			\multicolumn{2}{|c|}{0.487} &
			\multicolumn{2}{|c|}{0.509} \\
			$(0.15,0.15)$ & $(0.30,0.00)$ & 
			\multicolumn{2}{|c|}{0.317} &
			\multicolumn{2}{|c|}{0.448} & 
			\multicolumn{2}{|c|}{0.470} &
			\multicolumn{2}{|c|}{0.489} \\
			$(0.25,0.50)$ & $(0.50,0.25)$ & 
			\multicolumn{2}{|c|}{0.308} &
			\multicolumn{2}{|c|}{0.424} & 
			\multicolumn{2}{|c|}{0.436} &
			\multicolumn{2}{|c|}{0.457} \\
			\hline 
		\end{tabular}
	\end{table}
	
	Similarly, 
	we continue the simulation study
	with the Fr\'{e}chet distribution 
	$F \lp \beta = 5, \sigma = 2 \rp$,
	using the following specifications. 
	
	\begin{itemize}
		\item 
		Sample size: 
		$n = 100, 500, 1000.$
		
		\item 
		Estimators of $\theta$ and $\sigma$:
		\begin{itemize}
			\item 
			MLE, 
			
			\item  
			MTM using trimming proportions specified by either inequality~\eqref{eqn:abCondition2} or~\eqref{eqn:abCondition3}
		\end{itemize}
	\end{itemize}
	
	The simulation results are summarized 
	in Table~\ref{table:FrechetSim2}. 
	As in the normal case, all estimators 
	in the Fr{\'e}chet model accurately recover 
	both parameters \(\beta\) and \(\sigma\) 
	as the sample size increases and relative bias decreases. 
	The estimator for \(\beta\) becomes nearly unbiased 
	for sample sizes as small as \(n = 100\). 
	However,
	the relative bias of the \(\sigma\) 
	estimators is generally lower under trimming inequality~\eqref{eqn:abCondition2} compared to inequality~\eqref{eqn:abCondition3}, which is expected. 
	For instance, under inequality~\eqref{eqn:abCondition2}
	with \((a_1, b_1) = (0.05, 0.05)\) and \((a_2, b_2) = (0.00, 0.10)\), 
	a greater proportion of large observations 
	are excluded from the second moment calculation. 
	In contrast,
	under inequality~\eqref{eqn:abCondition3} with the same
	\((a_1, b_1)\) but \((a_2, b_2) = (0.10, 0.00)\),
	the second moment retains all larger values due
	to the absence of right-side trimming,
	resulting in higher bias for the scale parameter \(\sigma\). 
	The finite relative efficiencies (FREs)
	also converge toward their asymptotic counterparts,
	though at a slower rate, with some approaching from above.
	
	\section{Real Data Analysis}
	\label{sec:RealData}
	
	In this section,
	we apply both the MTM and MLE approaches to analyze 
	the normalized damage amounts from the 30 most damaging hurricanes 
	in the United States between 1925 and 1995, 
	as reported by \citet{pl98}. 
	This dataset has previously been examined using 
	the same trimming proportions for all moments 
	by \citet{MR2497558}, 
	and using the same winsorizing proportions 
	for all moments by \citet{MR3758788}. 
	The damage figures were normalized to 1995 dollars, 
	accounting for inflation, 
	changes in personal property values, 
	and coastal county population growth. 
	Our objective is to assess how initial 
	assumptions and parameter estimation 
	methods influence model fit.
	
	\begin{figure}[bht!]
		\centering
		\includegraphics[width=0.90\linewidth]{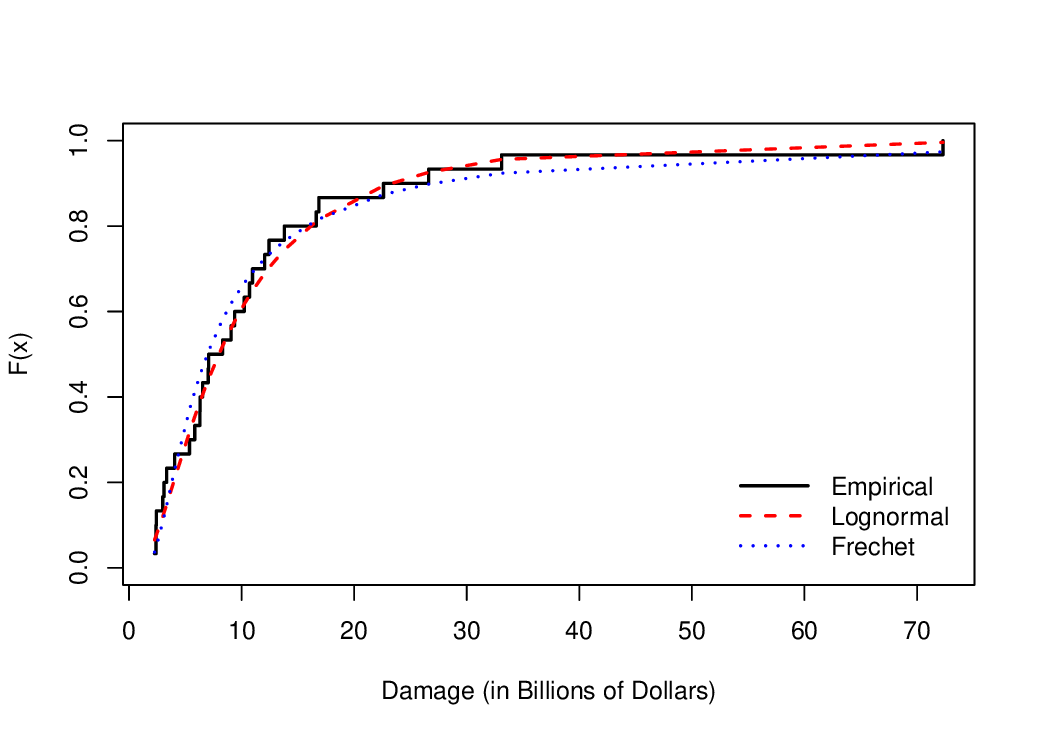}
		\caption{Empirical cumulative distribution function (CDF) of the hurricane damage data overlaid with fitted CDFs from the lognormal and Fr\'echet models, illustrating the diagnostic fit performance of each model.}
		\label{fig:Hurricane_CDFs}
	\end{figure}
	
	Both lognormal and Fr{\'e}chet models provide a good fit to the hurricane damage data, with maximum likelihood estimates summarized in the MLE row of Table~\ref{table:HurricaneDataOrgModBoth}. The corresponding $p$-values are 0.9678 and 0.7611, respectively, and the fitted CDFs are shown alongside the empirical CDF in Figure~\ref{fig:Hurricane_CDFs}. 
	At the 5\% significance level, neither model is rejected, indicating that both are plausible for modeling the data. 
	
	To further evaluate the robustness of the proposed flexible trimming approach, we follow \citet{MR2497558} and introduce a mild data modification by replacing the largest observation, 72.303, with 723.03. The resulting parameter estimates and goodness-of-fit (GOF) measures, using different trimming proportions defined by inequalities~\eqref{eqn:abCondition2} and~\eqref{eqn:abCondition3}, are presented in Table~\ref{table:HurricaneDataOrgModBoth} for both the original and modified datasets. 
	The goodness-of-fit is assessed using the mean absolute deviation
	\[
	\mbox{FIT}
	:=
	\frac{1}{30} 
	\sum_{j=1}^{30} \left|
	\log 
	\left( 
	\widehat{F}^{-1} \left( \frac{j - 0.5}{30} \right)
	\right) 
	- 
	\log
	\left( 
	X_{j:30}
	\right) 
	\right|,
	\]
	between the log-fitted and log-observed data \citep{MR2497558}, along with AIC and BIC.
	
	\begin{table}[hbt!]
		\caption{Parameter estimates and goodness-of-fit measures 
			for the lognormal and Fr\'echet models fitted to the original and modified hurricane loss data.}
		\label{table:HurricaneDataOrgModBoth}
		\centering
		{\footnotesize 
			\begin{tabular}{c|cc|ccccc|ccccc}
				\hline 
				Est. & 
				\multicolumn{2}{|c|}{Proportion} & 
				\multicolumn{5}{|c|}{Lognormal Model} & 
				\multicolumn{5}{|c}{Fr{\'e}chet Model} \\
				\hline
				{} & \multicolumn{1}{|c}{$(a_{1}, b_{1})$} &  
				\multicolumn{1}{c|}{$(a_{2}, b_{2})$} & 
				$\widehat{\theta}$ & $\widehat{\sigma}$ & 
				FIT & AIC & BIC &
				$\widehat{\beta}$ & $\widehat{\sigma}^{*}$ & 
				FIT & AIC & BIC \\
				\hline 
				\multicolumn{13}{c}{Original Data} \\
				\hline 
				MLE & 
				\multicolumn{2}{|c|}{--} & 
				22.80 & 0.83 & 0.1036 & 1446 & 1449 & 0.72 & 5.35 & 0.1277 & 1446 & 1448 \\
				\hline 
				$T_{1}$  & $(0, 0)$ & $(0, 0)$ & 22.80 & 0.83 & 0.1036 & 1446 & 1449 & 0.65 & 5.48 & 0.1168 & 1446 & 1449 \\
				$T_{2}$  & $(0, 1/30)$ & $(0, 1/30)$ & 22.79 & 0.80 & 0.1069 & 1446 & 1449 & 0.69 & 5.44 & 0.1199 & 1446 & 1449 \\
				$T_{3}$  & $(1/30, 1/30)$ & $(1/30, 1/30)$ & 22.77 & 0.85 & 0.1013 & 1446 & 1449 & 0.70 & 5.39 & 0.1222 & 1446 & 1449 \\
				$T_{4}$  & $(7/30, 7/30)$ & $(7/30, 7/30)$ & 22.78 & 0.77 & 0.1133 & 1447 & 1449 & 0.66 & 6.03 & 0.1219 & 1448 & 1451 \\
				\hline
				$T_{5}$  & $(0, 0)$ & $(0, 1/30)$ & 22.80 & 2.91 & 1.6402 & 1494 & 1497 & 0.54 & 5.83 & 0.1719 & 1454 & 1457 \\
				$T_{6}$  & $(1/30, 1/30)$ & $(0, 2/30)$ & 22.77 & 7.30 & 5.1024 & 1558 & 1561 & 0.58 & 5.73 & 0.1443 & 1450 & 1453 \\
				\hline
				$T_{7}$ & $(1/30, 1/30)$ & $(2/30, 0)$ & 22.77 & 0.88 & 0.1016 & 1446 & 1449 & 0.63 & 5.60 & 0.1186 & 1447 & 1450 \\
				$T_{8}$ & $(0, 3/30)$ & $(0, 0)$ & 22.80 & 0.90 & 0.1107 & 1447 & 1449 & 0.60 & 5.64 & 0.1298 & 1448 & 1451 \\
				$T_{9}$ & $(4/30, 5/30)$ & $(5/30, 2/30)$ & 22.76 & 0.87 & 0.1026 & 1446 & 1449 & 0.66 & 5.73 & 0.1114 & 1447 & 1449 \\
				$T_{10}$ & $(7/30, 15/30)$ & $(15/30, 7/30)$ & 22.78 & 0.78 & 0.1108 & 1447 & 1449 & 0.67 & 6.00 & 0.1250 & 1447 & 1450 \\
				\hline
				\multicolumn{13}{c}{Modified Data} \\
				\hline 
				MLE & 
				\multicolumn{2}{|c|}{--} & 
				{22.88} &  1.10 & 0.2932 & 1467 & 1470 & 0.77 & 5.47 & 0.1896 & 1456 & 1459 \\
				\hline 
				$T_{1}$  & {$(0, 0)$} & $(0, 0)$ &  {22.88} &  1.10 & 0.2932 & 1467 & 1470 & 0.86 & 5.26 & 0.2121 & 1457 & 1460 \\
				$T_{2}$  & $(0, 1/30)$ & $(0, 1/30)$ &  22.79 &  0.80 &  0.1838 & 1475 & 1478 & 0.69 & 5.44 & 0.1813 & 1457 & 1460 \\
				$T_{3}$  & {$(1/30, 1/30)$} & $(1/30, 1/30)$ & {22.77} &   0.85 &  0.1781 & 1472 & 1475 & 0.70 & 5.39 & 0.1818 & 1457 & 1460 \\
				$T_{4}$  & $(7/30, 7/30)$ & $(7/30, 7/30)$ & 22.78 & 0.77 & 0.1900 & 1478 & 1480 & 0.66 & 6.03 & 0.1846 & 1459 & 1462 \\
				\hline 
				$T_{5}$  & $(0, 0)$ & $(0, 1/30)$ & 22.88 & 3.86 & 2.3122 & 1515 & 1518 & 1.50 & 3.62 & 0.7212 & 1476 & 1479 \\
				$T_{6}$  & {$(1/30, 1/30)$} & $(0, 2/30)$ & {22.77}   &  7.30  & 5.0255 & 1552 & 1554 & 0.58 & 5.73 &  0.2211 & 1463 & 1466 \\
				\hline 
				$T_{7}$  & {$(1/30, 1/30)$} & $(2/30, 0)$ &   {22.77} &  1.41 &   0.4813 & 1471 & 1474 & 1.00 & 4.61 & 0.3133 & 1461 & 1464 \\
				$T_{8}$  & $(0, 3/30)$ & $(0, 0)$ &   22.87 &  1.27 &   0.3887 &  1468 & 1471 & 0.85 & 5.27 & 0.2112 & 1457 & 1460 \\
				$T_{9}$ & $(4/30, 5/30)$ & $(5/30, 2/30)$ & 22.76 & 0.87 & 0.1793 & 1472 & 1474 & 0.66 & 5.73 & 0.1766 & 1458 & 1461 \\
				$T_{10}$ & $(7/30, 15/30)$ & $(15/30, 7/30)$ & 22.78 & 0.78 & 0.1876 & 1477 & 1479 & 0.67 & 6.00 & 0.1844 & 1459 & 1461 \\
				\hline 
			\end{tabular}
			{\sc Note:} 
			Est. stands for Estimators. 
			Fr{\'e}chet estimated scale parameter
			$\widehat{\sigma} = \widehat{\sigma}^{*} \times 10^{9}$.
		}
	\end{table}
	
	Several conclusions emerge from this analysis. First, the goodness-of-fit (GOF) statistics for most robust MTM estimators remain stable under data modification, provided the trimming proportions are nonzero. In contrast, the MLE fit changes substantially. The GOF values are notably higher for $T_{5}$ and $T_{6}$ under both models. This may be attributed, as discussed in Note~\ref{note:DiffRel2}, to the strictly positive support of the sample data and the fitted lognormal model, suggesting that trimming based on inequality~\eqref{eqn:abCondition3} is more appropriate than inequality~\eqref{eqn:abCondition2}, which assumes approximate symmetry about the origin.
	
	Second, the increase in GOF is more pronounced for $T_{7}$ than for MLE when comparing original and modified data. This is expected, as $T_{7}$ involves no right-tail trimming for the second moment, allowing the inflated maximum value to influence the fit.
	
	Third, under $T_{2}$, $T_{3}$, $T_{4}$, $T_{9}$, and $T_{10}$, parameter estimates for both models remain unchanged between original and modified data, reflecting the robustness of MTM. Notably, $T_{10}$—which uses disjoint middle portions of the data for the two moments—produces low GOF values under the Fréchet model, even with data contamination.
	
	Fourth, the Fréchet tail index estimate $\widehat{\beta}$ is consistently below one, except for $T_{5}$ and $T_{7}$ under the modified data, suggesting a lighter right tail in the sample. This supports the higher $p$-value observed for the lognormal fit compared to the Fréchet fit. 
	Overall, the lognormal model appears more suitable for the original data, while the robust Fréchet model is better suited when the data contain a large outlier.
	
	\section{Conclusion}
	\label{sec:Conclusion}
	
	This paper introduced a general and flexible framework
	for robust parametric estimation based 
	on trimmed $L$-moments (MTM), 
	allowing distinct trimming proportions for different moments. 
	The proposed approach extends traditional 
	trimmed moment methods by enabling asymmetric 
	and moment-specific trimming strategies, 
	which improve robustness against outliers 
	and model misspecification without 
	sacrificing computational tractability.
	Estimators derived under this framework 
	maintain closed-form expressions
	and avoid iterative optimization, 
	making them suitable for large-scale data analysis. 
	We derived their asymptotic properties
	under the general theory of $L$-statistics 
	and provided analytical variance 
	expressions to support comparative efficiency analysis.
	
	Simulation studies across various 
	scenarios demonstrate that 
	the proposed estimators offer 
	strong finite-sample performance
	and effectively balance robustness and efficiency. 
	The flexibility to assign distinct trimming proportions
	to different moments enhances adaptability 
	to the data structure and contamination patterns. 
	This advantage is especially evident in asymmetric
	or heavy-tailed settings,
	where the general MTM consistently outperforms classical 
	and symmetric-trimming methods in
	terms of bias and mean squared error.
	
	The practical utility of the proposed methodology
	was further validated using a real-world dataset 
	of the 30 most damaging hurricanes in the United States.
	Both lognormal and Fréchet models were fitted using 
	MLE and several MTM variants. 
	The results confirmed that MLE is highly sensitive
	to data perturbations, 
	while MTM estimates remained stable and interpretable.
	When the largest loss was inflated by a factor of 10,
	the properly designed MTM estimators remained consistent, 
	whereas the MLE fit deteriorated considerably, 
	especially under the Fréchet model. 
	Moreover, goodness-of-fit metrics 
	(e.g., AIC, BIC, and empirical quantile deviation)
	reinforced the robustness and adaptability 
	of the MTM framework in both original and modified datasets.
	
	Overall, 
	the proposed general MTM approach offers
	a simple yet powerful extension of $L$-estimation techniques, 
	making it a valuable tool for robust inference 
	in the presence of contamination or heavy tails.
	Future work may include extending the framework
	to mixture models, 
	multivariate distributions, incorporating covariate information, 
	or applying the methodology to other domains
	such as environmental risk, 
	finance, and reliability analysis.
	
	\newpage 
	
	\baselineskip 4.80mm
	\bibliographystyle{apalike}

	\newpage 
	
	\begin{appendices}
		\section{Proofs}
		\label{sec:Appendix1}
		
		{\sc Proof of Proposition \ref{prop:EtaRealtion1}:}
		Define 
		\[
		U_{\tau} 
		\sim 
		\mbox{Uniform} \lp a_{\tau}, \ol{b}_{\tau} \rp
		\quad 
		\mbox{and} 
		\quad 
		W_{\tau} 
		:= 
		F_{0}^{-1} \lp U_{\tau} \rp, 
		\quad 
		\tau \in \{ i, j \}.
		\]
		Then, for any positive integer $k$, 
		it follows that 
		\[
		c_{k}
		\lp 
		a_{\tau}, \ol{b}_{\tau}
		\rp 
		= 
		\E 
		\lb 
		W_{\tau}^{k}
		\rb.
		\]
		
		\begin{itemize} 
			\item[(i)] 
			Thus, we get 
			\begin{align}
				\eta 
				\lp 
				a_{i},\ol{b}_{j}
				\rp 
				& = 
				c_{1}^{2} 
				\lp 
				a_{i}, \ol{b}_{i}
				\rp 
				-
				2 
				c_{1} 
				\lp 
				a_{i}, \ol{b}_{i}
				\rp 
				c_{1} 
				\lp 
				a_{j}, \ol{b}_{j}
				\rp 
				+ 
				c_{2} 
				\lp 
				a_{j}, \ol{b}_{j}
				\rp 
				\nonumber \\
				& = 
				\lp 
				\E 
				\lb 
				W_{i}
				\rb
				\rp^{2}
				-
				2 \, 
				\E 
				\lb 
				W_{i}
				\rb \, 
				\E 
				\lb 
				W_{j}
				\rb
				+ 
				\E 
				\lb 
				W_{j}^{2}
				\rb 
				\nonumber \\
				& = 
				\lp 
				\E 
				\lb 
				W_{i}
				\rb
				- 
				\E 
				\lb 
				W_{j}
				\rb
				\rp^{2}
				+
				\E 
				\lb 
				W_{j}^{2}
				\rb 
				-
				\lp 
				\E 
				\lb 
				W_{j}
				\rb
				\rp^{2} 
				\nonumber \\
				& = 
				\lp 
				\E 
				\lb 
				W_{i}
				\rb
				- 
				\E 
				\lb 
				W_{j}
				\rb
				\rp^{2}
				+
				\Var 
				\lb 
				W_{j}
				\rb. 
				\label{eqn:Eta12Positive1}
			\end{align}
			Since $W_{j}$ is a non-degenerate random variable
			giving $\Var \lb W_{j} \rb > 0$
			and 
			\(
			\lp 
			\E 
			\lb 
			W_{i}
			\rb
			- 
			\E 
			\lb 
			W_{j}
			\rb
			\rp^{2}
			\ge 
			0,
			\)
			Eq.~\eqref{eqn:Eta12Positive1} takes the form 
			\[
			\eta 
			\lp 
			a_{i},\ol{b}_{j}
			\rp 
			=
			\lp 
			\E 
			\lb 
			W_{i}
			\rb
			- 
			\E 
			\lb 
			W_{j}
			\rb
			\rp^{2}
			+
			\Var 
			\lb 
			W_{j}
			\rb
			>
			0.
			\]
			
			\item[(ii)]
			With the given inequality \eqref{eqn:abCondition1},
			$U_{j}$ is smaller than $U_{i}$ 
			in stochastic order
			\citep[see, e.g.,][Ch 1]{MR2265633}, 
			i.e., 
			$U_{j} \le_{\mbox{st}} U_{i}$.
			Being a quantile function of the standard location-scale
			family of distributions, $F_{0}^{-1}$ is strictly 
			increasing, then again it follows that 
			\citep[][Theorem 1.A.3]{MR2265633}, 
			$W_{j} \le_{\mbox{st}} W_{i}$.
			For odd positive integer $k$, 
			$g(x) = x^{k}$ 
			is strictly an increasing function,
			giving $W_{j}^{k} \le_{\mbox{st}} W_{i}^{k}$.
			Thus, 
			\begin{align}
				\label{eqn:cKRelation1}
				\E 
				\lb 
				W_{j}^{k}
				\rb 
				& \le 
				\E 
				\lb 
				W_{i}^{k}
				\rb, 
				\quad 
				\mbox{i.e.,}
				\quad 
				c_{k}
				\lp 
				a_{j}, \ol{b}_{j}
				\rp 
				\le 
				c_{k}
				\lp 
				a_{i}, \ol{b}_{i}
				\rp.
			\end{align} 
			
			\item[(iii)]
			From $(i)$, it immediately follows that 
			$0 < \eta_{r}$.
			Further, we know that  
			$\eta \left(a_{j}, \ol{b}_{j} \right) = \Var[W_{j}]$.
			Thus from Eq.~\eqref{eqn:Eta12Positive1}, 
			it follows that 
			\begin{align*}
				\eta_{r} 
				& = 
				\dfrac{\eta(a_j,\ol{b}_{j})}
				{\eta(a_{i},\ol{b}_j)}
				= 
				\dfrac{\Var[W_{j}]}
				{\lp 
					\E 
					\lb 
					W_{i}
					\rb
					- 
					\E 
					\lb 
					W_{j}
					\rb
					\rp^{2}
					+
					\Var 
					\lb 
					W_{j}
					\rb}
				\le 
				1.
			\end{align*}
			
			\item[(iv)]
			We have
			\begin{eqnarray*}
				& & 
				c_2
				\lp 
				a_j, \overline{b}_{j}
				\rp 
				\ge
				\eta_r \,
				c_1^2
				\lp 
				a_i, \overline{b}_i
				\rp \\
				\iff 
				& & 
				c_2
				\lp 
				a_j, \overline{b}_{j}
				\rp 
				\, 
				\eta
				\lp 
				a_i, \overline{b}_{j}
				\rp 
				\ge 
				c_1^2
				\lp 
				a_i, \overline{b}_{i}
				\rp 
				\, 
				\eta
				\lp 
				a_j, \overline{b}_{j}
				\rp \\
				\iff 
				& & 
				c_2
				\lp 
				a_j, \overline{b}_{j}
				\rp 
				\, 
				\lp 
				c_{1}^{2}(a_{i},\ol{b}_{i})
				- 
				2 c_{1}(a_{i},\ol{b}_{i}) c_1(a_{j},\ol{b}_{j}) 
				+ 
				c_{2}(a_{j},\ol{b}_j)
				\rp \\
				& & \qquad 
				\ge 
				c_1^2
				\lp 
				a_i, \overline{b}_{i}
				\rp 
				\, 
				\lp 
				c_2 \lp a_{j}, \ol{b}_{j} \rp
				-
				c_1^2 \lp a_{j}, \ol{b}_{j} \rp
				\rp \\
				\iff 
				& & 
				c_{2}^{2} \lp a_{j}, \ol{b}_{j} \rp
				-
				2 \, 
				c_{1} (a_{i},\ol{b}_{i}) \, 
				c_{1}(a_{j},\ol{b}_{j}) \, 
				c_{2}(a_{j},\ol{b}_{j})
				+ 
				c_{1}^{2}(a_{i},\ol{b}_{i}) \, 
				c_{1}^{2}(a_{j},\ol{b}_{j}) \\
				\iff 
				& & 
				\lp 
				c_{2} \lp a_{j}, \ol{b}_{j} \rp
				-
				c_{1}(a_{i},\ol{b}_{i}) \, 
				c_{1}(a_{j},\ol{b}_{j})
				\rp^{2}
				\ge 
				0,
			\end{eqnarray*}
			a valid inequality as desired. 
			
			\item[(v)]
			For $\theta = 0$ and $\sigma = 1$, 
			this converts to the inequality given in $(iv)$. 
			Further, if $\theta$ is positive,
			then this inequality again 
			follows immediately from $(iv)$. 
			But in general, it follows that 
			\begin{eqnarray*}
				& & \quad 
				\eta 
				\lp 
				a_{i}, \ol{b}_{j}
				\rp 
				T_{2} 
				- 
				\eta 
				\lp 
				a_{j}, \ol{b}_{j}
				\rp
				T_{1}^{2} \\
				& & 
				=
				\scalebox{2}{(}
				c_{2} 
				\lp 
				a_{j}, \ol{b}_{j} 
				\rp 
				\sigma 
				-
				c_{1} 
				\lp 
				a_{i}, \ol{b}_{i} 
				\rp 
				c_{1} 
				\lp 
				a_{j}, \ol{b}_{j} 
				\rp 
				\sigma 
				+ 
				c_{1} 
				\lp 
				a_{j}, \ol{b}_{j} 
				\rp 
				\theta 
				- 
				c_{1} 
				\lp 
				a_{i}, \ol{b}_{i} 
				\rp 
				\theta 
				\scalebox{2}{)}^{2} \\
				& & 
				= 
				\scalebox{2}{(}
				\sigma 
				\lp 
				c_{2}(a_{j}, \ol{b}_{j})
				- 
				c_{1}(a_{i}, \ol{b}_{i}) \, 
				c_{1}(a_{j}, \ol{b}_{j})
				\rp 
				+ 
				\theta 
				\lp
				c_{1}(a_{j}, \ol{b}_{j})
				-
				c_{1}(a_{i}, \ol{b}_{i})
				\rp
				\scalebox{2}{)}^{2} \\
				& & 
				\ge 
				0. 
			\end{eqnarray*}
			
			Thus, 
			\begin{eqnarray} 
				\label{eqn:T1T2Rel1}
				\eta 
				\lp 
				a_{i}, \ol{b}_{j}
				\rp 
				T_{2} 
				- 
				\eta 
				\lp 
				a_{j}, \ol{b}_{j}
				\rp
				T_{1}^{2}
				& & 
				\ge 
				0 
				\quad 
				\implies 
				\quad 
				T_{2} 
				- 
				\eta_{r}
				T_{1}^{2} 
				\ge 
				0,
			\end{eqnarray}
			as $\eta(a_{i}, \ol{b}_{j}) > 0$
			from $(i)$.
			\qed 
		\end{itemize}
		
		{\sc Proof of Corollary \ref{cor:ZetaRealtion1}:}
		$(i)$, $(iii)$, $(iv)$, and $(v)$  
		follow by similar arguments as in Proposition \ref{prop:EtaRealtion1}.  
		Note that $\log(-\log (u))$ is a 
		decreasing function of $u \in (0,1)$, 
		which establishes $(ii)$.  
		\qed 
		
		\section{Asymptotic Covariance Matrix Entries}
		\label{sec:Appendix2}
		
		Here we summarize the simplified single-integral
		expressions for the asymptotic variance-covariance 
		entries corresponding to the parametric examples 
		discussed in Section~\ref{sec:ParExamples}.
		
		By applying Theorem~\ref{thm:MTM_Var1}
		under the trimming inequality \eqref{eqn:abCondition2}, 
		the definitions of the notations
		\(\Lambda_{ijk}\), 
		for \(1 \le i, j \le 2\) 
		and \(1 \le k \le 3\), 
		as used in
		Eqs.~\eqref{eqn:MTM_Sigma112}–\eqref{eqn:MTM_Sigma222}, 
		are given below:
		\begin{eqnarray*}
			\Lambda_{111}
			& = & 
			\Gamma(1,1) 
			\int_{a_{1}}^{\ol{b}_{1}}\int_{a_{1}}^{\ol{b}_{1}}
			K(w, v) \, 
			d F_{0}^{-1}(v) \, 
			d F_{0}^{-1}(w)
			\, dv \, dw  \\
			& = & 
			\Gamma(1,1)
			\left\{ 
			a_{1} 
			\ol{a}_{1} 
			\lb 
			F_{0}^{-1}\left(a_{1}\right)
			\rb^{2}
			+
			b_{1} 
			\ol{b}_{1}
			\lb 
			F_{0}^{-1}\left(\ol{b}_{1}\right)
			\rb^{2}
			-
			2\,a_{1}\,b_{1}\,
			F_{0}^{-1}\left(a_{1}\right)
			F_{0}^{-1}\left(\ol{b}_{1}\right)
			\right. 
			\nonumber \\ 
			& & 
			\left.
			- 
			2
			\lp
			1 - a_{1} - b_{1}
			\rp 
			\lb 
			a_{1}\,F_{0}^{-1}\left(a_{1}\right)
			+ 
			b_{1}\,F_{0}^{-1}\left(\ol{b}_{1}\right)
			\rb
			c_{1}(a_{1},\ol{b}_{1}) 
			\right. 
			\nonumber \\ 
			& & 
			\left. 
			-
			\lb 
			\Gamma(1,1)
			\rb^{-1}
			c_{1}^{2}(a_{1},\ol{b}_{1}) 
			+ 
			\lp
			1 - a_{1} - b_{1}
			\rp
			c_{2}(a_{1},\ol{b}_{1})
			\right\}, \\ 
			\Lambda_{121}
			& = & 
			\Gamma(1,2) 
			\int_{a_{1}}^{\ol{b}_{1}}
			\int_{a_{2}}^{\ol{b}_{2}}
			K(w, v) \, 
			d F_{0}^{-1}(v) \, 
			d F_{0}^{-1}(w)
			\, dv \, dw \\
			& = & 
			\Gamma(1,2)
			\scalebox{2}{\{}
			a_{2}\, \ol{a}_{1} \, 
			F_{0}^{-1}\left(a_{1}\right)\,
			F_{0}^{-1}\left(a_{2}\right) 
			+
			b_{1} \,
			\ol{b}_{2} \,
			F_{0}^{-1}\left(\ol{b}_{1}\right)\,
			F_{0}^{-1}\left(\ol{b}_{2}\right) \\
			& & 
			-
			a_{1}\,b_{2}\,
			F_{0}^{-1}\left(a_{1}\right)\,
			F_{0}^{-1}\left(\ol{b}_{2}\right) 
			-
			a_{2}\,b_{1}\,
			F_{0}^{-1}\left(a_{2}\right) \, 
			F_{0}^{-1}\left(\ol{b}_{1}\right) \\
			& & 
			-
			\lp 
			1 - a_{1} - b_{2}
			\rp 
			\lb 
			2a_{1}\,F_{0}^{-1}\left(a_{1}\right)
			+
			b_{1}\,F_{0}^{-1}\left(\ol{b}_{1}\right)
			+
			b_{2}\,F_{0}^{-1}\left(\ol{b}_{2}\right)
			\rb 
			c_{1} \lp a_{1}, \ol{b}_{2} \rp  \\
			& & 
			-
			\lp 
			1 - a_{1} - b_{2}
			\rp^2
			c_{1}^{2} \lp a_{1}, \ol{b}_{2} \rp 
			+
			\lp 
			1 - a_{1} - b_{2}
			\rp
			c_{2} \lp a_{1}, \ol{b}_{2} \rp \\
			& & 
			+
			\lp 
			1 - a_{1} - b_{1}
			\rp 
			\lb 
			a_{1}\,F_{0}^{-1}\left(a_{1}\right)
			-
			a_{2}\,F_{0}^{-1}\left(a_{2}\right)
			\rb 
			c_{1} \lp a_{1}, \ol{b}_{1} \rp \\ 
			& &
			+
			\lp a_{1} - a_{2} \rp 
			\scalebox{1.5}{\{}
			\ol{a}_{1} \, 
			F_{0}^{-1}\left(a_{1}\right)
			-
			b_{1} 
			F_{0}^{-1}\left(\ol{b}_{1}\right)
			-
			\lp 1 - a_{1} - b_{1} \rp 
			c_{1} \lp a_{1}, \ol{b}_{1} \rp 
			\scalebox{1.5}{\}}
			c_{1} \lp a_{2}, a_{1} \rp\\ 
			& & 
			+ 
			\lp b_{2} - b_{1} \rp 
			\scalebox{1.5}{\{}
			\ol{b}_{2} \, F_{0}^{-1}\left(\ol{b}_{2}\right)
			-
			a_{1}
			F_{0}^{-1}\left(a_{1}\right)
			-
			\lp 
			1 - a_{1} - b_{2}
			\rp
			c_{1} \lp a_{1}, \ol{b}_{2} \rp 
			\scalebox{1.5}{\}} 
			c_{1} \lp \ol{b}_{2}, \ol{b}_{1} \rp
			\scalebox{2}{\}} , \\
			\Lambda_{122}
			& = & 
			\Gamma(1,2)
			\int_{a_{1}}^{\ol{b}_{1}}
			\int_{a_{2}}^{\ol{b}_{2}}
			K(w, v) \, 
			F_{0}^{-1}(v) \, 
			d F_{0}^{-1}(v) \, 
			d F_{0}^{-1}(w)
			\, dv \, dw \\
			& = & 
			\dfrac{\Gamma(1,2)}{2}
			\scalebox{2}{\{}
			a_{2}\, \ol{a}_{1} \,
			F_{0}^{-1}\left(a_{1}\right)\,
			\lb 
			F_{0}^{-1}\left(a_{2}\right)
			\rb^2 
			+
			b_{1}\, \ol{b}_{2} \, 
			F_{0}^{-1}\left(\ol{b}_{1}\right)\,
			\lb 
			F_{0}^{-1}\left(\ol{b}_{2}\right)
			\rb^2 \\
			& & 
			-a_{2}\,b_{1}\,F_{0}^{-1}\left(\ol{b}_{1}\right)\,
			\lb 
			F_{0}^{-1}\left(a_{2}\right)
			\rb^2 
			-
			a_{1}\,b_{2}\, 
			F_{0}^{-1}\left(a_{1}\right) \, 
			\lb 
			F_{0}^{-1}\left(\ol{b}_{2}\right)
			\rb^2 \\
			& & 
			-
			\lp 
			1 - a_{1} - b_{2}
			\rp 
			\lb 
			a_{1}\,
			\lb 
			F_{0}^{-1}\left(a_{1}\right)
			\rb^2
			+
			b_{2}\,
			\lb 
			F_{0}^{-1}\left(\ol{b}_{2}\right)
			\rb^2
			\rb 
			c_{1}(a_{1}, \ol{b}_{2}) \\
			& & 
			-
			\lp 
			1 - a_{1} - b_{2}
			\rp 
			\lb 
			a_{1}\,F_{0}^{-1}\left(a_{1}\right)
			+
			b_{1}\,F_{0}^{-1}\left(\ol{b}_{1}\right)
			\rb 
			c_{2}(a_{1}, \ol{b}_{2}) \\
			& & 
			- 
			\lp 
			1 - a_{1} - b_{2}
			\rp^{2}
			c_{1}(a_{1}, \ol{b}_{2}) \, 
			c_{2}(a_{1}, \ol{b}_{2}) \\
			& & 
			+
			\lp 
			1 - a_{1} - b_{2}
			\rp
			c_{3}(a_{1}, \ol{b}_{2}) \\
			& & 
			+
			\lp 
			1 - a_{1} - b_{1}
			\rp
			\lb 
			a_{1}\,
			\lb 
			F_{0}^{-1}\left(a_{1}\right)
			\rb^2
			-
			a_{2}\, 
			\lb 
			F_{0}^{-1}\left(a_{2}\right)
			\rb^2
			\rb
			c_{1}(a_{1}, \ol{b}_{1}) \\
			& & 
			+ 
			\lp a_{1} - a_{2} \rp
			\scalebox{1.5}{\{}
			\ol{a}_{1} \, 
			F_{0}^{-1}\left(a_{1}\right)
			-
			b_{1}\,F_{0}^{-1}\left(\ol{b}_{1}\right)
			-
			\lp 1 - a_{1} - b_{1} \rp
			c_{1}(a_{1}, \ol{b}_{1})
			\scalebox{1.5}{\}}
			c_{2} \lp a_{2}, a_{1} \rp \\
			& & 
			+
			\lp b_{2} - b_{1} \rp
			\scalebox{1.5}{\{} 
			\ol{b}_{2} \, 
			\lb
			F_{0}^{-1}\left(\ol{b}_{2}\right)
			\rb^2
			-a_{1}\,
			\lb 
			F_{0}^{-1}\left(a_{1}\right)
			\rb^2
			-
			\lp 1 - a_{1} - b_{2} \rp
			c_{2}(a_{1}, \ol{b}_{2})
			\scalebox{1.5}{\}} 
			c_{1}(\ol{b}_{2}, \ol{b}_{1})
			\scalebox{2}{\}} , \\
			\Lambda_{221}
			& = & 
			\Lambda_{111},
			\text{ with $a_{1}$ replaced by $a_{2}$ 
				and $b_{1}$ replaced by $b_{2}$}, \\
			\Lambda_{222}
			& = & 
			\Gamma(2,2)
			\int_{a_{2}}^{\ol{b}_{2}}
			\int_{a_{2}}^{\ol{b}_{2}}
			K(w, v) \, 
			F_{0}^{-1}(w) \, 
			d F_{0}^{-1}(v) \, 
			d F_{0}^{-1}(w)
			\, dv \, dw \\
			& = & 
			\dfrac{\Gamma(2,2)}{2}
			\scalebox{2}{\{}
			a_{2} \, \ol{a}_{2} 
			\left[ 
			F_{0}^{-1}(a_{2}) 
			\right]^3 
			+ 
			b_{2} \, \ol{b}_{2}
			\left[
			F_{0}^{-1}(\ol{b}_{2}) 
			\right]^3 \\
			& & 
			- a_{2} \, b_{2} 
			F_{0}^{-1}(a_{2}) F_{0}^{-1}(\ol{b}_{2}) 
			\left[
			F_{0}^{-1}(a_{2}) 
			+ 
			F_{0}^{-1}(\ol{b}_{2}) 
			\right] \\
			& & 
			- 
			(1 - a_{2} - b_{2}) 
			\left[ 
			a_{2} 
			\left[ 
			F_{0}^{-1}(a_{2})
			\right]^2 
			+
			b_{2} 
			\left[
			F_{0}^{-1}(\ol{b}_{2})
			\right]^2 \right] 
			c_{1}(a_{2}, \ol{b}_{2}) \\
			& & 
			-
			(1 - a_{2} - b_{2}) 
			\left[ 
			a_{2} F_{0}^{-1}(a_{2}) 
			+ 
			b_{2} F_{0}^{-1}(\ol{b}_{2})
			\right] 
			c_{2}(a_{2}, \ol{b}_{2}) \\
			& & 
			- 
			\lb 
			\Gamma(2,2)
			\rb^{-1}
			c_{1}(a_{2}, \ol{b}_{2}) \, 
			c_{2}(a_{2}, \ol{b}_{2})
			+ 
			(1 - a_{2} - b_{2}) \,
			c_{3}(a_{2}, \ol{b}_{2})
			\scalebox{2}{\}}, \\
			\Lambda_{223}
			& = & 
			\Gamma(2,2)
			\int_{a_{2}}^{\ol{b}_{2}}
			\int_{a_{2}}^{\ol{b}_{2}}
			K(w, v) \, 
			F_{0}^{-1}(w) \, 
			F_{0}^{-1}(v) \, 
			d F_{0}^{-1}(v) \, 
			d F_{0}^{-1}(w)
			\, dv \, dw \\ 
			& = & 
			\dfrac{\Gamma(2,2)}{4}
			\scalebox{2}{\{}
			a_{2} \, \ol{a}_{2}
			\left[ 
			F_{0}^{-1}(a_{2}) 
			\right]^4 
			+ 
			b_{2} \, \ol{b}_{2}
			\left[ 
			F_{0}^{-1}(\ol{b}_{2})
			\right]^4 
			- 
			2a_{2} \, b_{2} 
			\left[ 
			F_{0}^{-1}(a_{2}) 
			\right]^2 
			\left[ 
			F_{0}^{-1}(\ol{b}_{2}) 
			\right]^2 \\
			& & 
			- 
			2(1 - a_{2} - b_{2}) 
			\left[ 
			a_{2} 
			\left[ 
			F_{0}^{-1}(a_{2})
			\right]^2
			+ 
			b_{2} 
			\left[
			F_{0}^{-1}(\ol{b}_{2})
			\right]^2 
			\right] 
			c_{2}(a_{2},\ol{b}_{2}) \\
			& & 
			-
			\lb 
			\Gamma(2,2)
			\rb^{-1}
			c_{2}^{2}(a_{2},\ol{b}_{2}) 
			+ 
			(1 - a_{2} - b_{2}) \, 
			c_{4} \lp a_{2}, \ol{b}_{2} \rp
			\scalebox{2}{\}}.
		\end{eqnarray*}
		
		\begin{note}
			\label{note:TrimIneFlip1}
			The notations
			\(\Lambda_{ijk}\), 
			for \(1 \le i, j \le 2\)
			and \(1 \le k \le 3\),
			can be similarly evaluated
			under the trimming inequality 
			\eqref{eqn:abCondition1Flip},
			i.e.,  
			\eqref{eqn:abCondition3},
			as noted in 
			Note~\ref{note:ChangeInequality1}.
			\qed 
		\end{note} 
		
		\label{page:PsiSimplified1}
		It follows that 
		Eq.~\eqref{eqn:FH1} is equivalent to 
		Eq.~\eqref{eqn:LSH1}, and 
		Eq.~\eqref{eqn:FH2} is equivalent to 
		Eq.~\eqref{eqn:LSH2D}, under the substitutions 
		\(\log(\sigma) \mapsto \theta\), 
		\(\beta \mapsto -\sigma\), and 
		\(\Delta(u) \mapsto F_{0}^{-1}(u)\)
		for \(u \in (0,1)\). 
		Consequently, 
		the expressions for \(\Psi_{ijk}\), 
		with \(1 \le i, j \le 2\) and \(1 \le k \le 3\), 
		as defined in 
		Eqs.~\eqref{eqn:MTM_FSigma112}--\eqref{eqn:MTM_FSigma222},
		can be obtained by applying 
		Theorem~\ref{thm:MTM_Var1} under the trimming 
		inequality \eqref{eqn:abCondition2}, 
		following the structure of \(\Lambda_{ijk}\) 
		for the same index ranges, and are given below:
		\begin{eqnarray*}
			\Psi_{111}
			& = & 
			\Gamma(1,1) 
			\int_{a_{1}}^{\ol{b}_{1}}\int_{a_{1}}^{\ol{b}_{1}}
			\dfrac{K(w,v)}
			{v w \log(v) \, \log(w)}
			\, dv \, dw  \\
			& = & 
			\Gamma(1,1)
			\scalebox{2}{\{} 
			a_{1} \, \ol{a}_{1} 
			\lb 
			\Delta\left(a_{1}\right)
			\rb^2
			+ 
			b_{1} \, \ol{b}_{1} 
			\lb 
			\Delta\left(\ol{b}_{1}\right)
			\rb^2 
			-
			2 \, a_{1} \, b_{1} \, 
			\Delta \left(a_{1}\right) \, 
			\Delta \left(\ol{b}_{1}\right) \\
			& & 
			- 
			2
			\lp 
			1 - a_{1} - b_{1}
			\rp 
			\lb 
			a_{1}
			\Delta \left(a_{1}\right)
			+ 
			b_{1} 
			\Delta \left(\ol{b}_{1}\right)
			\rb 
			\kappa_{1} 
			\lp 
			a_{1}, \ol{b}_{1}
			\rp \\
			& & 
			-
			\lb 
			\Gamma(1,1)
			\rb^{-1}
			\kappa_{1}^{2} 
			\lp 
			a_{1}, \ol{b}_{1}
			\rp 
			+ 
			\lp 
			1 - a_{1} - b_{1}
			\rp
			\kappa_{2} 
			\lp 
			a_{1}, \ol{b}_{1}
			\rp
			\scalebox{2}{\}}, \\
			\Psi_{121}
			& = & 
			\Gamma(1,2) 
			\int_{a_{1}}^{\ol{b}_{1}} 
			\int_{a_{2}}^{\ol{b}_{2}} 
			\frac{K(w,v)}
			{v w \log{(v)} \log{(w)}}
			dv \, dw \\
			& = & 
			\Gamma(1,2)
			\scalebox{2}{\{}
			a_{2}\, \ol{a}_{1} \, 
			\Delta\left(a_{1}\right)\,
			\Delta\left(a_{2}\right) 
			+
			b_{1} \,
			\ol{b}_{2} \,
			\Delta\left(\ol{b}_{1}\right)\,
			\Delta\left(\ol{b}_{2}\right) \\
			& & 
			-
			a_{1}\,b_{2}\,
			\Delta\left(a_{1}\right)\,
			\Delta\left(\ol{b}_{2}\right) 
			-
			a_{2}\,b_{1}\,
			\Delta\left(a_{2}\right) \, 
			\Delta\left(\ol{b}_{1}\right) \\
			& & 
			-
			\lp 
			1 - a_{1} - b_{2}
			\rp 
			\lb 
			2a_{1}\,\Delta\left(a_{1}\right)
			+
			b_{1}\,\Delta\left(\ol{b}_{1}\right)
			+
			b_{2}\,\Delta\left(\ol{b}_{2}\right)
			\rb 
			\kappa_{1} \lp a_{1}, \ol{b}_{2} \rp  \\
			& & 
			-
			\lp 
			1 - a_{1} - b_{2}
			\rp^2
			\kappa_{1}^{2} \lp a_{1}, \ol{b}_{2} \rp 
			+
			\lp 
			1 - a_{1} - b_{2}
			\rp
			\kappa_{2} \lp a_{1}, \ol{b}_{2} \rp \\
			& & 
			+
			\lp 
			1 - a_{1} - b_{1}
			\rp 
			\lb 
			a_{1}\,\Delta\left(a_{1}\right)
			-
			a_{2}\,\Delta\left(a_{2}\right)
			\rb 
			\kappa_{1} \lp a_{1}, \ol{b}_{1} \rp \\ 
			& &
			+
			\lp a_{1} - a_{2} \rp 
			\scalebox{1.5}{\{}
			\ol{a}_{1} \, 
			\Delta\left(a_{1}\right)
			-
			b_{1} 
			\Delta\left(\ol{b}_{1}\right)
			-
			\lp 1 - a_{1} - b_{1} \rp 
			\kappa_{1} \lp a_{1}, \ol{b}_{1} \rp 
			\scalebox{1.5}{\}}
			\kappa_{1} \lp a_{2}, a_{1} \rp\\ 
			& & 
			+ 
			\lp b_{2} - b_{1} \rp 
			\scalebox{1.5}{\{}
			\ol{b}_{2} \, \Delta\left(\ol{b}_{2}\right)
			-
			a_{1}
			\Delta\left(a_{1}\right)
			-
			\lp 
			1 - a_{1} - b_{2}
			\rp
			\kappa_{1} \lp a_{1}, \ol{b}_{2} \rp 
			\scalebox{1.5}{\}} 
			\kappa_{1} \lp \ol{b}_{2}, \ol{b}_{1} \rp
			\scalebox{2}{\}} , \\
			\Psi_{122}
			& = & 
			\Gamma(1,2) 
			\int_{a_{1}}^{\ol{b}_{1}} 
			\int_{a_{2}}^{\ol{b}_{2}} 
			\frac{K(w,v) \log{( -\log(v))}}
			{v w \log{(v)} \log{(w)}
			}
			dv \, dw \\
			& = & 
			\dfrac{\Gamma(1,2)}{2}
			\scalebox{2}{\{}
			a_{2}\, \ol{a}_{1} \,
			\Delta\left(a_{1}\right)\,
			\lb 
			\Delta\left(a_{2}\right)
			\rb^2 
			+
			b_{1}\, \ol{b}_{2} \, 
			\Delta\left(\ol{b}_{1}\right)\,
			\lb 
			\Delta\left(\ol{b}_{2}\right)
			\rb^2 \\
			& & 
			-a_{2}\,b_{1}\,\Delta\left(\ol{b}_{1}\right)\,
			\lb 
			\Delta\left(a_{2}\right)
			\rb^2 
			-
			a_{1}\,b_{2}\, 
			\Delta\left(a_{1}\right) \, 
			\lb 
			\Delta\left(\ol{b}_{2}\right)
			\rb^2 \\
			& & 
			-
			\lp 
			1 - a_{1} - b_{2}
			\rp 
			\lb 
			a_{1}\,
			\lb 
			\Delta\left(a_{1}\right)
			\rb^2
			+
			b_{2}\,
			\lb 
			\Delta\left(\ol{b}_{2}\right)
			\rb^2
			\rb 
			\kappa_{1}(a_{1}, \ol{b}_{2}) \\
			& & 
			-
			\lp 
			1 - a_{1} - b_{2}
			\rp 
			\lb 
			a_{1}\,\Delta\left(a_{1}\right)
			+
			b_{1}\,\Delta\left(\ol{b}_{1}\right)
			\rb 
			\kappa_{2}(a_{1}, \ol{b}_{2}) \\
			& & 
			- 
			\lp 
			1 - a_{1} - b_{2}
			\rp^{2}
			\kappa_{1}(a_{1}, \ol{b}_{2}) \, 
			\kappa_{2}(a_{1}, \ol{b}_{2}) \\
			& & 
			+
			\lp 
			1 - a_{1} - b_{2}
			\rp
			c_{3}(a_{1}, \ol{b}_{2}) \\
			& & 
			+
			\lp 
			1 - a_{1} - b_{1}
			\rp
			\lb 
			a_{1}\,
			\lb 
			\Delta\left(a_{1}\right)
			\rb^2
			-
			a_{2}\, 
			\lb 
			\Delta\left(a_{2}\right)
			\rb^2
			\rb
			\kappa_{1}(a_{1}, \ol{b}_{1}) \\
			& & 
			+ 
			\lp a_{1} - a_{2} \rp
			\scalebox{1.5}{\{}
			\ol{a}_{1} \, 
			\Delta\left(a_{1}\right)
			-
			b_{1}\,\Delta\left(\ol{b}_{1}\right)
			-
			\lp 1 - a_{1} - b_{1} \rp
			\kappa_{1}(a_{1}, \ol{b}_{1})
			\scalebox{1.5}{\}}
			\kappa_{2} \lp a_{2}, a_{1} \rp \\
			& & 
			+
			\lp b_{2} - b_{1} \rp
			\scalebox{1.5}{\{} 
			\ol{b}_{2} \, 
			\lb
			\Delta\left(\ol{b}_{2}\right)
			\rb^2
			-a_{1}\,
			\lb 
			\Delta\left(a_{1}\right)
			\rb^2
			-
			\lp 1 - a_{1} - b_{2} \rp
			\kappa_{2}(a_{1}, \ol{b}_{2})
			\scalebox{1.5}{\}} 
			\kappa_{1}(\ol{b}_{2}, \ol{b}_{1})
			\scalebox{2}{\}}, \\
			\Psi_{221}
			& = & 
			\Psi_{111},
			\text{ with $a_{1}$ replaced by $a_{2}$ 
				and $b_{1}$ replaced by $b_{2}$}, \\ 
			\Psi_{222}
			& = & 
			\Gamma(2,2) 
			\int_{a_{2}}^{\ol{b}_{2}}
			\int_{a_{2}}^{\ol{b}_{2}} 
			\dfrac{ K(w,v) 
				\log{( -\log(v))}} 
			{{v w \log{(v)} \log{(w)}}} 
			dv \, dw \\
			& = & 
			\dfrac{\Gamma(2,2)}{2}
			\scalebox{2}{\{}
			a_{2} \, \ol{a}_{2} 
			\left[ 
			\Delta(a_{2}) 
			\right]^3 
			+ 
			b_{2} \, \ol{b}_{2}
			\left[
			\Delta(\ol{b}_{2}) 
			\right]^3 \\
			& & 
			- a_{2} \, b_{2} 
			\Delta(a_{2}) \Delta(\ol{b}_{2}) 
			\left[
			\Delta(a_{2}) 
			+ 
			\Delta(\ol{b}_{2}) 
			\right] \\
			& & 
			- 
			(1 - a_{2} - b_{2}) 
			\left[ 
			a_{2} 
			\left[ 
			\Delta(a_{2})
			\right]^2 
			+
			b_{2} 
			\left[
			\Delta(\ol{b}_{2})
			\right]^2 \right] 
			\kappa_{1}(a_{2}, \ol{b}_{2}) \\
			& & 
			-
			(1 - a_{2} - b_{2}) 
			\left[ 
			a_{2} \Delta(a_{2}) 
			+ 
			b_{2} \Delta(\ol{b}_{2})
			\right] 
			\kappa_{2}(a_{2}, \ol{b}_{2}) \\
			& & 
			- 
			\lb 
			\Gamma(2,2)
			\rb^{-1}
			\kappa_{1}(a_{2}, \ol{b}_{2}) \, 
			\kappa_{2}(a_{2}, \ol{b}_{2})
			+ 
			(1 - a_{2} - b_{2}) \,
			\kappa_{3}(a_{2}, \ol{b}_{2})
			\scalebox{2}{\}}, \\
			\Psi_{223}
			& = & 
			\Gamma(2,2) 
			\int_{a_{2}}^{\ol{b}_{2}}
			\int_{a_{2}}^{\ol{b}_{2}} 
			\dfrac{K(w,v) 
				\log{( -\log(v))} \log{( -\log(w))}} 
			{{v w \log{(v)} \log{(w)}}} 
			dv \, dw \\
			& = & 
			\dfrac{\Gamma(2,2)}{4}
			\scalebox{2}{\{}
			a_{2} \, \ol{a}_{2}
			\left[ 
			\Delta(a_{2}) 
			\right]^4 
			+ 
			b_{2} \, \ol{b}_{2}
			\left[ 
			\Delta(\ol{b}_{2})
			\right]^4 
			- 
			2a_{2} \, b_{2} 
			\left[ 
			\Delta(a_{2}) 
			\right]^2 
			\left[ 
			\Delta(\ol{b}_{2}) 
			\right]^2 \\
			& & 
			- 
			2(1 - a_{2} - b_{2}) 
			\left[ 
			a_{2} 
			\left[ 
			\Delta(a_{2})
			\right]^2
			+ 
			b_{2} 
			\left[
			\Delta(\ol{b}_{2})
			\right]^2 
			\right] 
			\kappa_{2}(a_{2},\ol{b}_{2}) \\
			& & 
			-
			\lb 
			\Gamma(2,2)
			\rb^{-1}
			\kappa_{2}^{2}(a_{2},\ol{b}_{2}) 
			+ 
			(1 - a_{2} - b_{2}) \, 
			\kappa_{4} \lp a_{2}, \ol{b}_{2} \rp
			\scalebox{2}{\}}.
		\end{eqnarray*}
		
		\begin{note}
			\label{note:TrimIneFlip2}
			As discussed in Notes~\ref{note:TrimIneFlip1} 
			and~\ref{note:ChangeInequality1}, 
			the notations \(\Psi_{ijk}\), 
			for \(1 \le i, j \le 2\) 
			and \(1 \le k \le 3\), 
			can be similarly evaluated under
			the trimming inequality \eqref{eqn:abCondition1Flip}, 
			that is, \eqref{eqn:abCondition3}.
			\qed
		\end{note}
		
	\end{appendices}
	
\end{document}